\documentclass[conference]{IEEEtran}
\IEEEoverridecommandlockouts 

\usepackage{cite}

\usepackage{url}

\usepackage[hidelinks]{hyperref}

\usepackage{amssymb}

\usepackage[utf8]{inputenc}
\usepackage{xspace}

\usepackage{enumitem}

\setlist[itemize]{noitemsep, topsep=0.2em, left=0.3em}

\usepackage{booktabs}
\usepackage{array}
\usepackage{marvosym}

\usepackage{amsmath}
\usepackage{bm}

\usepackage{hhline}

\usepackage[
  vlined
  ]{algorithm2e}

\usepackage{multicol}

\usepackage{colortbl}
\usepackage{tikz}
  \usetikzlibrary{arrows}
  \usetikzlibrary{calc}
  \usetikzlibrary{fit}
  \usetikzlibrary {shapes.symbols}

  \usetikzlibrary{shapes.geometric, arrows.meta, positioning, calc}

\usepackage{array}
\usepackage{multirow}

\usepackage{soul}
\usepackage{pgfplotstable}
\usepackage{tikz-cd}
\usepackage{pgfplots}
\pgfplotsset{compat=1.17}

\usepackage{subcaption}

\usepackage{soul}
\sethlcolor{black!10}

\usepackage{changes}

\def\BibTeX{{\rm B\kern-.05em{\sc i\kern-.025em b}\kern-.08em
    T\kern-.1667em\lower.7ex\hbox{E}\kern-.125emX}}

\IEEEpubid{
  \makebox[\columnwidth]{ 
    \parbox{\columnwidth}{ 
      \footnotesize
      This work has been accepted for publication at the {IEEE Conference on Secure and Trustworthy Machine Learning (SaTML 2026)}. The final version will be available on IEEE Xplore.
    }
  }
  \hspace{\columnsep} 
  \makebox[\columnwidth]{ } 
}

\begin{document}


\renewcommand{\paragraph}[1]{\subsubsection*{#1}}
\let\oldparagraph\paragraph 
\renewcommand{\paragraph}[1]{\vspace{0.5\baselineskip}\textbf{#1.}}
\newcommand{\para}[1]{\vspace{0.09\baselineskip}\textbf{#1.}}

\newcommand{\phasepara}[1]{\vspace{0.06\baselineskip}\textbf{#1:}}

\newtheorem{theorem}{Theorem}
\newtheorem{definition}{Definition}
\newtheorem{lemma}{Lemma}

\newcommand{\specialcell}[2][c]{%
  \begingroup
  \renewcommand{\arraystretch}{0.75} 
  \begin{tabular}[#1]{@{}c@{}}#2\end{tabular}
  \endgroup
}

\newcommand{\tocode}[1]{{\color{blue}[TOCODE: #1]}}

\newcommand{\question}[1]{{\color{orange!90!black}[Q: #1]}}
\newcommand{\answer}[1]{{\color{green!70!black}[A: #1]}}

\newcommand{\xx}[1]{{\iffalse\color{red}#1\fi}}
\newcommand{\xxx}[1]{{\color{red}#1}}
\newcommand{\tocheck}[1]{{\color{blue}#1}}
\newcommand{\tocheckhack}[1]{{\color{black}#1}}

\newcommand{\enx}[1]{\ensuremath{#1}\xspace}
\newcommand{\eps}{\ensuremath{\epsilon}\xspace}

\newcommand{\aggregator}{server\xspace}

\newcommand{\ttp}{Trusted Third Party\xspace}
\newcommand{\ttps}{Trusted Third Parties\xspace}

\newcommand{\adv}{\textit{A}\xspace}
\newcommand{\adversary}{adversary\xspace}

\newcommand{\tnss}{\ensuremath{(t,m)}-SS\xspace}
\newcommand{\tnssl}{\ensuremath{(t,m)}-secret sharing\xspace}

\newcommand{\nnss}{\ensuremath{(m,m)}-SS\xspace}
\newcommand{\nnssl}{\ensuremath{(m,m)}-secret sharing\xspace}

\newcommand{\sss}{Shamir's secret sharing\xspace}
\newcommand{\tss}{threshold secret sharing\xspace}

\newcommand{\cl}{CL\xspace}
\newcommand{\epcl}{CPCL\xspace}

\newcommand{\ot}{OL\xspace}
\newcommand{\ol}{OL\xspace}
\newcommand{\otl}{outsourced learning\xspace}
\newcommand{\otlcap}{Outsourced Learning\xspace}

\newcommand{\fl}{FL\xspace}

\newcommand{\gc}{{GC}\xspace}
\newcommand{\gcl}{garbled circuits\xspace}
\newcommand{\gcls}{garbled circuits\xspace}

\newcommand{\yao}{Yao's GC\xspace}
\newcommand{\yaol}{Yao's Garbled Circuits\xspace}

\newcommand{\nclients}{\enx{n}}
\newcommand{\nservers}{\enx{m}}

\newcommand{\cp}[1]{\enx{{CP_{#1}}}}
\newcommand{\cpl}{computing party\xspace}
\newcommand{\cpls}{computing parties\xspace}

\newcommand{\s}[1]{{\enx{{S_{#1}}}}}
\newcommand{\slong}{server\xspace}
\newcommand{\sln}[1]{server {#1}\xspace}
\newcommand{\sls}{servers\xspace}

\newcommand{\p}[1]{\enx{{P_{#1}}}}
\newcommand{\pl}{party\xspace}
\newcommand{\pln}[1]{\enx{party_{#1}}}
\newcommand{\pls}{parties\xspace}

\newcommand{\ip}[1]{\enx{{IP_{#1}}}}
\newcommand{\op}[1]{\enx{{OP_{#1}}}}
\newcommand{\ipl}{input party\xspace}
\newcommand{\ipln}[1]{\mathrm{input party {#1}}}
\newcommand{\ipls}{input parties\xspace}

\let\oldc\c
\renewcommand{\c}[1]{{\enx{{C_{#1}}}}}
\newcommand{\cln}[1]{client {#1}}
\newcommand{\cls}{clients\xspace}

\newcommand{\npartial}[1]{\enx{\psi_{#1}}}
\newcommand{\ncentral}{\enx{\psi}}

\let\olddo\do 
\renewcommand{\do}{\textit{DO}\xspace}
\newcommand{\dataowners}{data owners\xspace}
\newcommand{\dataowner}{data owner\xspace}

\newcommand{\Xmark}{{$\textcolor{red}{\mathbf{\times}}$}}
\newcommand{\Xmarktab}{\raisebox{-0.2ex}{\footnotesize$\textcolor{red}{\mathbf{\times}}$\normalsize}}

\newcommand{\doubledash}{{\normalsize$\mathbf{\textcolor{orange}{\--}}$\normalsize}}
\newcommand{\doubledashtab}{\raisebox{-0.3ex}{\normalsize$\mathbf{\textcolor{orange}{\--}}$\normalsize}}
\newcommand{\tickmarkertab}{\raisebox{-0.2ex}{\footnotesize$\mathbf{\textcolor{cyan!60!black}{\checkmark}}$\normalsize}}
\newcommand{\tickmarker}{{\normalsize$\mathbf{\textcolor{cyan!60!black}{\checkmark}}$\normalsize}}

\newcommand{\XmarkGood}{\large$\textcolor{cyan!60!black}{\mathbf{\times}}$\normalsize}
\newcommand{\questionmark}{\large$\mathbf{\textcolor{orange}{?}}$\normalsize}
\newcommand{\tickmarkerRed}{\large$\mathbf{\textcolor{red}{\checkmark}}$\normalsize}

\newcommand{\enhancemark}{{\normalsize$\textcolor{cyan!30!yellow}{\mathbf{\circledast}}$\normalsize}}
\newcommand{\enhancemarktab}{\raisebox{-0.2ex}{\normalsize$\textcolor{cyan!30!yellow}{\mathbf{\circledast}}$\normalsize}}
\newcommand{\enhancemarkexp}{}

\newcommand{\enhancemarkg}{\enhancemark}
\newcommand{\enhancemarkgtab}{\enhancemarktab}
\newcommand{\enhancemarkgexp}{\large\mathrel{\textcolor{cyan!30!yellow}{\circledast}}\normalsize}

\newcommand{\enhancemarko}{\enhancemark}
\newcommand{\enhancemarkotab}{\enhancemarktab}
\newcommand{\enhancemarkoexp}{\large\mathrel{\textcolor{cyan!30!yellow}{\circledast}}\normalsize}
\newcommand{\enhancemarkv}{\enhancemark}
\newcommand{\enhancemarkvtab}{\enhancemarktab}
\newcommand{\enhancemarkvexp}{}


\newcommand{\perturbinput}{\enx{\mathsf{\textcolor{blue}{PerturbInput}}}}
\newcommand{\perturbloss}{\enx{\mathsf{\textcolor{red}{PerturbLoss}}}}
\newcommand{\perturbgrad}{\enx{\mathsf{\textcolor{green!60!black}{PerturbGradient}}}}
\newcommand{\perturbout}{\enx{\mathsf{\textcolor{orange}{PerturbOutput}}}}
\newcommand{\perturblabel}{\enx{\mathsf{\textcolor{violet}{PerturbLabel}}}}

\newcommand{\pgrad}{\enx{\mathsf{\textcolor{green!60!black}{Grad.}}}}
\newcommand{\pout}{\enx{\mathsf{\textcolor{orange}{Out.}}}}
\newcommand{\plabel}{\enx{\mathsf{\textcolor{violet}{Label}}}}


\newcounter{numberedparagraph}
\newlength{\parindentoriginal}
\newenvironment{numberedparagraphs}
{
    \setcounter{numberedparagraph}{0}
    \renewcommand{\paragraph}[1]{\refstepcounter{numberedparagraph}\textbf{\thenumberedparagraph) \hspace{0.1em}##1.}}
}{
    \setcounter{numberedparagraph}{0}
}


\newcommand{\phaseformat}[1]{\enx{\mathsf{#1}}}
\newcommand{\setup}{\phaseformat{Setup}}
\newcommand{\compgrad}{\phaseformat{GradientCompute}}
\newcommand{\protection}{\phaseformat{Protect}}
\newcommand{\reveal}{\phaseformat{Reveal}}
\newcommand{\perturb}{\phaseformat{Perturb}}
\newcommand{\perturbbold}{\enx{\bm{\mathsf{Perturb}}}}

\newcommand{\noise}{\perturb}
\newcommand{\aggr}{\phaseformat{Aggregate}}
\newcommand{\updatemodel}{\phaseformat{Update}}
%
%
\newcommand{\centralnoise}{centralized noise sampling\xspace}
\newcommand{\partialnoise}{partial noise aggregation\xspace}
\newcommand{\mpcnoise}{distributed noise sampling\xspace}
%
\newcommand{\partnoise}{\enx{\mathsf{\textcolor{blue!60!black}{{PNoise}}}}}
\newcommand{\centrnoise}{\enx{\mathsf{\textcolor{blue!60!black}{CNoise}}}}
\newcommand{\funcnoise}{\enx{\mathsf{\textcolor{blue!60!black}{FNoise}}}}
%
\newcommand{\eg}[1]{e.g.,\;{#1}}
\newcommand{\ie}[1]{i.e.,\;{#1}}


\newcommand{\shares}[1]{[ {#1} ] }

\newcommand{\clipparam}{K}
\newcommand{\epoch}{\enx{k}}
\newcommand{\numepochs}{\enx{T}}

\newcommand{\colorboxcustom}[2][white]{%
  \begingroup
  \setlength{\fboxsep}{0.0pt}  
  \raisebox{0pt}[\height][0pt]{\colorbox{#1}{\strut #2}}%
  \endgroup
}

\newcommand{\local}[1]{\colorboxcustom[orange!40]{#1}}
\newcommand{\glob}[1]{\colorboxcustom[\mylighterblue]{#1}}

\newcommand{\colorboxcustomtext}[2][white]{%
  \begingroup
  \setlength{\fboxsep}{-0.1em}  
  \colorbox{#1}{#2}%
  \endgroup
}

\newcommand{\privguarantee}{DP Considerations}

\newcommand{\tabcolor}{pyblue}
\newcommand{\tabheadform}[1]{{\small\textsf{\color{\tabcolor}#1}}}
\newcommand{\tableheader}[1]{\tabheadform{\footnotesize#1}} 
\newcommand{\tablesubheader}[1]{\tabheadform{\scriptsize#1}} 
\newcommand{\tableemph}[1]{{\small\tabheadform{#1}}}
\newcolumntype{C}[1]{>{\centering\arraybackslash}m{#1}}
\newcolumntype{G}[1]{>{\columncolor{\mylighterblue}}C{#1}}

\newcommand{\theader}[1]{\textbf{\textsf{#1}}} 

\newcommand{\Enc}{\enx{\mathsf{Enc}}}
\newcommand{\Dec}{\enx{\mathsf{Dec}}}
\newcommand{\Share}{\enx{\mathsf{Shr}}}
\newcommand{\Recon}{\enx{\mathsf{Rec}}}
\newcommand{\Mask}{\enx{\mathsf{Mask}}}
\newcommand{\UnMask}{\enx{\mathsf{UnMask}}}
\newcommand{\Garble}{\enx{\mathsf{Garble}}}
\newcommand{\Eval}{\enx{\mathsf{Eval}}}
\newcommand{\clip}{\enx{\mathsf{Clip}}}
\newcommand{\sample}{\enx{\mathsf{Sample}}}

\newcommand{\LaplaceITS}{\textsf{LaplaceITS}}
\newcommand{\Laplace}{\textsf{Laplace}}
\newcommand{\DiscLaplace}{\textsf{DiscLaplace}}
\newcommand{\DiscGaussian}{\textsf{DiscGauss}}
\newcommand{\Gauss}{\textsf{Gaussian}}
\newcommand{\Skellam}{\textsf{Skellam}}
\newcommand{\BoxMuller}{\textsf{BoxMuller}}
\newcommand{\Poisson}{\textsf{Poisson}}
\newcommand{\Bern}{\textsf{Bern}}
\newcommand{\Geom}{\textsf{Geom}}

\newcounter{obscounter}
\setcounter{obscounter}{1}

\newcommand{\obsbox}[1]{%
  \medskip\noindent%
  \colorbox{black!10}{\parbox[t]{\linewidth}{#1}}%
  \medskip%
}

\newcommand{\obs}[1]{%
  \textbf{(O\theobscounter)~{#1}}%
  \stepcounter{obscounter}%
}

\newcommand{\hlobs}[1]{%
  \hl{\textbf{(O\theobscounter)}}\hl{~\textbf{#1}}%
  \stepcounter{obscounter}
}

\newcommand{\obsmanual}[2]{\textbf{(O#1)~{#2}}}
\newcommand{\obsn}[1]{\textbf{(O#1)}}

\newcounter{researchdirectioncounter}
\setcounter{researchdirectioncounter}{1}

\newcommand{\research}[1]{%
  \textbf{(D\theresearchdirectioncounter)~{#1}.}%
  \stepcounter{researchdirectioncounter}%
}

\newcommand{\researchmanual}[2]{\textbf{(D#1)~{#2}}}

\newcommand{\code}[1]{\textbf{{#1}}}

\newcommand{\encrypted}{encrypted\xspace}
\newcommand{\encrypt}{encrypt\xspace}

\newcommand{\coloreduline}[2]{\setulcolor{#1}\ul{#2}}

\newcommand{\clset}{\enx{\mathcal{C}}}
\newcommand{\slset}{\enx{\mathcal{S}}}

\newcommand{\pp}{\enx{\text{\,pp}}}

\newcommand{\negvspacealgo}{\vspace{-0.57em}}

\newcommand{\datasetsize}{\enx{N}}

\newcommand{\inlineIfElse}[3]{\textnormal{{#2} \textbf{{if}} {#1} \textbf{else} {#3}}}

\newcommand{\outparties}{\enx{\mathcal{O}}}

\newcommand{\roadmapref}[1]{\mbox{\hfill{$\blacktriangleright$} #1}}


\title{SoK: Enhancing Cryptographic Collaborative Learning with Differential Privacy}




\author{\IEEEauthorblockN{Francesco Capano}
\IEEEauthorblockA{\textit{SAP SE} \\
Karlsruhe, Germany \\
francesco.capano@sap.com}
\and
\IEEEauthorblockN{Jonas B{\"o}hler}
\IEEEauthorblockA{\textit{SAP SE} \\
Karlsruhe, Germany \\
jonas.boehler@sap.com}
\and
\IEEEauthorblockN{Benjamin Weggenmann\textsuperscript{*}\thanks{\textsuperscript{*}Work done while he was at SAP SE.}}
\IEEEauthorblockA{\textit{Technische Hochschule Würzburg-Schweinfurt} \\
Würzburg-Schweinfurt, Germany \\
benjamin.weggenmann@thws.de}
}

\maketitle

\IEEEpubidadjcol

\begin{abstract}
In collaborative learning (CL), multiple parties jointly train a machine learning model on their private datasets.
However, data can not be shared directly due to privacy concerns.
To ensure \emph{input confidentiality}, cryptographic techniques, e.g., multi-party computation (MPC), enable training on encrypted data. Yet, even securely trained models are vulnerable to inference attacks aiming to extract memorized data from model outputs.
To ensure \emph{output privacy} and mitigate inference attacks, differential privacy (DP) injects calibrated noise during training. 
While cryptography and DP offer complementary guarantees, combining them efficiently for cryptographic and differentially private CL (CPCL) is challenging. Cryptography incurs performance overheads, while DP degrades accuracy, creating a privacy-accuracy-performance trade-off that needs careful design considerations. 
This work systematizes the CPCL landscape. We introduce a unified framework that generalizes common phases across CPCL paradigms, and identify secure noise sampling as the foundational phase to achieve \epcl.
We analyze trade-offs of different secure noise sampling techniques, noise types, and DP mechanisms discussing their implementation challenges and evaluating their accuracy and cryptographic overhead across CPCL paradigms. 
Additionally, we implement identified secure noise sampling options in MPC and evaluate their computation and communication costs in WAN and LAN. 
Finally, we propose future research directions based on identified key observations, gaps and possible enhancements in the literature. 
\end{abstract}





\definecolor{mylightblue} {RGB}{109,196,233}
\definecolor{myblue}      {RGB}{ 76,144,186}
\definecolor{mybluegreen} {RGB}{ 43,194,194}
\definecolor{mydarkgreen} {RGB}{32,149,130}
\definecolor{mylightgreen}{RGB}{ 71,176,170}
\definecolor{myyellow}    {RGB}{247,212,140}
\definecolor{myorange}    {RGB}{254,153, 41}
\definecolor{mydarkorange}{RGB}{202, 96,  0}
\definecolor{mylightred}  {RGB}{229,120,114}
\definecolor{myred}       {RGB}{240, 59, 32}

\definecolor{pyblue}      {RGB}{ 32,119,180}
\definecolor{pyorange}    {RGB}{248,118,  0}
\definecolor{pygreen}     {RGB}{ 43,153, 43}
\definecolor{pyred}       {RGB}{207, 37, 37}

\definecolor{mylightblueb}{RGB}{ 65,182,196} 
\definecolor{myblueb}     {RGB}{ 34, 94,168}
\definecolor{mybluegreenb}{RGB}{161,218,180}
\definecolor{myyellowb}   {RGB}{254,217,142}
\definecolor{myorangeb}   {RGB}{254,153, 41}
\definecolor{myredb}      {RGB}{240, 59, 32}

\definecolor{mycyanbg}      {RGB}{201, 245, 252}
\definecolor{myorangebg}      {RGB}{252, 210, 131}
\definecolor{mygraybg}{gray}{0.93}

\def\mylighterblue{ mylightblue!30}
  \def\mylighterbluetwo{mylightblue!15}
\def\mylighterred{  mylightred!30}
\def\mylightergreen{mylightgreen!30}
\def\mylighterorange{myorange!20}

\def\mydarkerblue{ mylightblue!85}
  \def\mydarkerbluetwo{ mylightblue!70}
\def\mydarkerred{  mylightred!85}
\def\mydarkergreen{mylightgreen!85}
\def\mydarkerorange{myorange!85}

\def\clientbordercolor{gray}
\def\clientfillcolor{gray!25}
\def\mpcserverfillcolor{\mylighterbluetwo}

\def\regionfillcolor{mylightblue!25}
\def\regionbordercolor{\mydarkerblue}
\def\pipecolor{gray!5}
\def\awsclientfillcolor{\mpcserverfillcolor}
\def\awsclientbordercolor{\clientbordercolor} 

\def\tabcolorlight{mylightblue!75}
\def\tabcolordark{gray!18}

\def\origColor{myred!85!black}
\def\prunedColor{pyblue!95!black}

\def\ExpMechColor{pyblue!80!black}
\def\ExpMechSubrangeColor{mydarkgreen!72!white}
\def\SmoothSensColor{pyorange!70!white}
\def\SampleAggColor{pyred}

\def\leftAxisColor{\origColor}
\def\rightAxisColor{\prunedColor}

\def\smallDomainColor{myblue}
\def\mediumDomainColor{mylightgreen}
\def\largeDomainColor{mylightred}

\def\arrowColor{gray!75}

%
\section{Introduction}\label{sec:intro}
Strict privacy laws, e.g., GDPR \cite{GDPR}, 
along with concerns over data misuse and breaches hinder direct data sharing among multiple parties to collaboratively train machine learning models.
Cryptographic techniques, \eg{\emph{multi party computation} (MPC) \cite{evans2018pragmatic} and \emph{homomorphic encryption} (HE) \cite{gentry2009fully}}, enable joint training on private datasets by encrypting data during training.
However, they do not prevent models from leaking information on training data during inference.
Thus, models remain vulnerable to inference attacks, e.g, membership inference, which can reveal if a specific sample was in the training data \cite{shokri2017membership}. 
To mitigate inference attacks, \emph{differential privacy} (DP) \cite{dwork2006} bounds the information leakage by injecting carefully calibrated noise during training.
In the central DP model (CDP), users send raw data to a trusted third party~(TTP) to add noise to the computation output.
To avoid sharing raw data, in the local model (LDP), each user adds noise to its data. 
However, LDP yields lower accuracy than CDP. 
For example, Google's LDP telemetry system \cite{bittau2017prochlo} failed to detect a common signal among 1 million users, despite billions of user reports. 
Enhancing cryptographic collaborative learning with DP to realize \emph{cryptographic and differentially private collaborative learning} (\epcl) provides: 
(I) \emph{input confidentiality} via cryptography, (II) \emph{output privacy} via DP, and (III) \emph{high accuracy} via secure sampling of CDP noise without a TTP.
\epcl is an emerging topic with growing interest from academia 
\cite{kairouz2021distributed,jayaraman2018distributed,pea2023ruan,agarwal2018cpsgd,byrd2022collusion} and industry \cite{madrigal2023project, xu2023federated, paulik2021federated}.
For example, Google's large-scale deployment \cite{xu2023federated} enables next-word predictions for Gboard keyboards by aggregating masked noisy information from clients to satisfy DP. 
Apple uses DP and secure aggregation to learn popular scenes photographed by iOS users to create personalized Memories \cite{apple-iconic}.
%
While various SoKs \cite{cabrero2021sok,mansouri2023sok,sp/NgC23} cover cryptographic CL, the integration of DP in cryptographic CL lacks a comprehensive systematization of key techniques, design considerations, and trade-offs. 
This work bridges this gap by introducing a comprehensive framework for \epcl, analyzing secure noise sampling techniques, and evaluating performance-accuracy trade-offs across learning paradigms. 
From our systematization, we identify two main learning paradigms: \emph{federated learning} (\fl) and \emph{outsourced learning} (\ol). In \fl, data-holding clients iteratively encrypt and send local model updates to servers aggregating them into a global update (Sec.~\ref{subsec:fl}). In contrast, \ol \cls send their encrypted data to servers to train on global encrypted data (Sec.~\ref{subsec:ot}). 
While cryptography and DP offer complementary guarantees, their integration is challenging since both introduce performance overhead and accuracy trade-offs: 
%
\begin{description}[leftmargin=0.4em, labelsep=0.3em,itemsep=-0.1em, style=unboxed]
    \item[\emph{Cryptography performance overhead:}] cryptographic~techniques incur high communication/computation costs, e.g., computation-intensive HE or communication-intensive MPC. Overhead also depends on the learning paradigm: Sec.~\ref{sec:noisetradeoff} shows that \ol can be \enx{10^3\times} slower than plaintext, while \fl is \enx{10\times} faster than \ol but leaks intermediate model updates. 
    \item[\emph{Cryptography performance-accuracy trade-off:}]~cryptographic techniques rely on fixed-point arithmetic trading accuracy for efficiency \cite{evans2018pragmatic}, and introducing numerical errors. 
    \ol approximates non-linear operations, e.g., Softmax, affecting accuracy (Sec.~\ref{sec:noisetradeoff}), while \fl quantizes local updates to reduce communication, introducing approximation errors. 
    \item[\emph{DP performance overhead:}] DP requires noise sampling and per-example gradient clipping. Naive clipping can slow training by \enx{200\times} (for a 3-layer NN, Sec.~\ref{sec:noisetradeoff}), while secure noise sampling further increases overhead (Sec.~\ref{sec:noisegen}). 
    \item[\emph{DP privacy-accuracy trade-off:}] DP noise degrades accuracy and must be calibrated to the threat model. Accuracy depends on \emph{where} noise is injected (Alg.~\ref{alg:dp-training}), \emph{who} samples it (Alg.~\ref{alg:cmp}), \emph{how} it is sampled (Sec.~\ref{sec:noisegen}), and \emph{which} distribution (Tab.~\ref{tab:noisedistributions}).
\end{description}

\para{Contributions}
We provide a thorough and structured analysis of 
solutions combining cryptography and DP for \cl.
While related works (Sec.~\ref{sec:related}) focus on either cryptographic CL (\fl \cite{mansouri2023sok,cabrero2021sok} or \ol \cite{sp/NgC23}) or DP in isolation \cite{ponomareva2023dp,yang2023privatefl}, we systematize their holistic integration.  
Our contributions are: 
\begin{itemize}
    \item We introduce a comprehensive framework to model and generalize \epcl solutions; we identify common phases across learning paradigms, distinguishing steps on local versus (securely joint) global data and possible design choices (Fig.~\ref{fig:flow_main}, Alg.~\ref{alg:cmp}). This enables us to highlight emerging trends and gaps in the \epcl literature as well as possible enhancements to existing solutions (Tab.~\ref{tab:summary}). 
    %
    \item We focus on noise sampling as foundational phase in \epcl to integrate DP guarantees into cryptographic CL. 
    We provide an in-depth analysis of secure techniques for distributed noise generation (Sec.~\ref{sec:noisegen}), and detail noise distributions (Tab.~\ref{tab:noisedistributions}) as well as sampling approaches (Alg.~\ref{alg:cmp}). 
    \item We implement noise sampling in MPC, evaluate computation and communication costs for semi-honest and malicious schemes in LAN and WAN (Sec.~\ref{subsec:distributed-noise-sampling}).\footnote{Code available at: https://github.com/SAP/sok-cpcl \label{fn:github}}
    %
    \item We evaluate and compare privacy-accuracy-performance trade-offs across \epcl paradigms and noise sampling techniques (Sec.~\ref{sec:noisetradeoff}).\textsuperscript{\ref{fn:github}} Our analysis offers insights and guidelines to enhance cryptographic CL with DP (Sec.~\ref{sec:design-epcl}). 
    %
    %
    \item Throughout our work, we distill key \hl{observations (\textbf{O\#})} from which we derive research directions (\textbf{D\#}) (Sec.~\ref{sec:observation}). 
\end{itemize}
\section{Scope \& Methodology}\label{sec:scope}
Next, we outline our scope and methodology. 

\para{Scope} 
This work systematizes the landscape of \epcl, focusing on 
existing and possible techniques and trade-offs to combine 
DP and cryptography for \cl. We include only works co-designing solutions to integrate DP and cryptography in \cl. 
We do not systematize works relying solely on cryptography, \eg{secure aggregation \cite{shokri2017membership,yeom2018privacy}}, as they do not provide output privacy, and those using only DP, due to strong trust assumptions (CDP) or reduced accuracy (LDP). Rather, we leverage our proposed framework~(Sec.~\ref{sec:framework}) to discuss design choices on how those can be augmented to achieve \epcl (Sec.~\ref{sec:design-epcl}).
We focus on horizontally partitioned datasets, where clients hold disjoint records with the same features as it is the most common setting in \cl and aligns with related SoKs~\cite{mansouri2023sok,cabrero2021sok, sp/NgC23}. 
We do not detail vertical partitioning \cite{xu2021achieving,zhu2021pivodl}, and split learning \cite{pereteanu2022split, khan2023love, kanpak2024cure} since they introduce challenges beyond our scope,~e.g., entity alignment, hot to split the model, but we map them to our framework in~App.~\ref{app:framework_generality}.

\para{Methodology}
We started by performing a systematic search across top-ranked venues for security, cryptography, and ML, based on established rankings~\cite{securityranking1,securityranking2,mlranking1}. We performed an extensive search on Google Scholar and BASE with specific keywords referencing our scope, \eg{differential privacy, cryptography, collaborative, training}, from 2018.
We detail search keywords and venues in App.~\ref{app:methodology}. 
Our systematic search returned 650 papers. After examining the abstracts we identified 61 potentially in-scope works. After a thorough analysis, we selected 11 works integrating cryptography and DP in \cl.
For comprehensive coverage, we expanded our search by applying the above criteria to works citing or being cited by selected works. We included 11 further works, for a total of 22 relevant works categorized in Tab.~\ref{tab:summary}. 

\section{Preliminaries}\label{sec:preliminaries}
Before introducing preliminaries for DP, cryptography, and privacy attacks, we recall \emph{Gradient Descent} (GD) in ML training.
GD is an optimization technique to find a set of optimal parameters \enx{\theta^*} to minimize a loss \enx{L(\theta)} of the model over a training dataset.
For each training step \enx{\epoch}, GD updates the parameters in the opposite direction of the gradient of the loss: 
\enx{ \theta^{(\epoch+1)} = \theta^{(\epoch)} - \eta \nabla L(\theta^{(\epoch)}) }.
Here, \enx{\eta} is the learning rate. 
For efficiency on large datasets, \emph{stochastic gradient descent} (SGD) approximates GD by using only a random subset of the training data, or \emph{batch} of size \enx{B}, at each step \cite{prince2023understanding}.

\subsection{Differential Privacy}\label{sec:DP}
DP is a privacy definition that guarantees that the inclusion or exclusion of a record in an analysis does not significantly impact the result.
Formally, a randomized mechanism \enx{\mathcal{M}} satisfies ($\eps, \delta$)-DP 
if for any neighboring dataset \enx{D_1, D_2} (differing in at most one
record), and for any subset \enx{\mathcal{S} \subseteq \mathrm{Range}\mathcal{(M)}}:
{
    \setlength{\abovedisplayskip}{0.5em}
    \setlength{\belowdisplayskip}{0.5em}    
    \begin{equation}\label{def:epsdDP}
        \mathrm{Pr}[\mathcal{M}(D_1)\in \mathcal{S}] \le \exp(\eps) \times \mathrm{Pr}[\mathcal{M}(D_2)\in \mathcal{S}] + \delta.
    \end{equation}
    }
Here, \enx{\eps > 0} is the \emph{privacy budget} and bounds output difference over \enx{\mathcal{M}} on \enx{D_1, D_2}. 
Smaller \eps indicates stronger protection.
Parameter \enx{\delta} models the probability of violating this privacy guarantee. For \enx{\delta=0}, we get \emph{pure} DP ($\eps$-DP), whereas for \enx{\delta > 0} we get \emph{approximated} DP. Typically, for ML \enx{\delta \ll \frac{1}{N}}, where \enx{N} is the dataset size \cite{ponomareva2023dp}.
DP in Eq.~\eqref{def:epsdDP} ensures record-level protection, as \enx{D_1, D_2} differ by one record \cite{dwork2014algorithmic}. User-level protection must account for users contributing multiple records, i.e., \enx{D_1, D_2} differ by an entire user's data~(Sec.~\ref{sec:noisetradeoff}).

\para{DP Mechanisms}
To satisfy DP, a mechanism \enx{\mathcal{M}} can add noise to a function output. 
Formally, \enx{\mathcal{M}(D, f(\cdot)) = f(D) + \psi}, where \enx{f : \mathcal{D}^n \to \mathbb{R}} is a function applied over a dataset \enx{D}, and \enx{\psi} is the noise sampled from a suitable distribution (Tab.~\ref{tab:noisedistributions}). 
Typically, the noise is calibrated on the \enx{f}'s \enx{l_p}-sensitivity: \enx{\Delta_p = \max_{D_1, D_2} ||f(D_1) - f(D_2)||_p}, where \enx{D_1, D_2} are neighboring datasets. 
In CL with multiple data owners different variants of DP can be used. 
In \emph{central} DP (CDP), data owners send their data to a Trusted Third Party (TTP) that applies \enx{\mathcal{M}} on the data. 
To avoid a TTP, in \emph{local} DP (LDP)  \cite{kasiviswanathan2011can}, each data owner applies the perturbation independently (App.~\ref{app:localdp}). 
LDP is a stricter guarantee than CDP, as it requires \enx{\mathcal{M}} to give bounded indistinguishable output between any possible pair of data points \enx{x_1, x_2} from \enx{D} \cite{ponomareva2023dp}. 
However, this stronger privacy guarantee incurs an accuracy cost. For \enx{n}-party count queries, LDP yields \enx{O(\sqrt{n})} error, while CDP's error is \enx{O(1)} \cite{wagh2021dp}.
The \emph{shuffle model} \cite{bittau2017prochlo} improves over LDP by using a TTP to shuffle messages from data owners and break user-message correlation with \enx{O(\log n)} error \cite{wagh2021dp}. 
While we exclude this model as it has weaker accuracy than CDP, we include works that employ shuffling to achieve CDP guarantees \cite{froelicher2020drynx,byrd2022collusion}.

\para{DP Properties for ML} 
DP is immune to \emph{post-processing}, i.e., 
an attacker can not weaken the privacy of a DP output.
%
Furthermore, composition of multiple DP mechanisms remains DP, crucial for tracking privacy budget across ML training iterations. 
\emph{Basic composition} defines the worst (\enx{\eps, \delta})-bound for \enx{n} (\enx{\eps_i, \delta_i})-DP mechanisms applied on a dataset, i.e.,  equivalent to  applying a \enx{(\sum_{i=1}^{n} \eps_i, \sum_{i=1}^{n} \delta_i)}-DP mechanism.
However, basic composition is overly conservative, especially for ML. 
To improve composition relaxed DP definitions like zero-Concentrated DP (zCDP) \cite{bun2016concentrated} and Rényi DP (RDP) \cite{mironov2017renyi} provide tighter bounds using tools like \emph{moments accountant}~\cite{abadi2016deep}.

{
\setlength{\textfloatsep}{0.3em} 
\begin{algorithm}[t]
    \scriptsize
    \caption{Training perturb options (based on \cite{jayaraman2019evaluating})}\label{alg:dp-training}
    \KwIn{Training data $D = \{x_i,y_i\}_{i=1}^n$, 
        and initial parameters $\theta^{(0)}$
        }
    \KwOut{Model parameters $\theta$}

    \SetKwFunction{DifferentiallyPrivateTraining}{DifferentiallyPrivateTraining}
    \SetKwFunction{PerturbInputs}{\perturbinput}
    \SetKwFunction{PerturbLabel}{\perturblabel}
    \SetKwFunction{InitializeModel}{InitializeModel}
    \SetKwFunction{PerturbLoss}{\enx{\mathsf{\textcolor{red}{PerturbLoss}}}}

    \SetKwFunction{PerturbGradients}{\perturbgrad}
    \SetKwFunction{PerturbOutput}{\perturbout}

    \DontPrintSemicolon
        ${x} \gets \PerturbInputs(x)$ \;
        \vspace{-0.2em}
        \ForEach{training step \epoch}{
                $L(\theta^{(\epoch)}, D) \gets \PerturbLoss( \frac{1}{n} \sum_{i=1}^{n} \ell(\theta^{(\epoch)}_i,x_i,y_i) )$ \;
                $g \gets \PerturbGradients( \nabla_{\hspace*{-0.4ex}\theta} L(\theta^{(i)}, D)$ )\;
                $\theta^{(\epoch+1)} \gets \theta^{(\epoch)} - \eta  \cdot {g}$\;
                \vspace{-0.2em}
        }
        \Return $\PerturbOutput(\theta)$ \;
\end{algorithm}
}
\begin{figure}
\vspace{-2.5em}
\end{figure}
\para{DP Options in ML}\label{subsec:perturbationmechanisms}
As Alg.~\ref{alg:dp-training} shows, DP noise
can be applied at different training stages. 
\perturbinput randomizes input data \cite{duchi2013local}, \perturbloss perturbs the loss \cite{chaudhuri2008privacy}, \perturbgrad applies DP noise to gradients \cite{dwork2006}, and \perturbout perturbs model parameters after training \cite{pathak2010multiparty}.
Notably, \perturbinput can obscure patterns, hindering learning. \perturbloss assumes a strongly convex loss function which is typically not the case in neural networks (NN) \cite{jayaraman2019evaluating}. \perturbout is limited to simple models (\eg{linear regression}) since NNs have complex dependencies between data and weights, making sensitivity analysis infeasible \cite{zhang2012functional}.
Another approach, \perturblabel \cite{papernot2018scalable} (App.~\ref{app:collab-labeling}) adds noise to predicted labels during inference. 
However, training with DP labels only ensures label-DP \cite{ghazi2021deep}. 
Thus, \perturbgrad is the most practical approach suitable for any gradient-based optimization method like SGD \cite{ponomareva2023dp}, as discussed next.

\para{DP-SGD}\label{sec:dpsgd}
To make SGD differentially private (DP-SGD) \cite{song2013stochastic,abadi2016deep}, \perturbgrad modifies SGD as follows:
{
\begin{equation}\label{eq:dpsgd}
\begin{split}
\nabla{L'(\theta^{(\epoch)})} =& \nabla{L(\theta^{(\epoch)})} \cdot \min(1, \clipparam / ||\nabla{L(\theta^{(\epoch)})}||_2 ), \\[-0.4em]
\theta^{(\epoch+1)} &= \theta^{(\epoch)} - \eta(\nabla{L'(\theta^{(\epoch)})} + \textcolor{green!60!black}{\psi}).
\end{split}
\end{equation}
}

Here, \enx{\textcolor{green!60!black}{\psi}} is a noise sample drawn from a suitable distribution (Tab.~\ref{tab:noisedistributions}), and \enx{\clipparam} is the clipping parameter.
The noise magnitude depends on the sensitivity of the gradient computation, i.e., the norm. Since gradient norm can be unbounded, each gradient \enx{l_2} norm is clipped to \enx{\clipparam} \cite{abadi2016deep}.
Typically, clipping requires computing \enx{B} per-example gradients, instead of one gradient over the batch loss, increasing computational overhead~(Sec.~\ref{sec:noisetradeoff}).
To mitigate noise impact, \emph{gradient accumulation} averages gradients over \enx{B} samples, thereby averaging also the noise. 
Privacy amplification by \emph{subsampling} enhances DP-SGD's privacy-accuracy trade-off by sampling a subset of data per iteration, leveraging the uncertainty of sample inclusion (App.~\ref{app:privacy_amplification}).

\subsection{Cryptography}\label{sec:crypto}
{
First, we introduce security models, party roles, and notations, followed by an overview of cryptographic techniques.
}

\para{Security Models, Parties \& Notation}\label{sec:securitymodels}
Cryptographic security models define adversarial capabilities. 
In the \emph{semi-honest} model, a passive attacker follows the protocol but tries to extract private information. 
In contrast, in the \emph{malicious} model, an active attacker can deviate from the protocol, e.g., manipulate inputs to alter outputs. 
In both, up to
\enx{t} \emph{colluding} parties can jointly try to infer others' private input. 
We distinguish three roles: 
\emph{input parties} own and provide data; 
\emph{\cpls} execute cryptographic protocols; 
and \emph{output parties} \outparties receive results.
We refer to \cpls as \emph{\sls} \slset, while \emph{\cls} \clset are input and output parties. 
In specific settings, \clset also act as \cpls, and \slset as output parties. 
Unless noted otherwise, we consider \enx{n} semi-honest \cls, \enx{m} semi-honest \sls, and up to \enx{t} colluding parties (\clset, \slset).
In some settings, we consider a semi-trusted server \enx{\s{}'} which does not have access to computation outputs, and can, e.g., sample DP noise in cleartext (Sec.~\ref{subsec:centralnoise}). 
We denote encrypted values $x$ 
as \enx{\shares{x}}, 
and{\glob{highlight}} computations on joint, encrypted data. 

\para{Masking}
Pair-wise masking \cite{bonawitz2017practical} enables secure aggregation ({SecAgg}), using additions modulo \enx{r}, with a single server \s{} ($|\mathcal{S}|=1$). Each pair of \cls \enx{(\c{i}, \c{j})} shares a mask \enx{b_{ij}} via, e.g., Diffie-Hellman key exchange \cite{diffie2022new}. \c{i} adds \enx{b_{ij}} to its data, \ie{\enx{\shares{x_i}=(x_i+b_{ij})\bmod r}}, while \c{j} adds \enx{-b_{ij}}. 
When \s{} sums those values, the masks cancel out, and reveal only the aggregated output, i.e., \enx{\shares{x_i} + \shares{x_j} = (x_i + b_{ij}) + (x_j - b_{ij}) = x_i + x_j}.
An alternative solution is based on \emph{learning with error} (LWE) \cite{regev2009lattices}. In LWE, an error vector \enx{e} breaks the linearity of a set of equations \enx{b_i = As_i + e_i}. Each client \enx{\c{i}} holds \enx{b_i} as a mask, \enx{s_i} as a secret key, and publicly shares the matrix $A$. 
After aggregation, the clients reconstruct \enx{\sum_i{s_i}} to remove the mask \enx{\sum_i{b_i = \sum_i(As_i + e_i)}}, where \enx{e = \sum_i{e_i}} can additionally provide DP \cite{stevens2022efficient} as we illustrate in Sec.~\ref{subsec:partialnoise}.

\para{Homomorphic Encryption (HE)} 
HE \cite{gentry2009fully} enables computing on encrypted data with a single server \s{} ($|\mathcal{S}|=1$).
All \cls share a public key {\enx{pk}}, and each client \enx{\c{i}} encrypts its data, i.e., \enx{\shares{x_i} = \Enc_{pk}(x_i)}, and sends \enx{\shares{x_i}} to \s{} which computes on the encrypted data. Clients can decrypt the result with the corresponding private key {\enx{sk}}, i.e., \enx{\sum_{i}x_i = \Dec_{sk}(\sum_{i}\shares{x_i})}.
Variants include, \emph{additive} HE (AHE) \cite{paillier1999public} for encrypted sums; \emph{fully} HE (FHE) \cite{cheon2017homomorphic} for both multiplication and additions (to evaluate arbitrary functions); and threshold HE \cite{damgaard2001generalisation, shi2011privacy}, where \enx{sk} is secret shared among the \cls, and decryption requires at least $(t+1) \leq n$ \cls. 

\para{Multi-Party Computation (MPC)}\label{sec:ss}
MPC allows multiple parties to jointly compute a function while keeping their inputs private with multiple \sls ($|\mathcal{S}|>1$).
Typically, MPC is split in a slow, data-independent \emph{offline phase} for pre-computation, and a fast, data-dependent \emph{online phase} consuming offline material \cite{evans2018pragmatic}.
A common paradigm for MPC is \emph{threshold secret sharing} (\tnss), where each client \c{i} splits its data $x_i$ into \enx{m} parts, called \emph{shares}, \ie{\enx{\shares{x_i} =\Share(x_i)}}, distributed to \enx{m} \sls. The servers compute on shares locally (e.g., addition) or interactively (e.g., multiplication). 
At least \enx{t+1} \sls are required to \emph{reconstruct} results,~i.e., \enx{\Recon(\shares{x_1}, ..., \shares{x_m})}. The threshold \enx{t} ensures the \tnss scheme is secure against up to \enx{t} colluding \sls.
Another paradigm that can securely evaluate arbitrary functions is \emph{garbled circuits} (\gc, App.~\ref{app:gc}) \cite{yao1986generate}, which is more suited than SS for Boolean operations.

\para{Cryptography and DP}
Combining cryptography and DP requires accounting for computational security of cryptographic schemes in the DP guarantee.
%
%
Computational DP \cite{dwork2014algorithmic} (formalized in App.~\ref{app:compdp}) adapts the DP definition (Def.~\ref{def:epsdDP}) to consider a bounded polynomial-time adversary by adding the negligible failure probability of cryptographic schemes to DP's $\delta$.
Hence, only approximated DP ($\delta>0$) is achievable with cryptography.
Additionally, sampling DP noise via cryptographic protocols can fail due to finite-precision arithmetic. 
Keller et al.~\cite{keller2024secure} absorb this failure probability, e.g., due to overflows or precision mismatch, into DP's $\delta$. 
%
%
Although increasing $\delta$ theoretically affects privacy accounting, the impact of typical security parameters (\eg{$2^{-128}$}) is negligible.

\subsection{Privacy Attacks in \cl}\label{sec:prelim_attacks}
Next, we discuss attacks in the semi-honest setting.
We detail further security models, attacks, and mitigations in~App.~\ref{app:privacyattacks}


\para{Membership Inference Attacks (MIA)}
MIA aim to infer whether a specific record (or user) was part of the training data by exploiting differences in model behavior, e.g., confidence scores, between training and non-training data, often due to overfitting \cite{shokri2017membership}.
DP training mitigates MIA by bounding the influence of any data point on the trained model \cite{tramer2022truth, aerni2024evaluations}.


\para{Gradient Inversion Attacks (GIA)}
GIA aim to reconstruct training data from gradients computed during training.  
In \fl, an adversary corrupting a client or server, has access to aggregated gradients and can reconstruct training samples by optimizing a loss function to match the observed gradients \cite{wu2023learning}. GIA mitigations rely on cryptographic techniques that ensures gradient secrecy against \slset, e.g., HE with a single server \cite{sebert2023combining} or SS with multiple servers (see Sec.~\ref{subsec:fl}). 

\section{Encrypted and DP Collaborative Learning}\label{sec:secureandprivateML}
Next, we systematize the \epcl landscape in a top-down approach. We overview building blocks in Sec.~\ref{sec:sok}, identify common phases in Sec.~\ref{sec:framework}, and analyze trade-offs for \fl and \ol in Sec.~\ref{subsec:fl},~\ref{subsec:ot}, respectively.
%
We highlight forward references to detailed discussions with \roadmapref{}within the overview.

\begin{table*}[htb]
    \scriptsize
    \sffamily 
    \renewcommand{\arraystretch}{0.98} 
    \setlength{\tabcolsep}{0.1em} 
    \centering
	\caption{
        {
            Categorization of {\epcl} \tableemph{Papers} by \tableemph{Learning Paradigm}, DP \tableemph{Noise} and security properties. \tableemph{Perturb}ations are $\mathsf{\textcolor{green!60!black}{Grad(ient)}}$, $\mathsf{\textcolor{orange}{Out(put)}}$.
            Noise \tableemph{Mechanisms} include Dist(ributed) Lap(lace), Disc(rete) Gauss(ian), Poisson-Bin(omial).
            {\tickmarker/\Xmark} denote existing/missing features, {\doubledash} gaps
            , and
            {\enhancemark} potential enhancements. 
            \tableemph{Privacy Unit} is R(ecord) or U(ser) level. 
            \tableemph{DP Analysis} includes Mom(ents) Acc(ountant) and C(entral) L(imit) T(heorem) \cite{bu2020deep}.
            We indicate \tableemph{Malicious} $\mathcal{C}/\mathcal{S}$, and \tableemph{Collusion} thresholds.
        }
    }
    \label{tab:summary}
	\vspace{0.5em}
    \resizebox{\textwidth}{!}{
    \begin{tabular}{c@{\hskip 0.0ex}c@{\hskip 0.5ex}c@{\hskip 0.5ex}c@{\hskip 0.0ex}c@{\hskip 1.2ex}c@{\hskip 2.7ex}c@{\hskip 0.5ex}c@{\hskip 3.5ex}c@{\hskip 0.0ex}c@{\hskip 0.8ex}c@{\hskip 0.8ex}c@{\hskip 0.0ex}c@{\hskip 0.0ex}c@{\hskip -1.7ex}c@{\hskip 0.0ex}c@{\hskip 0.3ex}c}
        \multirow{2}{*}[2.1em]{
		\rotatebox[origin=bc]{65}{\specialcell{\tablesubheader{Learning} \\ \tablesubheader{Paradigm}}}}&
        \multicolumn{5}{c}{\hspace{-5.5ex} \vspace{-2.4ex} \tableheader{Noise}} &
        \multicolumn{2}{c}{\hspace{-5.5ex} \specialcell{\tablesubheader{Gradient} \\ \tablesubheader{Secrecy}}}&
        \multicolumn{2}{c}{\hspace{-2.5ex} \specialcell{\tablesubheader{Model} \\ \tablesubheader{Secrecy}}}&
        \multirow{2}{*}[3.2ex]{\hspace{0.5ex}\rotatebox[origin=bc]{65}{\specialcell{\tablesubheader{Oblivious} \\ \tablesubheader{Noise}}}}&
        \multirow{2}{*}[1.2ex]{\rotatebox[origin=bc]{65}{\centering \tablesubheader{Papers}}}&
		\multirow{2}{*}[1.8ex]{\rotatebox[origin=bc]{65}{\centering \specialcell{\tablesubheader{Privacy} \\ \tablesubheader{Unit}}}}&
        \multirow{2}{*}[-0.5ex]{\centering \specialcell{\tablesubheader{DP} \\ \tablesubheader{Analysis}}}&
        \multirow{2}{*}[-0.5ex]{\centering{\specialcell{\tablesubheader{Cryptographic} \\ \tablesubheader{Techniques}}}} &
		\multirow{2}{*}[2.8ex]{\rotatebox{65}{\tablesubheader{Malicious}}} &
        \multirow{2}{*}[2.5ex]{\rotatebox{65}{\tablesubheader{Collusion}}}
        \\  \rule{0pt}{1.6\normalbaselineskip}
        & 
        \specialcell{\tablesubheader{Type} \\ {\tiny(Alg.~\ref{alg:cmp})}} &
        \specialcell{\tablesubheader{Sampling} \\ {\tiny(Fig.~\ref{fig:dp_noise_generation})}} &
        \specialcell{\tablesubheader{Perturb} \\ {\tiny(Alg.~\ref{alg:dp-training})}} &
        \specialcell{\tablesubheader{Mechanism} \\ {\tiny(Tab.~\ref{tab:noisedistributions})}} &
		\raisebox{.5em}{\tablesubheader{Sampler}} &
        {\raisebox{-0.2em}{\footnotesize $\mathcal{C}$}} & 
		\raisebox{-0.2em}{\footnotesize $\mathcal{S}$} & 
        \raisebox{-0.2em}{\footnotesize $\mathcal{C}$} & 
		\raisebox{-0.2em}{\footnotesize $\mathcal{S}$} & 
        & 
        & 
		& 
		& 
        & 
		& 
        \\
        \cline{1-17}
		\noalign{\vskip 0.08em}
        
        %
        \multirow{20}{*}[-0.0ex]{{\textbf{\fl}}} &
        \multirow{14}{*}[-0.0ex]{{{\partnoise}}} &
        \multirow{14}{*}[-0.0ex]{{{Local}}} &
        \multirow{11}{*}[-0.0ex]{{{\pgrad}}} &
        {Binomial} &
        {\scriptsize $\mathcal{C}$} & 
        \Xmarktab  & 
        \enhancemarkotab & 
        \Xmarktab  & 
        \enhancemarkvtab & 
        \enhancemarkgtab& 
        \cite{agarwal2018cpsgd} & 
        {U} & 
        \enx{(\eps, \delta)} & 
        Masking  & 
        \Xmarktab  & 
        \Xmarktab  
        \\ 
        \noalign{\vskip -0.1em}
        \arrayrulecolor[gray]{0.78}\cline{5-17}
        \noalign{\vskip 0.08em}
        
        
        & & & & \multirow{3}{*}{{Disc. Gauss.}} & 
        {\scriptsize$\mathcal{C}$} & 
        \Xmarktab  & \enhancemarkotab & \Xmarktab  & \enhancemarkvtab & \enhancemarkgtab & 
        \cite{kairouz2021distributed} & {U} & RDP & Masking & \Xmarktab  & \Xmarktab \\ 
        
        & & & & & 
        {\scriptsize$\mathcal{C}$} & 
        \Xmarktab  & \enhancemarkotab & \Xmarktab  & \enhancemarkvtab & \enhancemarkgtab & 
        \cite{wang2020d2p} & {R} & RDP & Masking & \Xmarktab  & \Xmarktab \\ 
        
        & & & & & 
        {\scriptsize$\mathcal{C}$} & 
        \Xmarktab  & \enhancemarkotab & \Xmarktab  & \enhancemarkvtab & \enhancemarkgtab & 
        \cite{stevens2022efficient} & {R} & RDP & LWE & {\scriptsize $\mathcal{C},\mathcal{S}$} & {\scriptsize\enx{n/2}} \\ 
        
        \noalign{\vskip -0.1em}
        \cline{5-17}
        \noalign{\vskip 0.08em}
        
        & & & & \multirow{2}{*}{Skellam} & 
        {\scriptsize$\mathcal{C}$} & 
        \Xmarktab  & \enhancemarkotab & \Xmarktab  & \enhancemarkvtab & \enhancemarkgtab & 
        \cite{agarwal2021skellam} & {U} & RDP & Masking & \Xmarktab  & \Xmarktab \\ 
        
        & & & & & 
        {\scriptsize$\mathcal{C}$} & 
        \Xmarktab  & \enhancemarkotab & \Xmarktab  & \enhancemarkvtab & \enhancemarkgtab & 
        \cite{bao2022skellam} & {R} & RDP & Masking & \Xmarktab  & \Xmarktab \\ 
        
        \noalign{\vskip -0.1em}
        \cline{5-17}
        \noalign{\vskip 0.08em}
        
        & & & & Poisson-Bin. & 
        {\scriptsize$\mathcal{C}$} & 
        \Xmarktab  & \enhancemarkotab & \Xmarktab  & \enhancemarkvtab & \enhancemarkgtab & 
        \cite{chen2022poisson} & {U} & RDP & Masking & \Xmarktab  & \Xmarktab \\ 
        
        \noalign{\vskip -0.1em}
        \cline{5-17}
        \noalign{\vskip 0.08em}
        
        & & & & \multirow{5}{*}{Gaussian} & 
        {\scriptsize $\mathcal{C}$} & 
        \Xmarktab  & \enhancemarkotab & \Xmarktab  & \enhancemarkvtab & \enhancemarkgtab & 
        \cite{truex2019hybrid} & {R} & Mom. Acc. & (A)HE & \Xmarktab  & {\scriptsize\enx{n-1}} \\ 
        
        & & & & & 
        {\scriptsize $\mathcal{C}$} & 
        \Xmarktab  & \tickmarkertab & \Xmarktab  & \tickmarkertab & \enhancemarkgtab & 
        \cite{sebert2023combining} & {U} & Mom. Acc. & (F)HE & \Xmarktab  & {\scriptsize\enx{n-1}} \\ 
        
        & & & & & 
        {\scriptsize $\mathcal{C}$} & 
        \Xmarktab  & \enhancemarkotab & \Xmarktab  & \enhancemarkvtab & \enhancemarkgtab & 
        \cite{lyu2020lightweight} & {R} & RDP & (A)HE & \Xmarktab  & {\scriptsize \enx{n-1}} \\ 
        
        & & & & & 
        {\scriptsize $\mathcal{S}$} & 
        \Xmarktab  & \enhancemarkotab & \Xmarktab  & \enhancemarkvtab & \enhancemarkgtab & 
        \cite{gu2021precad} & {R} & CLT & SS & {\scriptsize $\mathcal{C}$} & {\scriptsize \enx{n/2,m-1}} \\ 
        
        & & & & & 
        {\scriptsize $\mathcal{C}$} & 
        \Xmarktab  & \enhancemarkotab & \Xmarktab  & \enhancemarkvtab & \enhancemarkgtab & 
        \cite{allouah2025towards} & {U} & {RDP} & Masking & \clset, \slset  & {\scriptsize \enx{n-1}} \\ 
        
        \noalign{\vskip -0.1em}
        \cline{4-17}
        \noalign{\vskip 0.08em}
        
        & & & \multirow{3}{*}{{{\pout}}} & \multirow{3}{*}{{{Dist. Lap.}}} & 
        {\scriptsize$\mathcal{C}$} & 
        \tickmarkertab & \tickmarkertab & \Xmarktab  & \enhancemarkvtab & \tickmarkertab & 
        \cite{mugunthan2019smpai} & {R} & \enx{\eps} & Masking & \Xmarktab  & {\scriptsize\enx{n-1}} \\  
        
        & & & & & 
        {\scriptsize$\mathcal{C}$} & 
        \tickmarkertab & \tickmarkertab & \Xmarktab  & \enhancemarkvtab & \tickmarkertab & 
        \cite{byrd2022collusion} & {R} & \enx{\eps} & A(HE) & \Xmarktab  & {\scriptsize\enx{n-1}} \\
        
        & & & & & 
        {\scriptsize$\mathcal{C}$} & 
        \tickmarkertab & \tickmarkertab & \Xmarktab  & \enhancemarkvtab & \tickmarkertab & 
        \cite{bindschaedler17star} & {R} & \enx{\eps} & A(HE) & {\scriptsize$\mathcal{S}$} & {\scriptsize\enx{n-1}} \\
        
        \noalign{\vskip -0.1em}
        \arrayrulecolor{black}\hhline{~|*{16}{-}}
        \noalign{\vskip 0.08em}
        
        & \multirow{6}{*}[-0.0ex]{{{\centrnoise}}} & \cellcolor{\mylighterblue} & \multirow{3}{*}{\pgrad} & \multirow{3}{*}{Gaussian} & 
        {\scriptsize$\mathcal{S}$} & 
        \Xmarktab  & \enhancemarkotab & \Xmarktab  & \enhancemarkvtab & \tickmarkertab & 
        \cite{chase2017} & {R} & Mom. Acc. & SS + GC & \Xmarktab  & \Xmarktab \\ 
        
        & & \cellcolor{\mylighterblue} & & & 
        {\scriptsize$\mathcal{S}$} & 
        \Xmarktab  & \enhancemarkotab & \Xmarktab  & \enhancemarkvtab & \tickmarkertab & 
        \cite{jayaraman2018distributed} & {R} & zCDP & SS + GC & {\scriptsize$\mathcal{S}$} & \Xmarktab \\
        
        & & \cellcolor{\mylighterblue} & & & 
        {\scriptsize$\mathcal{S}$} & 
        \Xmarktab  & \enhancemarkotab & \Xmarktab  & \enhancemarkvtab & \tickmarkertab & 
        \cite{iwahana2022spgc} & {R} & \enx{(\eps, \delta)} & SS + GC & \Xmarktab  & \Xmarktab \\
        
        \noalign{\vskip -0.1em}
        \arrayrulecolor[gray]{0.78}\cline{4-17}
        \noalign{\vskip 0.08em}
        
        & & \multirow{-4}{*}[-0.0ex]{\cellcolor{\mylighterblue}Distributed} & \pout & Laplace & 
        {\scriptsize$\mathcal{S}$} & 
        \tickmarkertab & \tickmarkertab & \Xmarktab  & \enhancemarkvtab & \tickmarkertab & 
        \cite{jayaraman2018distributed} & {R} & \eps & SS + GC & {\scriptsize$\mathcal{S}$} & \Xmarktab \\ 
        
        \noalign{\vskip -0.1em}
        \arrayrulecolor[gray]{0.78}\cline{3-17}
        \noalign{\vskip 0.08em}
        
        & & \multirow{2}{*}[-0.0ex]{\underline{Centralized}} & \pgrad & \doubledashtab & 
        \doubledashtab & 
        \Xmarktab  & \enhancemarkotab & \Xmarktab  & \enhancemarkvtab & \tickmarkertab & 
        \doubledashtab & \doubledashtab & \doubledashtab & \doubledashtab & \doubledashtab & \doubledashtab \\
        
        \noalign{\vskip -0.1em}
        \arrayrulecolor[gray]{0.78}\cline{5-17} \arrayrulecolor{black}
        \noalign{\vskip 0.08em}
        
        & & & \pout & Laplace & 
        {\scriptsize$\mathcal{S}$} & 
        \tickmarkertab & \tickmarkertab & \Xmarktab  & \enhancemarkvtab & \tickmarkertab & 
        \cite{froelicher2020drynx} & {R} & \eps & (A)HE & {\scriptsize$\mathcal{C},\mathcal{S}$} & {\scriptsize\enx{m-1}} \\
        
        \noalign{\vskip -0.1em}
        \cline{1-17}
        \noalign{\vskip 0.08em}
        
        \multirow{6}{*}[-0.0ex]{{\textbf{\ot}}} & \multirow{3}{*}{{{\partnoise}}} & \multirow{3}{*}{{Local}} & \multirow{2}{*}{\pgrad} & Binomial & 
        {\scriptsize$\mathcal{S}$} & 
        \tickmarkertab & \tickmarkertab & \tickmarkertab & \tickmarkertab & \enhancemarkgtab & 
        \cite{pea2023ruan} & {R} & RDP & SS & \Xmarktab  & {\scriptsize\enx{m/2}} \\ 
        
        \noalign{\vskip -0.1em}
        \arrayrulecolor[gray]{0.78}\cline{5-17}
        \noalign{\vskip 0.08em}
        
        & & & & Disc. Gauss. & 
        {\scriptsize$\mathcal{S}$} & 
        \tickmarkertab & \tickmarkertab & \tickmarkertab & \tickmarkertab & \enhancemarkgtab & 
        \cite{das2025communication} & {R} & Mom. Acc. & SS & \Xmarktab  & {\scriptsize\enx{m-1}} \\ 
        
        \noalign{\vskip -0.1em}
        \cline{4-17}
        \noalign{\vskip 0.08em}
        
        & & & \pout & \doubledashtab & 
        \doubledashtab & 
        \tickmarkertab & \tickmarkertab & \tickmarkertab & \tickmarkertab & \enhancemarkgtab & 
        \doubledashtab & \doubledashtab & \doubledashtab & \doubledashtab & \doubledashtab & \doubledashtab \\ 
        
        \noalign{\vskip -0.1em}
        \arrayrulecolor{black}\hhline{~|*{16}{-}}
        \noalign{\vskip 0.08em}
        
        & \multirow{3}{*}[-0.0ex]{{{\centrnoise}}} & \cellcolor{\mylighterblue} & {\pgrad} & \doubledashtab & 
        \doubledashtab & 
        \tickmarkertab & \tickmarkertab & \tickmarkertab & \tickmarkertab & \tickmarkertab & 
        \doubledashtab & \doubledashtab & \doubledashtab & \doubledashtab & \doubledashtab & \doubledashtab \\ 
        
        \arrayrulecolor[gray]{0.76}\cline{5-17}
        
        & & \multirow{-2}{*}[-0.0ex]{\cellcolor{\mylighterblue}Distributed} & {\pout} & Dist. Lap. & 
        {\scriptsize$\mathcal{S}$} & 
        \tickmarkertab & \tickmarkertab & \tickmarkertab & \tickmarkertab & \tickmarkertab & 
        \cite{pentyala2022training} & {R} & \enx{\eps} & SS & {\scriptsize$\mathcal{S}$} & {\scriptsize\enx{m-1}} \\ 
        
        \noalign{\vskip -0.1em}
        \cline{3-17} \arrayrulecolor{black}
        \noalign{\vskip 0.08em}
        
        & & {\underline{Centralized}} & \doubledashtab & \doubledashtab & 
        \doubledashtab & 
        \tickmarkertab & \tickmarkertab & \tickmarkertab & \tickmarkertab & \tickmarkertab & 
        \doubledashtab & \doubledashtab & \doubledashtab & \doubledashtab & \doubledashtab & \doubledashtab \\
    \end{tabular}
    }
    
    \vspace{-1.5em}
\end{table*}

\subsection{Systematization}\label{sec:sok}
Next, we introduce core building blocks of \epcl, via columns from Tab.~\ref{tab:summary}, discussing them from left to right.
Tab.~\ref{tab:summary} categorizes \epcl works, highlights trends, identifies unexplored combinations of techniques, i.e., gaps ({\doubledash}), and potential enhancements using existing techniques ({\enhancemark}). 
For convenience, we provide a summary of notation in Tab.~\ref{tab:notation}~(App.~\ref{app:notation}).

\para{Learning Paradigms}
We identify two main \emph{learning paradigms} in \epcl: federated learning (\fl) and outsourced learning (\ot).
In \fl, clients train local models on their private data and send encrypted updates to a server for aggregation. In \ot, clients outsource training to servers that compute on encrypted data. 
Tab.~\ref{tab:summary} shows that \fl is the main paradigm in \epcl, likely due to its efficiency, i.e.,~only requiring additions on encrypted data. 
We analyze \fl and \ot trade-offs in \roadmapref{Sec.~\ref{subsec:fl},~\ref{subsec:ot}}, and evaluate them in \roadmapref{Sec.~\ref{sec:noisetradeoff}}.

\para{Noise Generation}
For \emph{noise generation} we distinguish five aspects. 
We identify two noise \emph{types}: 
\centrnoise, a single CDP noise term sampled globally, and
\partnoise, a partial non-DP noise term sampled locally by each client or server. While a single \partnoise does not satisfy DP, the aggregation of multiple \partnoise terms satisfies CDP.
We further distinguish three \emph{sampling techniques}: centralized \centrnoise sampling via a semi-trusted \slong, local \partnoise sampling by individual clients/servers, and distributed \centrnoise sampling via MPC across multiple servers.
Notably, while \partnoise is sampled independently and added locally to encrypted local updates, \centrnoise is sampled once and applied to already-aggregated data.
The two main perturbation options are: \perturbgrad and \perturbout. 
Different \emph{mechanisms} are used by systematized works to sample noise from suitable distributions. 
We also distinguish the \emph{sampler}, which can be either \cls or \sls. 
Among noise types, \partnoise is the most common, since it is the most efficient requiring only local sampling. 
Furthermore, \partnoise has been implemented with a variety of mechanisms, while \centrnoise is limited to Laplace and Gaussian. 
None of the analyzed works adopt distributed \centrnoise sampling with \perturbgrad, likely due to performance overhead. 
Notably, centralized \centrnoise sampling is used by only one \fl work \cite{froelicher2020drynx}, as it requires a semi-trusted server.  
In the rest of this work, we focus mainly on \perturbgrad, the most flexible and common option, but also highlight differences with \perturbout. 
We detail noise mechanism in \roadmapref{Tab.~\ref{tab:noisedistributions}, Sec.~\ref{sec:noisegen}}, analyze in-depth noise generation techniques in \roadmapref{Sec.~\ref{subsec:centralnoise},~\ref{subsec:partialnoise},~\ref{subsec:distributed-noise-sampling}}, and evaluate privacy-accuracy-performance trade-offs in \roadmapref{Tab.~\ref{tab:combined-dp-overhead}-\ref{tab:accuracy}, Sec.~\ref{sec:noisetradeoff}}.

\para{Security Properties}
We distinguish \emph{gradient secrecy} which hides aggregated DP gradients preventing gradient inversion attacks, and \emph{model secrecy} which protects the parameters of the trained model, preserving IP.
\ot inherently guarantees both, while \fl requires cryptographic enhancements~(\enhancemark). 
%
%
Notably, \fl with \perturbout guarantees gradient secrecy by design, as only noisy model parameters are revealed.
We also identify \emph{oblivious noise} when output parties do not know the sampled noise, preventing its removal and preserving privacy guarantees while improving accuracy.
%
While distributed \centrnoise guarantees noise obliviousness by default, \partnoise requires cryptographic enhancements~(\enhancemark) to prevent corrupted parties from removing their noise contributions. 
%
We discuss security properties for \fl and \ol in \roadmapref{Sec.~\ref{subsec:fl},~\ref{subsec:ot}}; detail noise obliviousness in \roadmapref{Sec.~\ref{sec:noisegen}} and its implications for \centrnoise in \roadmapref{Sec.~\ref{subsec:centralnoise}} and \partnoise in \roadmapref{Sec.~\ref{subsec:partialnoise}}. 

\para{Privacy Unit}
The granularity of DP's protection is typically given at user or record level. 
In \epcl the choice of privacy unit is crucial but often overlooked. Only \enx{27\%} of systematized works provide user-level DP, i.e., where each user contributes up to \enx{z} records. Notably, none of those works leverage \ol. 
Protecting only individual records can harm users contributing multiple records. For example, in language modeling each user provides thousands of examples (i.e., words or phrases) \cite{mcmahan2017learning}. We discuss how to achieve user-level DP in \roadmapref{Sec.~\ref{sec:noisegen}}.

\para{DP Analysis}
We specify the privacy accounting method to track the privacy budget across training iterations. 
Most of the systematized works leverage advanced accountants, e.g., moments accountant \cite{abadi2016deep} or R\'enyi DP \cite{mironov2017renyi}, for tighter privacy bounds. 
We provide definitions, conversion lemmas, and comparisons of composition bounds in \roadmapref{App.~\ref{app:relaxedDP}}. 

\para{Cryptographic Techniques}
HE and masking work with a single server but require key management. Specifically, HE requires key-pair generation via a trusted dealer or a distributed key generation protocol \cite{truex2019hybrid}, while masking requires the exchange of pairwise masks (or seeds) \cite{bonawitz2017practical}.
In contrast, MPC requires multiple non-colluding servers but no keys, using only local randomness. 
The cryptographic techniques have different bottlenecks.
MPC is communication-intensive, since servers exchange messages during computations. 
In comparison, HE is computation-heavy but communication-efficient as it requires only a single server. 
\fl works employ all three techniques. However, masking is the most common likely since it is specific to \fl. 
\ot works rely mainly on MPC with optimized protocols. Notably, no \ol work uses HE, even though it is beneficial in network-constrained settings, or hybrid HE-MPC approaches to balance computation and communication.

\para{Threat Models}
Most \epcl works assume \emph{semi-honest security} with non-colluding parties for efficiency. 
Only few works consider \emph{malicious security} against \slset and \clset during the whole training, i.e., noise sampling, gradient computation and aggregation \cite{stevens2022efficient,froelicher2020drynx}, or only during noise sampling \cite{mugunthan2019smpai} to defend against noise tampering.
Notably, colluding parties can weaken privacy guarantees by removing their \partnoise contributions from the aggregate. Mitigating this requires either increasing the noise variance (degrading accuracy) or cryptographic enhancements that introduce performance overhead.
We analyze mitigations for collusion with \partnoise in \roadmapref{Sec.~\ref{subsec:partialnoise}}, and evaluate the impact on accuracy in \roadmapref{Tab.~\ref{tab:accuracy}, Sec.~\ref{sec:noisetradeoff}}.
Furthermore, we discuss privacy attacks and mitigations against malicious parties in \roadmapref{App.~\ref{app:privacyattacks}}.

\label{subsec:[phases]}
%
%
\begin{figure*}[ht!]
    \footnotesize

\begin{tikzpicture}[thick]
    \def\xdist{1.9em}
    \def\ydist{0.5em}
    \def\boxheight{2.8em}
    \def\boxwidth{7.0em}
	\def\smallboxheight{2.8em}
    \def\smallboxwidth{5.0em}
	\def\smallerboxwidth{4.0em}
    \def\updateboxwidth{2.6em}
    \def\revealboxwidth{2.6em}
    \def\perturbboxwidth{4.0em}
    \def\protectboxwidth{4.0em}
    \def\setupboxwidth{2.5em}

    \tikzset{
        box/.style={
            signal, 
            signal from=west, signal to=east,
            draw,
            text width=\boxwidth,
            text centered,
            minimum height=\boxheight
        },
		smallbox/.style={
            signal, 
            signal from=west, signal to=east,
            draw,
            text width=\smallboxwidth,
            text centered,
            minimum height=\smallboxheight
        },
		smallerbox/.style={
            signal, 
            signal from=west, signal to=east,
            draw,
            text width=\smallerboxwidth,
            text centered,
            minimum height=\smallboxheight
        },
        setupbox/.style={
            signal, 
            signal from=west, signal to=east,
            draw,
            text width=\setupboxwidth,
            text centered,
            minimum height=\smallboxheight
        },
        updatebox/.style={
            signal, 
            signal from=west, signal to=east,
            draw,
            text width=\updateboxwidth,
            text centered,
            minimum height=\smallboxheight
        },
        revealbox/.style={
            signal, 
            signal from=west, signal to=east,
            draw,
            text width=\revealboxwidth,
            text centered,
            minimum height=\smallboxheight
        },
        protectbox/.style={
            signal, 
            signal from=west, signal to=east,
            draw,
            text width=\protectboxwidth,
            text centered,
            minimum height=\smallboxheight
        },
        perturbbox/.style={
            signal, 
            signal from=west, signal to=east,
            draw,
            text width=\perturbboxwidth,
            text centered,
            minimum height=\smallboxheight
        },
        client/.style={
            thick,
            draw=orange!80, 
            fill=orange!40
        },
        server/.style={
            draw,thick,
            draw=green!80
        },
        mpc/.style={
            draw=cyan!80, 
            fill=\mylighterblue
        },
        local/.style={
            fill=white,
        }
    }

	\node[text width=\smallboxwidth*1.1, align=left] (fl_label) {\normalsize\textbf{\textsf{FL}}};

	\node[above of= fl_label, yshift=-(\boxheight)*0.7,text width=\smallboxwidth*1.8, align=left, xshift=4.5em] (fl_label_client) {option 1:\\client-side \perturb};
	\node[below of= fl_label_client, yshift=(\boxheight)*0.08, text width=\smallboxwidth*1.8, align=left] (fl_label_server) {option 2:\\server-side \perturb};


	\node[box, client, right of= fl_label_client, xshift=\xdist*2.8] (gradCompute_fl1) {\compgrad \\ \clset};
	\node[perturbbox, client, right of=gradCompute_fl1, xshift=\xdist*0.97,anchor=west] (perturb_fl1) {\perturb \\ \clset};
    \node[protectbox, client, right of=perturb_fl1, xshift=\xdist*0.17, anchor=west] (protect_fl1) {\protection \\ \clset on \enx{\tilde{g}_C}};
	\node[smallbox, mpc, right of=protect_fl1, xshift=\xdist*0.17,anchor=west] (aggregate_fl1) {\aggr \\ \slset on \enx{\shares{\tilde{g}_C}}};
	\node[revealbox, mpc, right of=aggregate_fl1, xshift=\xdist*0.44,anchor=west] (reveal_fl1) {\reveal \\ \outparties};
    \node[updatebox, client, right of=reveal_fl1, xshift=-\xdist*0.19, anchor=west] (update_fl1) {\updatemodel \\ \outparties};

	\node[box, client, right of= fl_label_server, xshift=\xdist*2.8] (gradCompute_fl2) {\compgrad \\ \clset};
    \node[protectbox, client, right of=gradCompute_fl2, xshift=\xdist*0.97, anchor=west] (protect_fl2) {\protection \\ \clset on \enx{{g}_C}};
	\node[smallbox, mpc, right of=protect_fl2, xshift=\xdist*0.17,anchor=west] (aggregate_fl2) {\aggr \\ \slset on \enx{\shares{{g}_C}}};
    \node[perturbbox, mpc, right of=aggregate_fl2, xshift=\xdist*0.44, anchor=west] (perturb_fl2) {\underline{\perturb} \\ \slset (or \enx{S'})};
	\node[revealbox, mpc, right of=perturb_fl2, xshift=\xdist*0.17,anchor=west] (reveal_fl2) {\reveal \\ \outparties};
    \node[updatebox, client, right of=reveal_fl2, xshift=-\xdist*0.19, anchor=west] (update_fl2) {\updatemodel \\ \outparties};

    \node[protectbox, client,below of=gradCompute_fl2, yshift=1.1em, anchor=north, xshift=-1.0em] (setup_ol) {$\setup^*$ \\ \clset on \enx{D_C}};
    \node[protectbox, client, right of=setup_ol, anchor=west, xshift=0.32em] (protect3) {\protection \\ \clset on \enx{D_C}};
	\node[below of=fl_label, text width=\smallboxwidth*1.1, align=left,yshift=-2.2em] (ol_label) {\normalsize\textbf{\textsf{OL}}};
    \node[box, mpc, right of=protect3, xshift=\xdist*0.17, anchor=west] (gradCompute2) {\compgrad \\ \slset};
    \node[perturbbox, mpc, right of=gradCompute2, xshift=\xdist*0.97,anchor=west] (perturb4) {\underline{\perturb} \\ \slset (or \enx{S'})};
    \node[updatebox, mpc, right of=perturb4, xshift=\xdist*0.17,anchor=west] (update2) {\updatemodel \\ \slset};
	\node[revealbox, mpc, right of=update2, xshift=-\xdist*0.19,anchor=west] (reveal2) {$\reveal^*$\\ \outparties};

\node[rectangle, text width=9.6em, minimum height=8.5em, right of=update_fl1, xshift=8.0em, yshift=-3.6em, draw=gray] (legend_frame) {
    \begin{tikzpicture}[scale=0.8]
        \node[signal, signal from=west, signal to=east, draw, text width=6.5em, minimum height=1.5em, text centered] (template) {\scriptsize Phase\\[-0.1em]Party~[on~Input]};
        \node[signal, signal from=west, signal to=east, draw, fill=orange!40, draw=orange!80, text width=6.5em, minimum height=1.5em, text centered, below of=template, node distance=3.0em] (local_sampling) {\scriptsize \perturb samples  \partnoise};
        \node[signal, signal from=west, signal to=east, draw, mpc, text width=6.5em, minimum height=1.5em, text centered, below of=local_sampling, node distance=3.0em] (global_sampling) {\scriptsize \underline{\perturb} samples \partnoise/\centrnoise};
    \end{tikzpicture}
};

\end{tikzpicture}
\caption{Execution flows of \fl and \ot phases with \perturbgrad, party roles, inputs where relevant, and \local{local} or\glob{global} computations. 
\setup is omitted, except OL-specific pre-processing. Optional phases have superscript $*$.
}
\label{fig:flow_main}
\vspace{-1.5em}
\end{figure*}
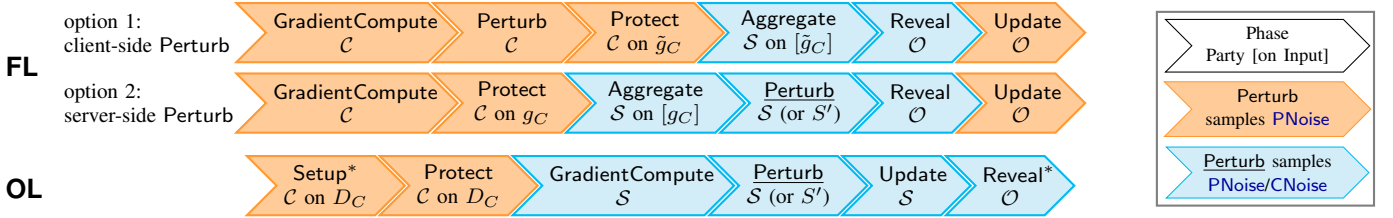
\subsection{\epcl Framework} \label{sec:framework}%
We propose a unified framework for \epcl by identifying seven common phases, detailed below for \perturbgrad.

\phasepara{\boldmath{\setup{$(\eps,\delta,\Lambda,\clset,\slset)$}}} \clset and \slset exchange cryptographic parameters $\Lambda$ (e.g., keys, seeds) and DP parameters (\eps, $\delta$). 
Depending on the cryptographic technique, this may involve generating keys (HE), exchanging seeds (masking), or pre-computing correlated randomness (MPC).
Optionally, \clset may perform local data pre-processing (e.g., feature extraction) to improve training efficiency, which we  discuss in \roadmapref{Sec.~\ref{subsec:ot}}.

\phasepara{\boldmath{\compgrad{(\enx{\theta^{(\epoch)}, D})}}} \clset or \slset compute per-example gradients on data \enx{D} at training step \epoch. 
The per-example gradients \enx{\mathbf{g}} are clipped to bound sensitivity for DP noise. While trivial on cleartext data, clipping requires non-linear operations (e.g., inverse square root, comparisons) which are costly in cryptographic protocols. We discuss the cryptographic performance overhead of \compgrad in \roadmapref{Sec.~\ref{subsec:ot}} and evaluate the performance impact in \roadmapref{Tab.~\ref{tab:combined-dp-overhead}, Sec.~\ref{sec:noisetradeoff}}.

\phasepara{\boldmath{\protection{(\enx{x, \mathcal{S}})}}} \clset \encrypt data or gradients \enx{x} to output encrypted $\shares{x}_{\scriptscriptstyle\s{}}$ to servers $\s{}\in\mathcal{S}$. 
Specifically, in MPC clients secret share (\Share) data among \enx{m} servers. Instead, in single-server settings, \cls leverage HE or masking to encrypt (\Enc) or mask the data creating a single ciphertext.

\phasepara{\boldmath{\perturb{(\enx{\tau,\mathbf{g}})}}} \clset or \slset sample noise \enx{\psi = \sample(\tau)} to output \enx{\mathbf{\tilde{g}} = \mathbf{g} + \psi}. The tuple \enx{\tau} specifies the noise type (i.e., \partnoise/\centrnoise), the sampling mechanism (Tab.~\ref{tab:noisedistributions}, Sec.~\ref{sec:noisegen}), and associated parameters (e.g., Gaussian \partnoise with variance \enx{\sigma^2} \enx{= {2\ln (1.25/\delta)} \Delta_2^2/\eps^2} to satisfy (\enx{\eps, \delta})-DP). 
Notably, we stylize sampling options as: \perturb for local \partnoise sampling,\glob{\perturb} for \centrnoise via MPC, and \underline{\perturb} for \centrnoise via a semi-trusted server.\glob{\underline{\perturb}} indicates any of these three may be used.
We identify \perturb as the foundational phase for integrating DP in cryptographic CL, and analyze all its components in depth in \roadmapref{Sec.~\ref{sec:noisegen}}. We formalize \sample algorithms in \roadmapref{Alg.~\ref{alg:mpc_noise} (App.~\ref{app:sampling_alg})}.

\phasepara{\boldmath{\aggr{(\enx{\mathbf{G}})}}} \slset aggregates the set \enx{\mathbf{G}} of locally computed gradients, typically via averaging. Specific to \fl, this phase relies on lightweight operations on \encrypted data (e.g., additions) as further discussed in \roadmapref{Sec.~\ref{subsec:fl}}.

\phasepara{\boldmath{\updatemodel{(\enx{\theta^{(\epoch)}, \mathbf{\tilde{g}}, \eta})}}} \clset or \slset update model parameters \enx{\theta^{(\epoch)}} at step \enx{\epoch} using noisy gradients \enx{\mathbf{\tilde{g}}} and learning rate \enx{\eta}.

\phasepara{\boldmath{\reveal{($\{\shares{x}_{\scriptscriptstyle\s{}}\}_{\mathcal{S}}$)}}} \clset or \slset decrypt cleartext $x$ from \encrypted \enx{\shares{x}_{\scriptscriptstyle\s{}}} received from \enx{\s{} \in \mathcal{S}} (i.e., via \Recon or \Dec).
We distinguish two types of reveal: \emph{implicit} which reveals results to \slset immediately upon aggregation, offering no gradient secrecy against \slset (e.g., pair-wise masking).
\emph{Explicit reveal} requires a distinct decryption protocol and if performed by \clset guarantees gradient and model secrecy against \slset (e.g., MPC). We discuss how reveal affects security of \fl in \roadmapref{Sec.~\ref{subsec:fl}}.

We provide cryptographic computation and communication complexities across phases in \roadmapref{Tab.~\ref{tab:complexity_phases}, App.~\ref{app:crypto_complexity}}.

%


\para{Framework Instantiations}
Our framework is general and covers different designs. 
This is achieved by adjusting the order of phases, the set of parties executing them (i.e., \cls \enx{\mathcal{C}}, \sls \enx{\mathcal{S}}, and output parties \enx{\mathcal{O} \subseteq \mathcal{C} \cup \mathcal{S}}), and the type of computations, i.e., local on cleartext data or global on encrypted data.
Fig.~\ref{fig:flow_main} provides a high-level overview of how to instantiate \fl and \ol with our framework, which we detail in \roadmapref{Sec.~\ref{subsec:fl}, \ref{subsec:ot}}, respectively.
Although we mainly focus on \fl and \ot, our framework can easily express other learning paradigms, e.g., split learning \cite{pereteanu2022split}, and vertical \fl \cite{xu2021achieving}, as detailed in \roadmapref{App.~\ref{app:framework_generality}} (see Fig.~\ref{fig:flows_per_paradigm}). 
%
Furthermore, we formalize the phases as pseudocode in Alg.~\ref{alg:cmp}. Specifically, phases are at the top and execution flow of \fl and \ol are given and compared at the bottom.
Overall, we can model single-server \fl (e.g., via HE) by setting \enx{|\mathcal{S}|=1}, where each \enx{\c{} \in \mathcal{C}} sends one ciphertext to a server \s{}, \ie{\enx{\shares{\tilde{\mathbf{g}}_{\scriptscriptstyle\c{}}}_{\scriptscriptstyle\s{}}} for \partnoise and \enx{\shares{{\mathbf{g}_{\scriptscriptstyle\c{}}}}_{\scriptscriptstyle\s{}}} for \centrnoise}.
By setting \enx{|\mathcal{S}|>1}, we model multi-server \fl (via MPC) where each \enx{\c{} \in \mathcal{C}} sends shares, e.g., \enx{\shares{\tilde{\mathbf{g}}_{\scriptscriptstyle\c{}}}_{\scriptscriptstyle\s{}}}, to each \enx{S \in \mathcal{S}}. 
Additionally, we can also model privacy amplification by \emph{subsampling} \cite{wang2019subsampled} (App.~\ref{app:privacy_amplification}): in \fl a random subset of \cls are sampled for each iteration, whereas in \ol, subsampling on \enx{D} can be implemented within \compgrad. 
Also, to model \perturbout, \perturb is applied on model parameters \emph{after} training, without clipping.

{
\SetAlgoSkip{negvspacealgo}
\setlength{\textfloatsep}{0.0em} 
\setlength{\floatsep}{0.0em} 
\setlength{\intextsep}{0.0em} 
\setlength{\algomargin}{0.1em} 
\begin{algorithm*}[t]
    
    \scriptsize
    \DontPrintSemicolon
    
    \SetAlgoNlRelativeSize{-2}
    
    \SetAlgoNlRelativeSize{-2}
    
    \SetAlgoNlRelativeSize{-2}

    \SetInd{0.7em}{0.7em} 
    \caption{
    {
        Comparison of \fl and \ot phases per training iteration using \perturbgrad with \partnoise or \centrnoise.
    }
    }\label{alg:cmp}

    \KwIn{Sets of clients $\mathcal{C}$, servers $\mathcal{S}$, and output parties $\mathcal{O}$ where $\clset \subseteq \mathcal{O} \subseteq \clset \cup \slset$ (i.e., we assume all clients are output parties, servers optionally); clipping parameter \enx{\clipparam}, per client training dataset \enx{D_C =\{x_i,y_i\}_{i=1}^{N_C}} with size \enx{N_C}, model parameters \enx{\theta^{(\epoch)}} at iteration \enx{\epoch}.
    }

    \SetKwProg{Fn}{Function}{:}{}
    \SetKwProg{Fl}{Noise}{:}{}
    \SetKwProg{Ph}{Phase}{:}{}
    \SetKwProg{Ot}{Noise}{:}{}
    \SetKwFunction{sendTo}{send}
    \SetKwFunction{To}{to}
    \SetKwFunction{Sample}{{\sample}}
    \SetKwFunction{Clip}{{\clip}}
    \SetKwFunction{ModelUpdate}{ModelUpdate}
    \SetKwFunction{PartialNoise}{{\partnoise}}
    \SetKwFunction{MPCNoise}{{\centrnoise}}
    \SetKwFunction{Rec}{Rec}
    \SetKwFunction{Shr}{Shr}

    \SetKwFunction{PartialNoiseLocal}{Partial}
    \SetKwFunction{MPCNoiseT}{Central}

    \SetKwFunction{Grad}{{\compgrad}}
    \SetKwFunction{Perturb}{{\noise}}
    \SetKwFunction{Protect}{{\protection}}
    \SetKwFunction{Reveal}{{\reveal}}
    \SetKwFunction{Aggregate}{{\aggr}}
    \SetKwFunction{Setup}{{\setup}}
    \SetKwFunction{ModelUpdate}{{\updatemodel}}

    \SetKw{KwNotation}{Notation:}
    \KwNotation{
        \textnormal{
        \hspace{-0.8em}
        Cleartext phases apply to \encrypted data when executed\glob{globally}. 
        \enx{\tau} in {{\perturb}} defines noise type (\partnoise/\centrnoise), mechanism (Tab.~\ref{tab:noisedistributions}) and distribution parameters (e.g., \enx{\sigma^2} for (\enx{\eps, \delta})-DP). \partnoise is sampled locally via \perturb, while \centrnoise is sampled via MPC\glob{\perturb} or by semi-trusted \enx{S'}(\underline{{\perturb}}).
        %
        %
        $\{\}_S$ denotes a set for each $\s{} \in \mathcal{S}$.
        }
    }

    \KwOut{Updated model parameters \enx{\theta^{(\epoch+1)}}.}

\vspace*{0.3ex}
    \begin{minipage}[t]{0.25\textwidth}
        \Fn{\Clip{$\mathbf{G}, \clipparam$}}{
        \Return $\{ \mathbf{g} \cdot \min(1, \clipparam / ||\mathbf{g}||_2  )\}_{g \in \mathbf{G}}$
        }
        \Ph{\Perturb{$\tau$, $\mathbf{g}$}}{
            $\psi \gets \Sample(\tau)$\ \tcp{See Alg.~\ref{alg:mpc_noise}}
            \Return $\mathbf{g} + \psi$\;
        }
    \end{minipage}
\hspace*{\fill}
    \begin{minipage}[t]{0.27\textwidth}
        \Ph{\Grad{$\theta^{(\epoch)}, D, \clipparam$}}{
            $ \mathbf{L}(\theta^{(\epoch)}, D) \gets \{ \ell(\theta^{(\epoch)},x_i,y_i) \}_{i \in [1,N]}$\;
            $ \mathbf{G} \gets \nabla_{\theta} \mathbf{L}(\theta^{(\epoch)}, D)$\;
            $ \mathbf{G'} \gets \Clip(\mathbf{G}, \clipparam)$\;
        \Return $\frac{1}{|\mathbf{G'}|} \sum_{\mathbf{g'} \in \mathbf{G'}}\mathbf{g'}$
        }
    \end{minipage}
\hspace*{\fill}
    \begin{minipage}[t]{0.2\textwidth}
        \Ph{\Protect{$\mathbf{x}, \mathcal{S}$}}{
            \Return $\{\shares{{\mathbf{{x}}}}_{\scriptscriptstyle\s{}}\}_{\mathcal{S}}$
        }
        \Ph{\Reveal{$\{\shares{x}_{\scriptscriptstyle\s{}}\}_{\mathcal{S}}$}}{
            \Return $\mathbf{x}$
        }
    \end{minipage}
\hspace*{\fill}
    \begin{minipage}[t]{0.21\textwidth}
        \Ph{\Aggregate{$\mathbf{G}$}}{
            \Return $\frac{1}{|\mathbf{G}|} \sum_{\mathbf{g} \in \mathbf{G}}\mathbf{g}$
        }
        \Ph{\ModelUpdate{$\theta^{(\epoch)}$, $\mathbf{\tilde{g}}$, $\eta$}}{
            \Return $\theta^{(\epoch)} - \eta  \cdot \mathbf{\tilde{g}}$
        }
    \end{minipage}

    \begin{minipage}[t]{0.54 \textwidth}
        \vspace{0.5ex}
        \small \textsc{Federated Learning:} \scriptsize\;
        \ForEach(\hspace*{-10em}\tcp*[f]{For each training iteration \enx{k}}\hfill){ $\c{} \in \mathcal{C}$}{
            $ \mathbf{g_c} \gets \Grad{$\theta^{(\epoch)}_c, D_c, \clipparam$}$\;
            \begin{minipage}[t]{0.491\textwidth}
                    \tcp{Client-side noise sampling}
                        ${\mathbf{\tilde{g}_c}} \gets \Perturb{$\tau, \mathbf{g_{\scriptscriptstyle\c{}}}$}$\;
                        $ \{\shares{{\mathbf{\tilde{g}_{\scriptscriptstyle\c{}}}}}_{\scriptscriptstyle\s{}}\}_{\mathcal{S}} \gets \Protect{$\mathbf{\tilde{g}_{\scriptscriptstyle\c{}}}, \mathcal{S}$}$\;
                        \lFor{$ \s{} \in \mathcal{S}$}{ \sendTo $\shares{ {\mathbf{\tilde{g}_{\scriptscriptstyle\c{}}}}}_{\scriptscriptstyle\s{}}$ \To $\s{}$}
            \end{minipage}
            \hspace*{-0.45em} 
            \begin{minipage}[t]{0.49\textwidth}
                    \tcp{Server-side noise sampling}
                            $ \{\shares{{\mathbf{{g}_{\scriptscriptstyle\c{}}}}}_{\scriptscriptstyle\s{}}\}_{\mathcal{S}} \gets \Protect{$\mathbf{{g}_{\scriptscriptstyle\c{}}}, \mathcal{S}$}$\;
                            \lFor{$ s \in \mathcal{S}$}{ \sendTo $\shares{ {\mathbf{{g}_{\scriptscriptstyle\c{}}}}}_{s}$ \To $\s{}$}
            \end{minipage}
        }
        \begin{minipage}[t]{0.491\textwidth}
                \ForEach{ $\s{} \in \mathcal{S}$}{
                            \glob{$ \shares{\tilde{\mathbf{g}}}_{\scriptscriptstyle\s{}} \gets \Aggregate{$\{ \shares{{\mathbf{\tilde{g}_{\scriptscriptstyle\c{}}}}}_{\scriptscriptstyle\s{}}\}_{\mathcal{C}}$}$}\;
                            \lFor{$O \in \mathcal{O}$}{\sendTo $\shares{\tilde{\mathbf{g}}}_{\scriptscriptstyle\s{}}$ \To $O$}
                }
        \end{minipage}
        \hspace*{-0.45em} 
        \begin{minipage}[t]{0.49\textwidth}
            \vspace{-2.5em}
                \ForEach{ $\s{} \in \mathcal{S}$}{
                                \glob{$\shares{{\mathbf{g}}}_{\scriptscriptstyle\s{}} \gets \Aggregate{{$\{ \shares{{\mathbf{{g}_{\scriptscriptstyle\c{}}}}}_{\scriptscriptstyle\s{}}\}_{\mathcal{C}}$}}$}\;
                                {\glob{ \underline{$ \shares{\tilde{\mathbf{g}}}_{\scriptscriptstyle\s{}} \gets \Perturb(\tau,\shares{{\mathbf{g}}}_{\scriptscriptstyle\s{}})$}}\;}
                                \vspace{0.2em}
                            \lFor{$O \in \mathcal{O}$}{\sendTo $\shares{\tilde{\mathbf{g}}}_{\scriptscriptstyle\s{}}$ \To $O$}
                }

        \end{minipage}
        \ForEach{ $O \in \mathcal{O}$}{
                    \glob{$ \tilde{\mathbf{g}} \gets \Reveal{$\{\shares{\tilde{\mathbf{g}}}_{\scriptscriptstyle\s{}}\}_\slset$}$}\;
                    $\theta_{\scriptscriptstyle O}^{(\epoch+1)} \gets \ModelUpdate{$\theta_{\scriptscriptstyle O}^{(\epoch)}$, $\tilde{\mathbf{g}}, \eta$}$\;
                }
    \end{minipage}
    \hspace*{\fill}
    \begin{minipage}[t]{0.42\textwidth}
        \vspace{0.5ex}
        \small
        \textsc{\otlcap:}\;
        \scriptsize
        \ForEach(\hspace*{-5.5em}\tcp*[f]{Only first iteration}\hfill){ $\c{} \in \mathcal{C}$  }{
                        $ \{\shares{D_{\scriptscriptstyle\c{}}}_{\scriptscriptstyle\s{}}\}_{\mathcal{S}} \gets \Protect(D_{\scriptscriptstyle\c{}}, \mathcal{S})$\;
                        \lFor{$ S \in \mathcal{S}$}{ \sendTo $\shares{D_{\scriptscriptstyle\c{}}}_{\scriptscriptstyle\s{}}$ \To $\s{}$}
        }
        \vspace{-0.2em}
        \ForEach(\hspace*{-1em}\tcp*[f]{For each training iteration \enx{k}}\hfill){ $\s{} \in \mathcal{S}$}{
            \glob{
                    $ \shares{ {\mathbf{g}}}_{\scriptscriptstyle\s{}} \gets \Grad{$\shares{\theta^{(k)}}_{\scriptscriptstyle\s{}}, \{\shares{D_{\scriptscriptstyle\c{}}}_{\scriptscriptstyle\s{}}\}_{\mathcal{C}}, \clipparam$} $\;
            }\;
                    \tcp{Server-side noise sampling}
                    \vspace{-0.15em}
                    \glob{
                        \underline{
                            $ \shares{ \tilde{\mathbf{g}} }_{\scriptscriptstyle\s{}} \gets \Perturb{$\tau, \shares{\mathsf{g}}_{\scriptscriptstyle\s{}}$}$
                        }
                    }\;
        \vspace{0.15em}
        \glob{
            $ \shares{ \theta^{(\epoch+1)} }_{\scriptscriptstyle\s{}} \gets \ModelUpdate{$\shares{ \theta^{(\epoch)} }_{\scriptscriptstyle\s{}}$, $\shares{ \tilde{\mathbf{g}} }_{\scriptscriptstyle\s{}}$, $\eta$}$
        }\;
        }
        \ForEach(\hspace*{-1.0em}\tcp*[f]{Only last iteration \enx{E} (Optional)}\hfill){ $S \in \slset$}{
            \lFor{ $O \in \mathcal{O}$}{\sendTo $\shares{ \theta^{(E)} }_{\scriptscriptstyle\s{}}$ \To $O$}
        }
        \vspace{-0.2em}
        \ForEach{ $O \in \mathcal{O}$}{
                    \glob{$ \theta_{\scriptscriptstyle O}^{(E)} \gets \Reveal{$\{\shares{ \theta^{(E)} }_{\scriptscriptstyle\s{}}\}_\slset$}$}\;
                }
    \end{minipage}
\end{algorithm*}
}
\begin{figure}
\vspace{-2.5em}
\end{figure}

\subsection{Federated Learning}
\label{subsec:fl}
Fig.~\ref{fig:flow_main} identifies two design options for \fl : client-side and server-side \perturb. 
Independently of the design choice, \cls \clset iteratively compute DP-SGD's per-example clipped gradients \enx{\mathbf{g}_{\scriptscriptstyle\c{}}} via \compgrad locally. 
In client-side \perturb, clients \clset sample local \partnoise via \perturb before \protection. After receiving the encrypted noisy local updates, \sls \slset merge them in\glob{\aggr}.
Instead, in server-side \perturb, \clset only execute \protection, while \slset perform\glob{\aggr} and\glob{\underline{\perturb}}, sampling \partnoise locally or \centrnoise globally.
In both option, output parties \outparties receive noisy aggregated gradients \enx{\tilde{\mathbf{g}}} and perform \updatemodel locally. 
Before \protection, \clset quantize updates to representations suitable for encrypted computations, \eg{fixed-point}.\glob{\reveal} decrypts (and dequantizes) the noisy aggregated \enx{\shares{\tilde{\mathbf{g}}}_s}. We detail quantization approaches in App.~\ref{sec:quantization}.

\para{Privacy Guarantees}
\fl does not inherently guarantee gradient and model secrecy against \clset, since \clset locally compute gradients on cleartext data. 
However, gradient and model secrecy can be guaranteed against \slset by leveraging cryptographic techniques. Specifically, using explicit \reveal performed by \clset (i.e., \enx{\mathcal{O} = \clset}), e.g., via HE, MPC, or LWE-based masking with distributed \Dec/\Recon.
Additionally, \fl has to account for dropouts, i.e., \cls leaving during training, which can disrupt \reveal if all clients are needed, and can introduce privacy risks, e.g., leaking local updates as discussed in App.~\ref{sec:dropouts}. 

\para{Performance}
In \fl, \clset are actively involved executing \compgrad and \protection locally. While naive per-example clipping can increase \compgrad runtime by up to \enx{10\times}, cleartext optimizations, e.g., ghost clipping \cite{li2021large}, effectively reduce overhead.
The main bottleneck in \fl is \aggr, which combines local updates from \clset introducing a communication bottleneck, i.e., up to \enx{95\%} of total runtime in our evaluation (Tab.~\ref{tab:combined-dp-overhead}, Sec.~\ref{sec:noisetradeoff}). 
Here, \clset iteratively send \encrypted messages of size \enx{O(|\theta|)} to \slset.  
For HE and masking each client sends \enx{O(1)} messages; with MPC each client sends \enx{O(m)} messages. 
Instead, \slset perform \enx{O(n)} local, i.e., interaction free, additions for all techniques, and send back the aggregated result, optionally performing \reveal.  
For details on cryptographic techniques for SecAgg, we refer to SoK~\cite{mansouri2023sok}.

\subsection{\otlcap}
\label{subsec:ot}
Fig.~\ref{fig:flow_main} show that \cls \enx{\mathcal{C}} apply \protection on their local data, 
while \sls \enx{\mathcal{S}} run\glob{\compgrad} globally. 
\ol supports only server-side\glob{\underline{\perturb}} on encrypted gradients. 
Here, \enx{\mathcal{S}} execute \glob{\underline{\perturb}}, which can locally sample \partnoise; or sample \centrnoise via cryptographic protocols or via a semi-trusted \enx{\s{}'}. 
Finally, \enx{\mathcal{S}} run\glob{\updatemodel}, aggregating \partnoise. 
Optionally, \cls can pre-process local data in \setup, and \encrypt the result.

\para{Privacy Guarantees}
By design \ot does not require \aggr and \reveal, unlike \fl. Thus, 
\hlobs{OL inherently guarantees gradient and model secrecy against both clients and servers,}preventing GIA.
Clients can opt to reveal the final DP model to enable local inference without cryptographic overhead.
Additionally, for strong convex losses, hiding intermediate updates improves DP composition bounds~\cite{ye2022differentially}, potentially offering tighter guarantees than \fl.

\para{Performance}
In \ot, \cls are not actively involved during the whole learning process. They execute \protection only once and can go offline, unlike \fl. 
Here, \sls execute \compgrad over encrypted data \enx{\shares{D}_S} and model parameters \enx{\shares{\theta}_S} via cryptographic protocols. 
This requires costly approximations for non-linear functions, e.g., activations and comparisons, which introduce an accuracy-efficiency trade-off and can make \ol up to \enx{30\times} slower than \fl in non-DP settings (Tab.~\ref{tab:combined-dp-overhead}, Sec.~\ref{sec:noisetradeoff}). 
For a comprehensive overview of cryptographic \ol solutions, we refer to SoK \cite{sp/NgC23}.
Additionally, DP-SGD requires per-example gradient clipping. While \fl clients perform \clip locally and can leverage cleartext optimizations (\eg{ghost clipping \cite{li2021large}}), in \ot cryptographic \clip presents a dual challenge. 
For OL, \hlobs{there are no secure per-example clipping optimizations via cryptographic protocols.}
Also, \clip requires computation of the inverse of the square root, multiplications and comparison, 
incurring significant overhead, i.e., up to \enx{50\%} of total runtime (Tab.~\ref{tab:combined-dp-overhead}, Sec.~\ref{sec:noisetradeoff}). 
While recent advancements exist \cite{panda2022polynomial,lu2020faster}, only PEA \cite{pea2023ruan} proposes an optimized protocol for the inverse of square root based on polynomial approximations in \epcl. 
Notably, \compgrad performance can be improved indirectly, as 
\hlobs{client local pre-processing in OL reduces cryptographic overhead while maintaining high accuracy.}
Here, \cls can perform, e.g., dimensionality reduction, feature extraction, or train local models to serve as an initial global model \cite{pea2023ruan}.
Thus, \sls can train simpler models, like logistic regression, on extracted features instead of complex models, such as neural networks (NN), on raw data.
For example, PEA \cite{pea2023ruan} effectively combines these techniques, achieving a \enx{100\times} speedup by training a logistic regression on {CIFAR-10} with the accuracy of a CNN \cite{tan2021cryptgpu}. 

\section{Noise Sampling Techniques}\label{sec:noisegen}
%
This section provides a comprehensive analysis of \perturb, the foundational phase of CPCL. We first overview common design choices, then detail the sampling techniques for \partnoise and \centrnoise (Sec.~\ref{sec:sok}) by increasing cryptographic overhead. 

\para{Noise Mechanisms}
We discuss the suitability of noise distributions in Tab.~\ref{tab:noisedistributions} for noise types, perturbation options and DP guarantees. We formalize sampling algorithms (\sample) in Alg.~\ref{alg:mpc_noise} (App.~\ref{app:sampling_alg}) and provide conversion between DP and distribution parameters in {App.~\ref{app:noisedistributions}}.
Note that all noise mechanisms are model agnostic and the choice between different mechanisms is based on analytical properties of the distributions rather than model architecture.
Tab.~\ref{tab:summary} shows that most works use Gaussian-like mechanisms for \perturbgrad, and Laplace for \perturbout. 
Laplace is applied once in \perturbout providing pure DP (\enx{\delta=0}).
In contrast, \perturbgrad applies noise iteratively, where Gaussian is preferred as it is closed under summation, i.e., \enx{\sum_{i=1}^{N}\mathcal{N}_i(\mu_i,\sigma^2_i) = \mathcal{N}(\sum_{i=1}^{N}\mu_i, \sum_{i=1}^{N}\sigma^2_i)}. Moreover, Gaussian has faster tail decay than Laplace, 
and the noise scales with the \enx{l_2}-sensitivity, often smaller than the \enx{l_1} for Laplace \cite{mironov2017renyi}.
Both noise mechanisms can be adapted to be aggregated from \partnoise (Sec.~\ref{subsec:partialnoise}): Laplace via the difference of \enx{N} i.i.d. Gamma random variables \cite{mugunthan2019smpai,byrd2022collusion}, and Gaussian leveraging its closure under summation.
%
Notably, secure sampling of continuous noise typically requires quantization as cryptography uses fixed-point representations \cite{agarwal2018cpsgd}. Alternatively, discrete distributions enable sampling from discrete domains. Next, we review discrete distributions that mimic Gaussian guarantees for \perturbgrad, which systematized works apply for \partnoise \cite{agarwal2021skellam,kairouz2021practical,agarwal2018cpsgd}. 
The \emph{binomial} mechanism \cite{agarwal2018cpsgd,pea2023ruan} satisfies \enx{(\eps, \delta)}-DP, and it is closed under summation, i.e., \enx{\mathsf{Bin}(m,p) = \sum_{i}^{n}\mathsf{Bin}(m_i,p)}.
The sum of \emph{Discrete Gaussian} \cite{kairouz2021distributed} satisfies \enx{(\alpha, \eps)}-RDP, despite not being closed under summation. 
%
%
The \emph{Skellam} mechanism \cite{agarwal2021skellam,bao2022skellam}, realized as the difference of two independent Poisson random variables, satisfies \enx{(\alpha, \eps)}-RDP and is closed under summation. 
Note that all mechanisms in Tab.~\ref{tab:noisedistributions} add noise to gradients
except for Poisson-Binomial
\cite{chen2022poisson} which encodes local gradients into a parameter of the binomial and outputs a discrete noisy value.
\begingroup 
\begin{table}[t]
    \scriptsize
    \centering
    \setlength{\tabcolsep}{0.7em}
    \renewcommand{\arraystretch}{1.4}
    \caption{DP noise distributions.
    Random variables are denoted like PDFs without dependence on \enx{x}.
    }
    \label{tab:noisedistributions}
    \vspace{-1.0em}
    \begin{tabular}{ll}
        \theader{Distribution} &
        \theader{Probability Density Function (PDF)}
        \\
        \cline{1-2}
        Laplace  &
        \enx{\mathsf{Lap}(x; \lambda) = \frac{1}{2\lambda}\exp (|x|/\lambda )} 
        \\
        Dist.~Laplace &
        $\mathsf{Lap}(\lambda) =  \sum_{k=1}^{N}  g_k^1 - g_k^2$;\; $g_k^1,g_k^2\sim \mathsf{Gamma}(x;\frac{1}{N},\lambda)$
        \\
        Gaussian &
        \enx{\mathcal{N}(x;\sigma^2) = \frac{1}{\sigma\sqrt{2\pi}}\exp(-x^2/2\sigma^2)}
        \\
        Binomial &
        \enx{ \mathsf{Bin}(x;m,p) = \binom{m}{x}p^x(1-p)^{m-x} }
        \\
        Disc.~Gaussian &
        $\mathcal{N}_\mathbb{Z}(x;\sigma^2) = e^{-x^2/2\sigma^2}/\sum_{y\in \mathbb{Z}} e^{-y^2/2\sigma^2}$
        \\
        Skellam  &
        $\mathsf{Sk}(\mu)= p_1 - p_2$;\; $p_1,p_2 \sim \mathsf{Poisson}(x;\mu/2)$
        \\
        Poisson-Binomial &
        $\mathsf{PoiBin}(x;\mathbf{g},b,p) =  \mathsf{Bin}(x; b, \mathbf{g} \theta/K + 1/2)$\;
        \\
    \end{tabular}%
    
    \renewcommand{\arraystretch}{1.0}
    \vspace{-2.0em}
\end{table}

\para{Obliviousness}
We distinguish noise obliviousness at selection and sampling level. 
\emph{Oblivious selection}~\cite{froelicher2020drynx} can be realized by a client or server locally sampling a list of candidate noises, encrypting and shuffling them; then another party selects a value from the encrypted list.
\emph{Oblivious sampling} \cite{jayaraman2018distributed} requires the noise to be sampled via cryptographic protocols, ensuring it remains encrypted and hidden from all \pls.

\para{Privacy Unit}
While for record-level DP, \clset or \slset clip per-record gradients, user-level DP requires bounding the contribution of all records from a user.
In \fl, each user holds only its own data and the record-level guarantee can be extended to user level by clipping local model updates as in FedAvg \cite{mcmahan2017learning}. Instead, in \ol, local datasets are pooled in a global dataset. Here, \slset need to securely identify user records and bound per-user contribution via cryptographic protocols. Furthermore, \hlobs{no established best practice exists for user-level DP in multi-user datasets.}A straightforward way is to emulate FedAvg in \ol: \sls sample records per user from the global dataset, and then compute and clip per-user gradients \cite{chuamind}. 
Alternatively, DP satisfies \emph{group privacy} enabling the translation from record- to user-level DP: data containing at most \enx{z} records per user satisfies user-level \enx{(z \eps,z \exp(z\eps) \delta)}-DP~\cite{dwork2014algorithmic}. However, this requires strict contribution bounding and rapidly weakens privacy guarantees for large \enx{z}.

%
%

\subsection{\centrnoise: Centralized Sampling}\label{subsec:centralnoise}
In centralized sampling, a semi-trusted server \enx{S'} samples \centrnoise \enx{\psi} in clear. This noise is then added to the \encrypted aggregated gradient, either by sending encrypted noise to the computing servers (\underline{\perturb}) or by adding it directly if \enx{S'} is oblivious to the gradient values.

\para{Rationale \& Trade-offs}
This technique offers the optimal privacy-accuracy trade-off, as in CDP, since noise is added once to global data, e.g., aggregated gradients in FL. 
Computationally, it is highly efficient because sampling occurs in clear. 
In our evaluation (Sec.~\ref{sec:noisetradeoff}), we estimate this adds negligible runtime overhead to \ot and small latency to \fl ($\approx 10\%$ runtime increase due to communication). 
However, it introduces a strong trust assumption: \enx{S'} can learn either the noisy output \enx{\tilde{\mathbf{g}}} or the noise \enx{{\psi}}, \emph{never} both. 
%

\para{Obliviousness Considerations}
Obliviousness to the noisy output \enx{\tilde{\mathbf{g}}} is enforced with different techniques, depending on the learning paradigm and the set of output parties \enx{\outparties}.
In \fl, if \enx{\slset \subset \mathcal{O}}, separation of duties can be applied: one server \enx{S' \notin \slset} can sample and \encrypt \centrnoise $\psi$, then distribute the encrypted noise term $\shares{\psi}_s$ to each \enx{S \in \mathcal{S}} for decentralized noise addition. 
Alternatively, one server \enx{S' \notin \slset} samples and adds \centrnoise \enx{\psi} to the \encrypted gradients $\shares{{\mathbf{g}}}_s$, while \enx{\slset} decrypt the noisy aggregate ${\tilde{\mathbf{g}}}$. The latter solution suits, e.g., HE-based FL, where a decryption key is required. Instead, MPC-based solutions require additional masking or re-sharing to prevent \enx{S'} from learning the gradients \cite{choquette2021capc}.
%
%
In \ot, servers only see random-looking, encrypted data, i.e., \enx{\slset \not\subseteq \outparties}, inherently guaranteeing obliviousness to \enx{\tilde{\mathbf{g}}}. Thus, a server \enx{S'\in \mathcal{S}} can sample and add \centrnoise \enx{\psi} to its local \encrypted gradients \enx{\shares{\mathbf{g}}_{s'}}.

Noise obliviousness can be achieved via oblivious noise selection, regardless of the learning paradigm. Here, \enx{S'} samples and encrypts a list of \enx{l} \centrnoise terms. The list is then shuffled, one term is obliviously selected and added to the \encrypted gradients.
For example, Drynx's \cite{froelicher2020drynx} \sls run a verifiable shuffle protocol \cite{neff2003verifiable}, and select the first encrypted \centrnoise in the shuffled list. 
The probability of guessing the chosen noise decreases linearly with \enx{l}, i.e.,~{\enx{\delta \propto \enx{1/l}}}.
%

\para{Obliviousness Overhead}
With output obliviousness (\enx{S' \notin \outparties}), \enx{S'} either locally adds noise over encrypted data, or distributes \encrypted noise \enx{\shares{\psi}_s} in \enx{O(m)} messages. 
Noise obliviousness via oblivious selection introduces a privacy-performance trade-off: increasing the number \enx{l} of noise terms enhances privacy, but also increases cryptographic overhead as computation and communication scale linearly with \enx{l} \cite{froelicher2020drynx}. 

\xx{
\para{Observations}
To ensure DP, the noise provider \enx{S'} can not learn either the noisy output or noise.
Noise obliviousness can be enforced by separation of duties, \ie{one server adds noise to the encrypted value and another server decrypts only the noisy aggregate \cite{choquette2021capc}}.
Alternatively, oblivious noise selection can be implemented to hide the noise from \enx{S'}. For example, Drynx \cite{froelicher2020drynx} proposes a collaborative oblivious noise selection protocol. In more detail, a selected \slong samples a list of candidate noises in clear text. Then, all the other servers in the protocol execute a verifiable shuffle protocol sequentially \cite{neff2003verifiable}. The first noise in the shuffled list is selected and added to the encrypted update. The probability of an adversary guessing the chosen noise decreases linearly with the size of the noise list, \ie{\enx{\delta} is inversely proportional to the size of the list} \cite{froelicher2020drynx}.
}

%
%
%

\xx{
\para{Overhead}
Since noise is sampled in clear, the main overhead comes from secure aggregation of labels in CAPC \cite{choquette2021capc}, and oblivious noise selection in Drynx \cite{froelicher2020drynx}. In CAPC, the communication cost increases linearly in the number of parties \enx{S'} receives the input from. Instead, in Drynx, the communication cost increases linearly in the number of shuffling rounds and noise candidates, resulting in a privacy-performance trade-off.
Additionally, each shuffling round involves a number of exponentiations linear in the list's length \cite{neff2003verifiable}.
}


\subsection{\partnoise: Local Sampling And Aggregation}\label{subsec:partialnoise}
In partial noise aggregation, each client \enx{\c{} \in \mathcal{C}} locally samples non-DP \partnoise \npartial{\c{}}. Notably, while \npartial{\c{}} alone does not satisfy DP, their sum \enx{\sum_{\scriptscriptstyle\c{} \in \mathcal{C}} \npartial{\c{}}} satisfies CDP.
As Alg.~\ref{alg:cmp} shows, in \fl, \cls add \npartial{\c{}} on local gradients \enx{\mathbf{g_{\scriptscriptstyle\c{}}}} and then \protection ensures local gradient remain hidden from servers. 
Alternatively, each server \enx{\s{} \in \mathcal{S}} adds \partnoise \npartial{\s{}} on local \encrypted gradients \enx{\shares{\mathbf{g}}_{\scriptscriptstyle\s{}}} in both \fl and \ol. 
%
%
%

\para{Rationale \& Trade-offs}
\partnoise is the dominant technique in \fl as it allows local sampling in cleartext, avoiding cryptographic protocols. 
In our evaluation, the runtime overhead is only around $2\%$ for \fl and negligible in \ot (Tab.~\ref{tab:combined-dp-overhead}, Sec.~\ref{sec:noisetradeoff}). 
%
To satisfy \enx{(\eps, \delta)}-DP, a specific variance \enx{\sigma^2_{\mathrm{DP}}} is required (App.~\ref{app:noisedistributions}). For example, each \enx{\c{} \in \mathcal{C}} (or \enx{\s{} \in \slset}), samples Gaussian noise with variance \enx{\sigma^2_{\scriptscriptstyle\c{}} \ge \sigma^2_{\scriptscriptstyle\mathrm{DP}} / n}. 
Importantly, in \fl, noise-sampling \cls learn aggregated noisy updates \enx{\tilde{\mathbf{g}} = 1/n\sum_{{\scriptscriptstyle\c{}} \in \clset} (\mathbf{g}_{\scriptscriptstyle\c{}} + \psi_{\scriptscriptstyle\c{}})}. A passive adversary corrupting a client, or up to \enx{t} colluding \cls, can remove their noise \enx{\psi_{\scriptscriptstyle\c{}}} from the noisy aggregated \enx{\tilde{\mathbf{g}}}, thereby weakening the targeted privacy guarantee and potentially launching a GIA (App.~\ref{app:privacyattacks}). 
Thus, each client must account for potential collusion by choosing a higher minimum variance \enx{\sigma^2_{\scriptscriptstyle\c{}}} adjusted for \enx{t}:
{
\setlength{\abovedisplayskip}{0.4em}
\setlength{\belowdisplayskip}{0.6em}
\begin{equation}\label{eq:partgaussvariance}
    \enx{\sigma^2_{\scriptscriptstyle\c{}} \ge \frac{1}{n-t} \sigma^2_{\scriptscriptstyle\mathrm{DP}}}. 
\end{equation}
}

The adjusted variance ensures \enx{\tilde{\mathbf{g}}} satisfies CDP even if \enx{t} colluding \cls remove their noise.
With large \enx{n} (e.g., 100) and high \enx{t} (e.g., \enx{n-1}), total noise increases beyond LDP variance, degrading accuracy to near random guessing (Tab.~\ref{tab:accuracy}, Sec.~\ref{sec:noisetradeoff}). 
To also tolerate up to $s$ dropouts, the denominator of Eq. \eqref{eq:partgaussvariance} can be set to \enx{n - (t + s)}.
Most \perturbgrad-based \fl works do not account for collusion (Tab.~\ref{tab:summary}), since it either increases the total noise variance (degrading accuracy) or introduces cryptographic overhead, as discussed below.

\xx{
\para{Noise Amount Considerations}
We illustrate the required per-party \partnoise variance via the Gaussian mechanism. 
To satisfy \enx{(\eps, \delta)}-DP, a specific variance \enx{\sigma^2_{\mathrm{DP}}} is required (App.~\ref{app:noisedistributions}). 
Each \enx{\c{} \in \mathcal{C}} in \fl (or \enx{\s{} \in \slset}), samples Gaussian noise with variance \enx{\sigma^2_{\scriptscriptstyle\c{}} \ge \sigma^2_{\scriptscriptstyle\mathrm{DP}} / n}. 
%
%
%
Importantly, in \fl, noise-sampling \cls learn aggregated noisy updates \enx{\tilde{\mathbf{g}} = 1/n\sum_{{\scriptscriptstyle\c{}} \in \clset} (\mathbf{g}_{\scriptscriptstyle\c{}} + \psi_{\scriptscriptstyle\c{}})}. A passive adversary corrupting a client, or up to \enx{t} colluding \cls, can remove their noise \enx{\psi_{\scriptscriptstyle\c{}}} from the noisy aggregated \enx{\tilde{\mathbf{g}}}, thereby weakening the targeted privacy guarantee and potentially launching a GIA (App.~\ref{app:privacyattacks}). 
Thus, each client must account for potential collusion by choosing a higher minimum variance \enx{\sigma^2_{\scriptscriptstyle\c{}}} adjusted for \enx{t}:
{
\begin{equation}\label{eq:partgaussvariance}
    \enx{\sigma^2_{\scriptscriptstyle\c{}} \ge \frac{1}{n-t} \sigma^2_{\scriptscriptstyle\mathrm{DP}}}. 
\end{equation}
}
%
The adjusted variance ensures \enx{\tilde{\mathbf{g}}} satisfies CDP even if \enx{t} colluding \cls remove their noise.
With large \enx{n} (e.g., 100) and high \enx{t} (e.g., \enx{n-1}), total noise increases beyond LDP variance, degrading accuracy to near random guessing (Tab.~\ref{tab:accuracy}, Sec.~\ref{sec:noisetradeoff}). 
%
To also tolerate up to $s$ dropouts, the denominator of Eq. \eqref{eq:partgaussvariance} can be set to \enx{n - (t + s)}.
Most \perturbgrad-based \fl works do not account for collusion (Tab.~\ref{tab:summary}), since it either increases the total noise variance (degrading accuracy) or introduces cryptographic overhead, as discussed below. 
%
}

\para{Oblivious Selection}
To reduce the noise variance in Eq.~\eqref{eq:partgaussvariance}, 
\hlobs{local $\mathsf{\textcolor{blue!60!black}{{PNoise}}}$ sampling can be enhanced with noise obliviousness.}We distinguish \emph{client-side} and \emph{server-aided} protocols.
In client-side protocols, each client receives \enx{l} encrypted \partnoise terms and obliviously selects one. The \cls can obtain noise terms from topological neighbors \cite{mugunthan2019smpai}, or a \slong can shuffle and forward \enx{l} encrypted noises per client \cite{byrd2022collusion}. 
For the latter, \enx{t} colluding \cls do not know who generated their noise but may still average their noise contributions to weaken the DP guarantee. However, even \enx{l=2} and \enx{t=n-1} is not enough to infer honest \cls' private information \cite{byrd2022collusion}.
Alternatively, in server-aided protocols, clients and servers jointly select noise terms. 
Specifically, each client locally samples and encrypts \enx{l} \partnoise terms. The server generates and encrypts an \enx{l}-bit selection vector to select one \partnoise per client (i.e., with 1 for selection, 0 for omission). 
The encrypted selection vector and noise terms are multiplied element-wise to produce oblivious \partnoise samples. For example, the server sends HE-encrypted selection vectors to clients, which are multiplied with \enx{l} encrypted \partnoise~\cite{bindschaedler17star}. 
%

%

%

\para{Obliviousness Selection Overhead}
Unlike local \partnoise sampling, oblivious noise selection incurs cryptographic overhead. 
%
In client-side protocols \cite{byrd2022collusion,mugunthan2019smpai}, each client samples \enx{l} noise terms for any other client or a subset of neighboring clients, resulting in \enx{O(n l)} local sampling and encryptions. 
The encrypted noise can be sent to a central server acting as a relay which can batch the encrypted lists in \enx{O(n)} messages \cite{byrd2022collusion}, or exchanged between clients in \enx{O(n^2)} messages \cite{mugunthan2019smpai}.
Computation and communication costs scale linearly in \enx{n} and \enx{l}. Runtime is mainly dominated by network latency, though its impact diminishes as \enx{n} increases \cite{byrd2022collusion}. 
%
Server-assisted protocols \cite{bindschaedler17star} reduce communication cost to \enx{O(n)} messages containing \enx{l} selection bits. 
%
Due to the cryptographic overhead, only \perturbout works use oblivious selection \cite{byrd2022collusion,mugunthan2019smpai,bindschaedler17star}, as the oblivious protocol is run once.

\para{Noise Mechanisms Considerations}
\hlobs{The choice of noise distribution affects utility and performance.}Distributions not closed under summation can diverge significantly at small noise levels, \ie{large number of \cls} \cite{agarwal2021skellam}. For discrete Gaussian with \enx{n=10^4} and \enx{\sigma_i = 0.5}, the RDP bound is $10^6\times$ larger than for Skellam \cite{agarwal2021skellam}.
%
%
For the binomial \cite{agarwal2018cpsgd}, we are not aware of advanced accounting methods (\eg{moments accountant}), which may result in injecting suboptimal noise.
Currently, ML libraries \cite{tensorflow2015-whitepaper,paszke2019pytorch} support efficient sampling for certain distributions, \eg{Gaussian and Skellam}, but not for the discrete Gaussian \cite{kairouz2021distributed} which can be up to \enx{40\%} slower than Skellam \cite{agarwal2021skellam}. 
%
Among discrete distributions, Skellam is the most efficient and accurate choice for \partnoise reaching the accuracy of continuous Gaussian using half-precision (16 bits) \cite{agarwal2021skellam}.
%
%
Notably, inherent errors from cryptographic approximations and quantization can serve as DP noise, suggesting that 
\hlobs{DP~can~be~embedded~in~phases~besides~\perturbbold,}i.e., in \protection or during quantization. 
Non-additive mechanisms, like Poisson-Binomial \cite{chen2022poisson}, directly map continuous gradients to discrete noisy values. Also, non-additive discrete mechanisms have bounded support and their communication costs decrease with \eps \cite{chen2022poisson}.
Remarkably, noise-based cryptographic techniques can integrate DP without sampling additional noise. 
For example, in LWE-based masking \cite{stevens2022efficient} a noise vector \enx{e} breaks the linearity of a set of equations. The vector \enx{e} itself can guarantee DP, \eg{by being sampled from a discrete Gaussian distribution \cite{stevens2022efficient}}. Thus, the LWE decryption does not remove all noise, as typically desired, but DP-suitable noise remains.

\xx{
In partial noise aggregation, each client \c{i} is in charge of sampling non-DP noise \npartial{i} in clear via \partnoise. Then, \npartial{i} is added to the encrypted local update. On its own, each \npartial{i} does not satisfy LDP, but their sum \enx{\sum_{i}^{n} \npartial{i}} satisfies CDP\footnote{
   Notably, \citet{mugunthan2019smpai} satisfy LDP and not CDP via partial noises (App.~\ref{app:partialldp}).
}. To ensure that only the aggregated noisy update can be decrypted, \protection uses threshold SS or HE.
}

%

\xx{
\para{Noise Mechanisms}
%
%
For \perturbout each client samples \npartial{i} from a gamma distribution with shape \enx{1/N} and scale \enx{\lambda}.
For \perturbgrad, there are more options.
\todo{This is already all explained in Table 1, right? So to shorten it further, can't we just briefly reference which distribution is applicable and that parameterization is defined in table 1?}
For the Gaussian mechanism, each client samples \npartial{i} from a Gaussian with variance \enx{\sigma^2_i} \cite{truex2019hybrid}.
The same applies to the discrete Gaussian.
The Binomial mechanism \cite{dwork2006our} can be implemented with verifiable secret sharing \cite{damgaard2006unconditionally} to generate shares of random bits (0,1), and convert those to integer shares to sum them.
For the Skellam mechanism \cite{agarwal2021skellam,bao2022skellam} each client samples a Skellam noise with variance \enx{\mu_i} as a difference of two Poisson random variables with variance \enx{\mu_i/2}.
For the Poisson-Binomial mechanism \cite{chen2022poisson} each client encodes the local gradient (\enx{g_i}) into the parameter \enx{p_i = \frac{\theta/}{K} g_i + 1/2}, with \enx{\theta \in [0,1/4]}, of a binomial distribution \enx{\mathsf{Binom}(m_i, p_i)}, and output a sample from this distribution.
Since this mechanism is optimized for \enx{l_\infty} geometry, each client computes the Kashin representation \cite{lyubarskii2010uncertainty} of \enx{g_i}, \ie{transforms the geometry of the data from the \enx{l_2} to \enx{l_\infty}}, before applying the mechanism. The \slong reverts the Kashin representation to \enx{l_2} geometry after the aggregation.
}
%

\xx{
\para{Noise Mechanisms}
\todo{First distributions, then details}
Distributed Laplace can be expressed as difference of \enx{N} i.i.d. \emph{gamma} random variables
\cite{mugunthan2019smpai,byrd2022collusion,bindschaedler17star} (Tab.~\ref{tab:noisedistributions}).
Gaussian distribution is closed under summation and its distributed version is realized by summing \enx{N} i.i.d. Gaussians \enx{\mathcal{N}_i(\mu_i,\sigma^2_i)}, obtaining \enx{\mathcal{N}(\sum_{i=1}^{N}\mu_i, \sum_{i=1}^{N}\sigma^2_i)} \cite{truex2019hybrid}.
{Continuous} noise needs quantization as cryptographic techniques typically use fixed-point instead of floating-point numbers for efficiency.
Discrete distributions provide an alternative by directly sampling from discrete domains.
Next, we focus on discrete mechanisms that mimic the Gaussian: binomial, discrete Gaussian, Skellam (Tab.~\ref{tab:noisedistributions}).
The \emph{binomial} mechanism \cite{agarwal2018cpsgd,pea2023ruan} satisfies \enx{(\eps , \delta)}-DP, if \cls hold enough shares \enx{m} of unbiased coins \enx{p=0.5}, \ie{bits} (details in Tab.~\ref{tab:noisedistributions}).
The straightforward solution by \citet{dwork2006our} uses verifiable secret sharing \cite{damgaard2006unconditionally} to generate shares of random bits (0,1), then converts those to integer shares to sum them.\footnote{However, to our knowledge, the binomial has not been analyzed with advanced accounting methods, \eg{moments accountant}, and its composition might be suboptimal.}
%
While not closed under summation, sum of \emph{discrete Gaussians} still satisfies \enx{(\alpha, \eps)}-RDP and has been analyzed with moments accountant \cite[Section 3]{kairouz2021distributed}. The upper bound of the sum of discrete Gaussians leads to privacy guarantees extremely close to continuous Gaussian
\footnote{The sum of two discrete Gaussian with \enx{\sigma^2=3} leads to an error of \enx{10^{-12}} on the max divergence with the sum of continuous Gaussian \cite[Theorem 9]{kairouz2021distributed}}. However, the discrete Gaussian can not be sampled efficiently according to \cite{agarwal2021skellam}.
On the other hand, the \emph{Skellam} mechanism \cite{agarwal2021skellam,bao2022skellam} can be efficiently computed as the difference of two independent Poisson random variables, and matches the RDP composition bound of the discrete Gaussian \cite[Fig.~3]{agarwal2021skellam}. Furthermore, the Skellam distribution is closed under summation.
\todo{Disc Gaussian not an issue if centrally sampled, since the noise}
}


\xx{
\para{Oblivious Selection}
To reduce the variance of Eq. \eqref{eq:partgaussvariance}, client has to be oblivious to its noise \npartial{i}.
To guarantee oblivious selection, some \fl works \cite{mugunthan2019smpai,byrd2022collusion} let each client receive two encrypted noise terms and obliviously select one. Specifically, \cls receive two encrypted gamma samples for the distributed Laplace. Then, each client samples its own gamma noise, in clear, and subtracts one of the encrypted gamma noises obliviously. The two gammas can come either from a topological neighbor \cite{mugunthan2019smpai} {(Fig.~\ref{fig:onlyclientsnoise})} or from any client in the protocol, using the \slong as a shuffler \cite{byrd2022collusion} {(Fig.~\ref{fig:clientservernoise})}.
In the latter, \enx{t} colluding \cls do not know who generated their noise, but can still average their noise contributions sent to the server to reduce the DP guarantee. However, \citet{byrd2022collusion} observe that this is not enough to infer honest parties' private information and can guarantee resistance against \enx{n-1} colluding \cls.

\citet{bindschaedler17star} implement a server-aided oblivious selection 
guaranteeing resistance against \enx{n-1} colluding \cls and a malicious \slong (\s{}). To achieve CDP, \s{} {obliviously} selects \enx{N/n} gammas, {from each client candidate noise vector}, via a set of binary {selector} vectors, \ie{1 indicates selection and 0 omission}. The vectors are
encrypted with AHE \cite{damgaard2001generalisation} and sent to \cls who multiply them with their noise vector and mask each entry of the resulting vector before sending it to \s{}.
Then, \s{} sums all entries and masks cancel out. 
AHE and masking hide the noise selection and noise value from clients and server, respectively.
%
Also, ZKPs ensure selector vectors contain only 0 and 1 (App.~\ref{app:vercomp}). 
}
\xx{
\para{Noise Mechanisms Considerations}
%
The underlying noise distribution itself is an important consideration.
\obs{Choice of noise distribution affects utility and performance}.
%
Distributions not closed under summation, \eg{discrete Gaussian}, can diverge significantly at small noise levels, \ie{large number of \cls} \cite{agarwal2021skellam}. For the discrete Gaussian, with \enx{n=10000} and \enx{\sigma_i = 0.5}, the RDP bound is around $10^6$ times larger than for Skellam \cite{agarwal2021skellam}.
Among discrete distributions, Skellam and discrete Gaussian achieve the same accuracy as continuous Gaussian with bit-width of only 16 bits 
\cite{agarwal2021skellam}.\footnote{
    For a small CNN (\enx{2^{20}} parameters) for the EMNIST dataset, with \enx{\eps \in [1,20]}.
    }
%
%
%
For some noise mechanisms, \eg{the binomial \cite{agarwal2018cpsgd}}, we are not aware of advanced accounting methods (\eg{moments accountant}), which typically results in loose bounds and hence may inject suboptimal noise.
%
%
Currently, ML libraries \cite{tensorflow2015-whitepaper,paszke2019pytorch} support efficient local sampling for certain distributions, \eg{Gaussian and Skellam}, but not for the discrete Gaussian \cite{agarwal2021skellam} which is up to \enx{40\%} slower than Skellam in Tensorflow \cite{agarwal2021skellam}. 
However, implementing efficient sampling routines, \eg{\cite{dgs}}, mitigates this issue. 
Accuracy and efficiency makes Skellam the ready choice for \partnoise for quantized models.

Notably, approximation of non-linear functions for cryptography, and quantization inherently introduce errors during computation which can be leveraged as DP noise. Specifically, \obs{DP can be embedded in phases besides \perturbbold}, \ie{in \protection or during quantization in \textbf{\compgrad}}. 
Non-additive mechanisms, \eg{the Poisson-Binomial mechanism  \cite{chen2022poisson}}, take as input a continuous gradient and directly output a discrete noisy value. This also allows reducing communication costs, since non-additive mechanisms have bounded support and the communication cost decreases as \eps goes to 0 \cite{chen2022poisson}.
Moreover, noise-based cryptographic techniques can integrate DP to avoid sampling additional noise, \eg{in \emph{learning with error} (LWE) (App.~\ref{app:lwe})}.
LWE requires a noise vector to break the linearity of a set of equations. The noise vector itself can guarantee DP, \eg{by being sampled from a discrete Gaussian distribution \cite{stevens2022efficient}}. Thus, the LWE decryption does not remove all noise, as typically desired, but DP-suitable noise remains.

}

\xx{
\para{Overhead}
%
The overhead in \partialnoise comes from the sampling algorithm and the oblivious selection protocol. Local sampling is efficient for certain distributions, \eg{Gaussian and Skellam}, but not for discrete Gaussian as it lacks efficient routines in ML libraries \cite{tensorflow2015-whitepaper,paszke2019pytorch}.
The utility loss in \partialnoise depends on the noise distribution and security assumptions, \ie{colluding parties}.
Although discrete Gaussian sampling is up to \enx{40\%} slower than Skellam in Tensorflow \cite{agarwal2021skellam}, there exist efficient routines for it that can be implemented in ML libraries, \eg{\cite{dgs}}.

\todo{Put connection, Notably Poi-Bin takes another route}
\obs{DP can be embedded in phases besides \perturb}, \ie{in \protection or quantization for \compgrad}. Non-additive discrete mechanisms take as input a continuous gradient and directly output a discrete noisy value, \eg{the Poisson-Binomial mechanism  \cite{chen2022poisson}}\todo{Simplify shorten, simplify shorten}. This allows to also reduce the communication cost, since non-additive mechanisms have bounded support and the communication cost decreases as \eps goes to 0 \cite{chen2022poisson}.
Noise-based cryptographic techniques can integrate DP to avoid sampling additional noise, \eg{the \emph{learning with error} problem \cite{regev2009lattices} requires a noise vector \enx{e} to break the linearity of a set of equations \enx{b = As + e}. \citet{stevens2022efficient} samples \enx{e} as DP noise from a discrete Gaussian.} \todo{remove name, generalize ...}

\obs{Oblivious selection protocols introduce overhead}, since it introduces \enx{O(1)} communication rounds and \enx{O(l)} additional encryptions per \cl.
}

\xx{
\para{Utility Loss}
The utility loss in \partialnoise depends on the noise distribution and security assumptions, \ie{colluding parties}.
\obs{The choice of the noise distribution affects utility}. Distributions not analyzed with advanced accounting methods, \eg{moments accountant}, might inject suboptimal noise, \eg{the binomial} \cite{agarwal2021skellam}.
Furthermore, distributions not closed under summation, \eg{discrete Gaussian}, can diverge significantly at small noise levels, \ie{large number of \cls} \cite{agarwal2021skellam}. For the discrete Gaussian, with \enx{n=10000} and \enx{\sigma_i = 0.5}, the RDP is around $10^6$ times larger than the Skellam \cite{agarwal2021skellam}.
Skellam and discrete Gaussian achieve the same accuracy as continuous Gaussian with bit-width \enx{b=16}\footnote{For a small convolutional NN (\enx{2^{20}} parameters) for the EMNIST dataset.} \cite{agarwal2021skellam}.
The Poisson-Binomial matches the results of discrete Gaussian and Skellam for distributed mean estimation with \enx{b=14}.
For small bit-widths, \eg{8-10} bits, \citet{bao2022skellam} provide a new analysis for Skellam which includes also the noise in the quantization. This allows to reduce the amount of noise, and thus manages to achieve a meaningful utility.
Regarding utility loss from security assumptions (Eq. \eqref{eq:partgaussvariance}), 
for \enx{t=0}, 
the sum of the partial noises achieves CDP,
but for \enx{t=n-1}, \cls can just add LDP noise directly \cite{mugunthan2019smpai}.
%
Depending on \enx{t}, partial noises lie between those two extremes, i.e., LDP and CDP.
Assuming at most half of the clients collude, \ie{(\enx{t \le n/2})}, the variance of \enx{\sum_i \npartial{i}} doubles to \enx{2\sigma^2_{DP}}, leading to a non-negligible impact on the model utility.

\obs{Oblivious selection protocols introduce overhead}, since it introduces \enx{O(1)} communication rounds and \enx{O(l)} additional encryptions per \cl.

}

\subsection{\centrnoise: Distributed Sampling} \label{subsec:distributed-noise-sampling}
In distributed sampling, \sls \slset collaboratively run \glob{\perturb} to sample \centrnoise \enx{\shares{\psi}} via cryptographic protocols, e.g., MPC.
As Alg.~\ref{alg:cmp} shows, each $\s{}\in\mathcal{S}$ adds a noise share \enx{\shares{\psi}_{\scriptscriptstyle\s{}}} to the local \encrypted \enx{\shares{\mathbf{g}}_{\scriptscriptstyle\s{}}}.
%
%
%

%
%
%

\para{Rationale \& Trade-offs}
This approach provides the best accuracy and security guarantees. It neither requires a semi-trusted \s{}' as in centralized \centrnoise, nor increases noise variance to handle collusion as in \partnoise.
However, sampling via cryptographic protocols incurs significant performance overhead. 
Our evaluation shows that generating a single Gaussian sample via MPC is about \enx{10^3\times} slower than local sampling (Tab.~\ref{tab:combined-dp-overhead}, Sec.~\ref{sec:noisetradeoff}). 
While this overhead is small for \ot it increases \fl runtime by nearly \enx{10\times}, making it impractical. 
\para{MPC Sampling Challenges}
Basic continuous sampling algorithms like {Inverse Transform Sampling} \cite{law2007simulation} and {Box-Muller} \cite{box1958note} enable {\Laplace} and {\Gauss} mechanisms in MPC by transforming uniform samples \enx{u \sim U(0,1)} via fixed arithmetic operations (\eg{\enx{\log}, \enx{\sin}}) \cite{chase2017,jayaraman2018distributed,pentyala2022training} as detailed in Alg.~\ref{alg:mpc_noise}, App.~\ref{app:sampling_alg}.
To generate a truly random $u$, \sls can sample and combine local values, e.g., via \emph{XOR} \cite{jayaraman2018distributed}.
Recent works \cite{keller2024secure, fu2024benchmarking, wei2023securely, meisingseth2025practical} explore MPC sampling of discrete Laplace and Gaussian distributions, but not in \epcl (gaps {\doubledash} in Tab.~\ref{tab:summary}). Typically, \epcl uses quantized representations with fixed bit-width, e.g., fixed point, which require scaling the noise variance according to the quantization scale \enx{s}, to avoid output only multiple of \enx{s} \cite{agarwal2021skellam}. This leads to unexplored high variance  affecting performance and accuracy. 
Note that while continuous Gaussian scales with $\sigma^2$ and $\mu$, i.e., \enx{\sigma\cdot\mathcal{N}(0,1)+\mu = \mathcal{N}(\mu,\sigma^2)}, this does not hold for discrete Gaussian \cite{canonne2020discrete}. 
Additionally, no existing work samples Skellam via MPC \cite{agarwal2021skellam}. 
A key challenge is that often \hlobs{sampling algorithms for discrete distributions have non-constant runtimes}due to rejection sampling, which samples iteratively until a specific condition is met. This introduces data-dependent loops, \eg{\code{while}}, whose runtime leaks information on the noise values, e.g., the loop count in {\Poisson} reveals the sampled noise $p$ (Alg.~\ref{alg:mpc_noise}).
To prevent leaks, loops with a fixed iteration count, \eg{\code{for}}, must replace data-dependent ones. The iteration count is set to ensure, except with negligible failure probability, that at least one correct sample is returned.
%
%
Additionally, fixed bit-width can lead to underflow or overflow, e.g., $\exp(-1/\mu)$ in \textsf{Poisson} can cause underflow for large $\mu$. 
%

\para{MPC Sampling Overhead} 
The runtime of iterative algorithms for discrete noise sampling in MPC depends mainly on the fixed iteration count set to ensure constant-time execution, which is affected by DP parameters, bit-width, and failure probability \cite{keller2024secure, wei2023securely}.
To evaluate the overhead, we implement MPC versions of {\DiscGaussian} \cite{kairouz2021distributed} and \Skellam \cite{knuth1997art} (Alg.~\ref{alg:mpc_noise}).
We scale the variance of {\DiscGaussian} (\enx{s^2\sigma^2 = 29698}) and {\Skellam} (\enx{s^2\mu = 30840}) for \enx{\eps=1}, \enx{\delta=10^{-6}} and \enx{\clipparam=0.1}, according to \cite{kairouz2021distributed, agarwal2021skellam}.
We empirically set the iteration count to satisfy a failure probability $<10^{-6}$ (App.~\ref{app:samplingthreshold}).
We evaluate in WAN (Tab.~\ref{tab:eval}) and LAN (App.~\ref{app:online}) using the MPC framework MP-SPDZ \cite{spdz}, with setup details in App.~\ref{app:evalsetup}.
We select three SS schemes: 3-party semi-honest and malicious Shamir with honest-majority (\enx{t < m/2}), and 2-party malicious Mascot \cite{keller2016mascot} with dishonest-majority (\enx{t > m/2}).
Tab.~\ref{tab:eval} reports average runtime over 10 runs with 95\% confidence intervals.
To address the underflow in {\Skellam} for \enx{\exp{(-1/\mu)}} with large \enx{\mu}, we leverage Skellam's closure under summation, by sampling and summing \enx{1542} {\Skellam} samples with small \enx{\mu=20}.
\Skellam's runtime depends on \enx{\mu} and requires \enx{2(\mu+1)} iterations on average.
From our results, semi-honest and malicious Shamir have comparable times, whereas Mascot is at least $2\times$ slower.
Notably, discrete sampling algorithms are at least $10\times$ slower than continuous ones due to their iterative nature, potentially explaining the gap in Tab.~\ref{tab:summary} on distributed discrete \centrnoise sampling. 
We omit {\Skellam} in Tab.~\ref{tab:eval}, as its runtime is disproportionate, i.e., about $20$ minutes for all schemes in a LAN.
LAN runtimes are 10$\times$ faster than WAN, with online/offline phase and communication detailed in App.~\ref{app:mpc_sampling}. 
Overall, {\Laplace} is the fastest but mainly used for \perturbout (Sec.~\ref{sec:noisegen}). {\Gauss} via {\BoxMuller} is the most efficient for \perturbgrad, generating two samples per run.

\begin{table}[t]
    \scriptsize
    \renewcommand{\arraystretch}{1.0}
    \setlength{\tabcolsep}{0.5em}
    \caption{Distributed \centrnoise sampling runtime for WAN.}
    \label{tab:eval}
    \vspace{-0.8em}
    \def\seps{/}
    \def\colwfirst{0.24\columnwidth}
    \def\colw{0.17\columnwidth}
    \def\colwalt{0.19\columnwidth}
    \centering
    \begin{tabular}{p{\colwfirst}C{\colw}C{\colw}C{\colwalt}}
        \textbf{WAN, seconds} & \cellcolor{\mylighterblue}{\Laplace} & \cellcolor{\mylighterblue}{\Gauss} & \cellcolor{\mylighterblue}{\DiscGaussian} \\
        \toprule
        \noalign{\vskip -0.3em}
        Shamir & \cellcolor{\mylighterblue}{$ 6.21 \pm 0.03 $} & \cellcolor{\mylighterblue}{$ 18.57 \pm 0.13 $} & \cellcolor{\mylighterblue}{$ 232.59 \pm 0.36 $} \\
        Malicious Shamir & \cellcolor{\mylighterblue}{$ 7.36 \pm 0.03 $} & \cellcolor{\mylighterblue}{$ 18.76 \pm 0.09 $} & \cellcolor{\mylighterblue}{$ 261.64 \pm 0.44 $} \\
        Mascot & \cellcolor{\mylighterblue}{$ 34.91 \pm 0.54 $} & \cellcolor{\mylighterblue}{$ 107.2 \pm 2.11 $} & \cellcolor{\mylighterblue}{$ 1638.12 \pm 8.80 $} \\
    \end{tabular}
    
    \vspace{-2.0em}
\end{table}

%
\xx{
Note that here breaking the non-collusion assumption would not only reveal the noise values but also the secret-shared data.
}

%
%
%
%
\xx{
We formalize and overview noise sampling techniques in Alg.~\ref{alg:mpc_noise}
and discuss basic sampling concepts next.
\todo{all the stuff on how noise can be sample is a systematization effort so should be highlighted as a paragraph, together with sampling via GC, SS etc. in  the observation paragraph}
%
\emph{Inverse transform sampling} (ITS) allows sampling random variables \enx{\mathcal{Z} \sim f{(x)}} from any probability distribution \enx{f(x)} based on the uniform distribution. ITS computes \enx{\mathcal{Z} = CDF^{-1}(u)}, where \enx{CDF^{-1}(\cdot)} is the inverse \emph{cumulative density function} of \enx{f(x)}, and \enx{u \sim U(0,1)} is a uniformly distributed random variable in \enx{(0,1)} \cite{law2007simulation}.
For example, \xspace{\LaplaceITS} (Alg.~\ref{alg:mpc_noise}) computes \enx{-\frac{1}{\lambda}\; \mathsf{sign}(u-0.5) \log u}
with sign function $\mathsf{sign}(x)$ outputting $+1$ (resp., $-1$) for positive (resp., negative) inputs $x$.
%
However, the Gaussian's \enx{CDF^{-1}} does not have a closed form and needs to be approximated, e.g., via Taylor polynomials \cite{sabater2022private}. An alternative method is the \emph{Box-Muller transform} \cite{box1958note}
({\BoxMuller} in Alg.~\ref{alg:mpc_noise})
which receives two uniformly random variables 
and outputs two samples from the standard Gaussian \enx{\mathcal{N}(0,1)}.\footnote{Note that $\sigma\cdot\mathcal{N}(0,1)+\mu = \mathcal{N}(\mu,\sigma^2)$, i.e., standard Gaussian samples can be easily transformed to desired $\sigma^2,\mu$.}
%
ITS and Box-Muller require sampling \enx{u\sim U(0,1)}.
To ensure truly random $u$ in distributed settings, \pls can sample local values that are combined, \eg{\emph{XOR}ed and scaled to \enx{(0,1)} \cite{jayaraman2018distributed}}.
}
%
%
%
%

\xx{
Box-Muller and ITS can be directly implemented in MPC,
\ie{via GC \cite{chase2017,jayaraman2018distributed} or SS \cite{pentyala2022training}}.
Note, that ITS and Box-Muller require sampling \enx{u\sim U(0,1)}. To ensure truly random $u$, 
\sls can sample local values that are combined, \eg{\emph{XOR}ed and scaled to \enx{(0,1)} \cite{jayaraman2018distributed}}.
Mechanisms with non-discrete inputs, i.e., Poisson-Binomial \cite{chen2022poisson}, and Skellam from \cite{bao2022skellam}, are not suitable for straightforward distributed sampling.
The discrete Gaussian sampled via MPC, minimizes approximation errors from summing small noise values across numerous \cls.\footnote{The sum of two discrete Gaussian with \enx{\sigma^2=3} leads to an error of \enx{10^{-12}} on the max divergence with the sum of continuous Gaussian \cite[Thm. 9]{kairouz2021distributed}.}  
%
%
%
%
MPC sampling of continuous distributions 
via ITS or Box-Muller is straightforward, mainly requiring a fixed number of arithmetic operations, \eg{\enx{\log}, \enx{\sin}}. 
Discrete ones, however, introduce challenges.
%
\obs{Discrete distributions have non-constant runtimes} 
as they use rejection sampling, \ie{multiple iterations to sample a single value, until a certain condition is met}. This procedure is implemented via data-dependent loops, \ie{\code{while}}, and its runtime can leak the noise values, \eg{the loop count in Poisson 
corresponds to the noise sample $p$ (see Alg.~\ref{alg:mpc_noise}}).
Overall, data-dependent loops
have to be replaced with fixed iteration counts, \eg{\code{for}}. Here, exit conditions have to be evaluated and stored for each iteration. 
To set fixed iteration counts, a public threshold must be selected with a negligible failure probability,
ensuring at least one correct sample is returned.
Another point to consider is that using a fixed bit-width can lead to under/overflows. For instance, computing $\exp(-1/\mu)$ 
for Skellam causes underflows for large $\mu$.
}

%
%
%
%

\xx{
    Additionally, we evaluate runtime and communication of required, basic MPC protocols (e.g., $\log, \exp$) provided by MP-SPDZ in App.~\ref{app:basic_eval}.
    Briefly, focusing on the online phase,
    square root is the most costly operation in terms of runtime
    ($\approx$351ms in LAN with $\approx$21KB)
    followed by \enx{\log_2} ($\approx$278ms, 20KB),
    \enx{2^{(\cdot)}} ($\approx$278ms, 6KB),
    and $\sin, \cos$ ($\approx$155ms, 6KB).
}
\section{Empirical Evaluation \& Design Trade-Offs}\label{sec:noisetradeoff}
%
%
%
Next, we evaluate key trade-offs and considerations to enhance cryptographic CL with DP. Specifically, we analyze DP's impact on runtime (Tab.~\ref{tab:combined-dp-overhead}) and accuracy (Tab.~\ref{tab:accuracy}) across paradigms and noise techniques. 
Using standard benchmarks (MNIST, Fashion MNIST) and setups from prior work \cite{das2025communication,abadi2016deep}, we provide representative empirical evidence on the practical implications of design choices in \epcl.



\para{Evaluation Setup} 
We train a 3-layer NN ($\approx 90\text{K}$ parameters and ReLU activations) on MNIST and Fashion MNIST datasets, as in \cite{das2025communication,abadi2016deep}. 
Training runtimes (Tab.~\ref{tab:combined-dp-overhead}) are averaged over 10 runs for batch size \enx{B}, split into \emph{mini-batches} of size \enx{b} for efficiency with gradient accumulation. We evaluate the impact of naive \clip with \enx{b=1} and \noise compared to non-DP training with \enx{b=1} and $b>1$ to estimate the speed-up of batching. 
For accuracy (Tab.~\ref{tab:accuracy}), we vary \enx{\eps \in \{1,3,8\}}, \enx{\delta = 10^{-5}}, use Gaussian noise \enx{\mathcal{N}(0, (\sigma\clipparam)^2)}, where \enx{\sigma \in \{1.18, 0.73, 0.54\}}, and evaluate collusion thresholds \enx{t \in \{n/2, n-1\}} for \partnoise. 
For single-server \fl, we select \enx{n=100} clients per round from~\enx{1000}, and hyperparameters according to our search (App.~\ref{app:hyperparameter_fl}). For \ot, we use hyperparameters from \cite{abadi2016deep}. 
We implement FL with PFL \cite{granqvist2024pfl}. For \ot, we extend CrypTen \cite{knott2021crypten} with missing DP-SGD components.
App.~\ref{app:evalsetup} details the full evaluation setup.
\para{Clipping Impact}
Naive per-example clipping (\enx{b=1}) significantly increases runtime and memory usage by computing (and clipping) \enx{B} gradients instead of one per batch (Tab.~\ref{tab:combined-dp-overhead}).
In \fl, \enx{b=1} increases non-DP \compgrad runtime by \enx{8\times}, and \clip makes DP \compgrad \enx{2\times} slower. 
%
%
However, \aggr, which accounts for \enx{75\%} of DP FL runtime due to communication and encrypted aggregation overhead, minimizes the impact of non-batched computation, keeping \fl runtimes in the tens-of-seconds range.
Instead, MPC-based inverse square root and comparisons in \clip make DP \ol \enx{2\times} slower than non-DP \ol (\enx{b=1}). Here, non-batched computations increase non-DP runtime by \enx{100\times}. 
%
Notably, accuracy depends on \enx{\clipparam}:
a large \enx{K} increases noise variance, obscuring the gradients, while a small \enx{K} hinders convergence.
For example, in \ol, \enx{K=3} yields \enx{82\%} accuracy on EMNIST, \enx{K=1} drops accuracy by $3$ percentage points (pp), while \enx{K=10} results in near-random guessing. (App.~\ref{app:hyperparameter}).
In \partnoise, {\clipparam} needs to account for increased noise variance from collusion, e.g., a sub-optimal \enx{\clipparam=0.5} leads to a \enx{17\pp} accuracy drop on MNIST in \fl, vs. \enx{\clipparam=0.3}~(\enx{\eps=3, t=n-1}).

\begin{table}[t]
    \centering
    \scriptsize
    \renewcommand{\arraystretch}{1.0}
    \setlength{\tabcolsep}{0.08em}
    \setlength{\tabcolsep}{0.25em}
    \caption{Runtime of \fl/\ot over a batch in LAN, with $\lfloor$ denoting the subprotocols of the above runtime.} 
    \label{tab:combined-dp-overhead}
    \vspace{-0.8em}
        \def\colwfirst{0.44\columnwidth}
        \def\colw{0.25\columnwidth}
        \def\colwalt{0.15\columnwidth}
        \begin{tabular}{p{\colwfirst}C{\colw}C{\colw}} 
            \textbf{Runtime (seconds)} &
            {\fl (\enx{B=60, n=100})} &
            \cellcolor{\mylighterblue}{\ol (\enx{B=500, m=2})} 
            \\
            \noalign{\vskip -0.1em}
            \toprule
            \noalign{\vskip -0.3em}
            Total non-DP (\enx{b_{\text{FL}}=10, b_\text{OT}=128}) &
            $ 0.19 \pm 0.01 $ & 
            \cellcolor{\mylighterblue}{$ 6.49 \pm 0.05$} 
            \\
            \noalign{\vskip -0.1em}
            \quad $\lfloor$ \compgrad non-DP &
            $ 0.0047 \pm 0.0008 $ & 
            \cellcolor{\mylighterblue}{$ 6.45 \pm 0.04$} 
            \\
            \noalign{\vskip -0.2em}
            \quad $\lfloor$ \aggr &
            $ 0.18 \pm 0.011 $ & 
            {$ \-- $}
            \\
            \noalign{\vskip -0.2em}
            \midrule
            \noalign{\vskip -0.3em}
            Total non-DP (\enx{b=1}) &
            $ 0.22 \pm 0.0149 $ & 
            \cellcolor{\mylighterblue}{$ 627.73 \pm 3.38 $} 
            \\
            \noalign{\vskip -0.1em}
            \quad $\lfloor$ \compgrad non-DP &
            $ 0.036 \pm 0.004 $ & 
            \cellcolor{\mylighterblue}{$ 627.72 \pm 3.38$} 
            \\
            \noalign{\vskip -0.2em}
            \midrule
            \noalign{\vskip -0.3em}
            Total DP (\enx{b=1}) with \partnoise &
            $ 0.25 \pm 0.024 $ & 
            \cellcolor{\mylighterblue}{$ 1236.81 \pm 55.12$ } 
            \\
            \noalign{\vskip -0.1em}
            \quad $\lfloor$ \compgrad DP &
            $ 0.060 \pm 0.0130 $ & %
            \cellcolor{\mylighterblue}{$ 1236.79 \pm 55.12 $}    
            \\
            \noalign{\vskip -0.1em}
            \quad \quad $\lfloor$ \clip ($K_{\text{FL}}=0.3, K_{\text{OL}}=4.0$) &
            $ 0.023 \pm 0.0043 $ & 
            \cellcolor{\mylighterblue}{ $ 718.93 \pm 32.55 $} 
            \\
            \noalign{\vskip -0.1em}
            \quad $\lfloor$\noise (\partnoise, Gaussian) & 
            \multicolumn{2}{c}{$ 0.0051 \pm 0.0007 $} 
            \\
            \noalign{\vskip -0.2em}
            \midrule
            \noalign{\vskip -0.3em}
            Total DP (\enx{b=1}) with \centrnoise &
            $  3.11 \pm  0.125 $ & 
            \cellcolor{\mylighterblue}{$ 1238.87 \pm 55.23$ } 
            \\
            \noalign{\vskip -0.1em}
            \rowcolor{\mylighterblue}
            \quad $\lfloor$ \noise (\centrnoise, Gaussian) &
            \multicolumn{2}{c}{$ 2.87 \pm 0.10 $} 
            \\

        \end{tabular}

    \vspace{-2.0em}
\end{table}

\para{Noise Techniques}
Our evaluations (Tab.~\ref{tab:combined-dp-overhead}, Tab.~\ref{tab:accuracy}) highlight the performance-accuracy trade-offs of different noise sampling techniques.
\partnoise sampling is the most efficient secure sampling technique, as it relies on local sampling, with a minimal impact on \fl runtime (\enx{2\%}) and negligible on \ot (\enx{\ll 1\%}). 
However, \partnoise's utility is compromised under high collusion (\enx{t=n-1}), reducing accuracy by up to \enx{70\pp} (Tab.~\ref{tab:accuracy}).
Mitigating this requires cryptographic enhancements: oblivious protocols for \partnoise, or sampling \centrnoise, which introduce performance overhead.  
While centralized \centrnoise sampling achieves optimal utility (as in CDP), it requires a semi-trusted \enx{S'} and obliviousness (Sec.~\ref{subsec:centralnoise}).
If \enx{S' \notin \outparties}, communication latency to send encrypted noises is the main overhead, which is negligible for \ol, but more significant for \fl (\enx{10\%} of \partnoise runtime). 
Distributed \centrnoise sampling removes the need for \enx{S'} and maintains optimal utility. However, it incurs large MPC sampling overhead, e.g., MPC sampling of Gaussian noise is almost \enx{10^3\times} slower than local sampling (Sec.~\ref{subsec:distributed-noise-sampling}). 
While this overhead is negligible for \ol, it increases \fl runtime by nearly \enx{10\times}.
Furthermore, the choice of the mechanism significantly impacts performance for both \partnoise (\textbf{O5}) and \centrnoise (Tab.~\ref{tab:eval}), e.g., with discrete mechanisms being slower than continuous ones.
\para{Privacy-Accuracy Trade-Off}
We use plain training (no DP or cryptography) as a baseline with an accuracy of $96.6\%$ on MNIST and $86.0\%$ on Fashion MNIST, while non-DP FL reaches $92.3\%$ and $85.4\%$, respectively.
Introducing DP with \centrnoise (i.e., non-crypto \centrnoise) reduces accuracy by up to \enx{3.2\pp} from plain training. Smaller \eps causes further drops (as \enx{\sigma^2} increases), by up to \enx{1.1\pp} from \enx{\eps=8} to \enx{\eps=1}.
Cryptography further reduces accuracy due to fixed-point representations and quantization. \fl with \centrnoise shows a more significant accuracy drop (up to \enx{8.7\pp} against non-DP \fl) compared to \ot with \centrnoise, which remains within \enx{0.5\pp} of non-crypto \centrnoise accuracy.
As expected, secure \centrnoise sampling in \fl outperforms LDP, which loses up to \enx{66\pp} in accuracy with \enx{\eps=1}.
FL with \centrnoise accuracy is lower than OL since it relies on aggregation of local updates, whereas OL with \centrnoise can closely match non-DP accuracy due to global dataset updates.
This gap increases with heterogeneous (i.e., non-IID) data, e.g., if each client holds only samples of a single class, or classes with conflicting features \cite{kairouz2021advances}. Addressing non-IID settings requires specialized optimization methods, e.g., alternating direction multipliers~\cite{miao2024privacy}. 
Accuracy under \partnoise is highly sensitive to \eps and collusion threshold \enx{t}, since higher \enx{t} increases the adjusted noise variance (Eq.~\eqref{eq:partgaussvariance}, Sec.~\ref{subsec:partialnoise}).
In \fl with \partnoise (with \enx{n=100}), half colluding clients (\enx{t=n/2}) cause up to \enx{8.5\pp} accuracy reduction compared to \centrnoise. 
Here, high collusion (\enx{t=n-1}) increases the total noise variance above LDP levels. This causes over \enx{70\pp} accuracy drop compared to \centrnoise with \enx{\eps=1}, resulting in even lower accuracy than LDP, i.e., up to \enx{26\pp} with \enx{\eps=3} on FashionMNIST.
In contrast, \ot typically involves fewer servers than \fl clients, reducing the impact of collusion. \ot with \partnoise, \enx{m=2} and high collusion (\enx{t=m-1}) has a minimal accuracy drop, i.e., at most \enx{1\pp} below \ol with \centrnoise. Increasing \enx{m} to \enx{10} only drops accuracy by \enx{\approx5\pp} compared to \ol with \centrnoise\cite{das2025communication}. Our extended evaluation (App.~\ref{app:collusion}) shows that in \ot with \partnoise and \enx{m=100}, accuracy drops by \enx{\approx 10\pp}. 
Notably, multi-server \fl with \partnoise could similarly benefit from server-side \partnoise, as \enx{m < n}, improving accuracy in case of collusion.

\begin{table}[tbp]
    \centering
    \scriptsize
    \renewcommand{\arraystretch}{1.0}
    \setlength{\tabcolsep}{0.6em}
    \caption{
        Accuracy on MNIST and FashionMNIST for \fl, \ot and non-crypto(graphic) DP training, combined with \centrnoise, \partnoise with different collusion thresholds (in parentheses) and LDP. We set \enx{n=100} for \fl and \enx{m=2} for \ol.
    }
    \label{tab:accuracy}
    \vspace{-0.8em}
    \begin{tabular}{lcccccc}
        \multirow{2}{*}{\textbf{Accuracy (\%)}} & \multicolumn{3}{c}{MNIST} & \multicolumn{3}{c}{Fashion MNIST} \\
        \noalign{\vskip -0.1em}
         & \enx{\eps = 1} & \enx{\eps = 3} & \enx{\eps = 8} & \enx{\eps = 1} & \enx{\eps = 3} & \enx{\eps = 8}
        \\
        \noalign{\vskip -0.2em}
        \toprule
        \noalign{\vskip -0.3em}
        Non-crypto \centrnoise & {95.8} & {96.3} & {96.4} & {82.8} & {83.8} & {83.9} \\
        FL with LDP & {39.6} & {63.0} & {69.6} & {16.7} & {57.5} & {63.7} \\
        \fl \centrnoise & {87.1} & {89.7}& {90.0}& {78.0}& {81.6} & {82.7} \\
        \fl \partnoise \enx{(n/2)} & {80.0} & {87.1}& {87.7}& {69.5}& {79.9}& {82.3}\\
        \fl \partnoise \enx{(n-1)} & {15.2} & {47.5}& {72.7}& {13.1}& {31.4} & {59.2} \\ 
        \ol \centrnoise & {95.6} & {96.2}& {96.5} & {82.7} & {83.6}& {83.8}\\
        \ol \partnoise \enx{(m-1)} & {95.4} & {95.8}& {96.0}& {82.5}& {83.5}& {83.6} \\

    \end{tabular}
    
    \vspace{-2.0em}
\end{table}

\subsection{Designing \epcl Solutions}\label{sec:design-epcl}
%
Drawing from our systematization, Tab.~\ref{tab:summary}, Alg.~\ref{alg:cmp} and evaluations above, we find that 
\hlobs{efficient and accurate CPCL solutions require careful design considerations for learning paradigms, cryptographic protocols, and noise sampling techniques.}Enhancing cryptographic CL with DP requires integrating the \perturb phase, \clip and subsampling within \compgrad. 
This creates a privacy-utility-performance trade-off space with multiple design choices. Next, we outline key considerations for designing \epcl solutions. 
%

\para{Noise Sampling Strategies}
For \perturb, if priorities are performance and ease of implementation, local \partnoise sampling and aggregation is the most efficient sampling technique, as it relies solely on cleartext operations. 
For mechanism choice, Skellam \cite{agarwal2021skellam} is currently the preferred additive option, as it matches the accuracy of continuous Gaussian using low-precision integers. 
However, \partnoise leads to significant accuracy drops under high collusion (\enx{t=n-1}), i.e., up to \enx{70\pp} in \fl (Tab.~\ref{tab:accuracy}).  
Oblivious \partnoise sampling \cite{byrd2022collusion,mugunthan2019smpai} mitigates this, but requires cryptographic protocols, which introduce performance overhead.
If a semi-trusted \enx{S'}, that is oblivious to output, is an option, centralized \centrnoise sampling introduces minimal overhead and achieves optimal utility (as in CDP) \cite{froelicher2020drynx}.  
If strong security is the main requirement and performance is a secondary concern, distributed \centrnoise sampling is the best choice \cite{pentyala2022training}, as it achieves optimal utility without relying on a semi-trusted server. 

\para{Cryptographic Constraints}
Cryptographic choices constrain noise strategies: techniques with implicit \reveal (\eg{pairwise masking}) require the use of \partnoise as the aggregated result must already be DP \cite{bonawitz2017practical}. Single-server solutions (\eg{HE, LWE-based masking \cite{stevens2022efficient}}) are suited for \partnoise or centralized \centrnoise, while MPC supports all techniques. 
Furthermore, infrastructure bottlenecks guide design: MPC requires non-colluding servers and is communication-intensive, whereas HE is computation-intensive.

\para{Learning Paradigm Considerations}
While \fl clients locally run \compgrad (and \clip), \ot requires server-side cryptographic protocols with high overhead (e.g., \enx{200\times} slowdown with naive \clip, Tab.~\ref{tab:combined-dp-overhead}). Notably, the lack of secure clipping optimizations for \ot represents a significant gap that hinders performance.
Despite \fl being the most efficient paradigm (up to \enx{10^4\times} faster than \ot, Tab.~\ref{tab:combined-dp-overhead}) it has several drawbacks: it exposes (noisy) gradients to GIA (App.~\ref{app:privacyattacks}); may show lower accuracy (e.g., up to \enx{8.5\pp} drop vs. \ol with \centrnoise, Tab.~\ref{tab:accuracy}); and requires online clients and cryptographic protocols to handle dropouts (App.~\ref{sec:dropouts}).
Conversely, OL approaches with local pre-processing, such as PEA \cite{pea2023ruan}, offer a promising middle ground, balancing the strong security of \ot with the efficiency of local computations.

\xx{
\para{Trade-offs in Learning Paradigms}
    \todo{ensure to remove similar text in previous sections if not really required (potentially referencing this section)}
Going from \fl to \ot, computation complexity shifts towards the server.
While in \fl each client executes on-device training sharing local updates, in \ot the clients do not need high computation power, since they only secret share their data.
\fl requires interactivity, i.e., clients are online during the whole training, or handling of dropouts (App.~\ref{sec:dropouts}).
In \ot, however, \cls can go offline after having shared their data.
The intermediate outputs of \fl, \ie{gradients and model parameters}, pose a security threat since the model can be subjected to gradient inversion attacks (Sec.~\ref{sec:securityanalysis}), and lead to a non-optimal DP composition bound as hiding intermediates outputs provides a lower bound\footnote{This holds only for strongly convex loss functions, but is an active area of research for non-convex ones} \cite{ye2022differentially}.
%
%
}

\xx{
\para{Utility-Efficiency-Privacy Trade-Off}
Next, we discuss how different approaches balance the trade-off between utility and efficiency to improve privacy guarantees.
\ot with distributed noise sampling is an approach where all the operations are executed in an MPC protocol. While this introduces performance overhead, it allows all training steps to remain hidden, reducing the attack surface, \ie{mitigating gradient inversion attacks}, and achieving an optimal composition bound for strongly convex loss functions \cite{ye2022differentially}. Furthermore, the \cls can be lightweight devices, with constrained computational and energy resources, since can go offline after \setup and \protection. Importantly, the \sls are oblivious to the DP noise.
On the other hand, FL with partial noise aggregation distributes the computational load among the \cls, with each client executing training iterations locally. Only the aggregation (and potential verification measures) requires cryptography. However, each client executes \compgrad and \updatemodel on-device, and clients must be online or dropouts have to be specifically handled (App. \ref{sec:dropouts}). In contrast to \ot, FL reveals intermediate gradients, making the model vulnerable to gradient inversion attacks (Sec.~\ref{sec:securityanalysis}), and leads to a non-optimal DP composition bound \cite{ye2022differentially}.
On the positive side, clipping is local, as well as the noise sampling, but requires additional noise to mitigate the collusion of \enx{t} \cls. This results in more noise than the CDP  which impact the model accuracy, especially for small numbers of clients.
There is a spectrum between those to extremes, namely FL with partial noise and \ot with distributed noise sampling. For example, FL can be augmented with oblivious noise selection, which increases accuracy (\ie{reduces noise requirements}) but also reduces efficiency. Alternatively, assuming multiple \sls in FL, the \sls can perform \mpcnoise. In \ot, \partialnoise can improve its efficiency, but the clipping still needs to be executed collaboratively. 
}

\xx{
\para{Privacy Unit}
Another consideration is privacy unit,
i.e., at what level the DP protection applies.
Def.~\ref{def:epsdDP} considers \emph{record-level} DP which assumes that each user is associated with one record in the dataset.
\emph{User-level} DP further {considers users who contribute up to \enx{k} records}.
For user-level DP, the neighboring datasets of Def. \ref{def:epsdDP} differ in at most $k$ records.

\obs{Privacy units matter},
when a user contributes multiple records, protecting a single record can leak data from that user, \eg{in \fl each client computes local update on more than one record}.
For example, in language modeling, each user may provide thousands of examples (i.e., words or phrases) to the training data \cite{mcmahan2017learning}.
%
DP satisfies \emph{group privacy} enabling the translation from record-level to user-level 
Specifically, \enx{(\eps, \delta)}-DP for record-level becomes \enx{(k \eps,k \exp(k\eps) \delta)}-DP for user-level protection.
This cost is prohibitive when each user contributes to thousands of records.
}

%
%


\xx{
\para{User-level DP}
Def.~\ref{def:epsdDP} considers \emph{record-level} DP which assumes that each user is associated with one record in the dataset. \emph{User-level} DP further
{considers users who contribute up to \enx{k} records}
%
%
DP satisfies \emph{group privacy} enabling the translation from record-level to user-level privacy. Specifically, \enx{(\eps, \delta)}-DP for record-level becomes \enx{(k \eps,k \exp(k\eps) \delta)}-DP for user-level protection.
Although some FL works guarantee user-level DP \cite{agarwal2018cpsgd,agarwal2021skellam,kairouz2021distributed}, none of the works in \ot explicitly mention providing user-level DP.
%
}

\xx{
Next, we detail the foundational phase for DP learning, \ie{\noise}.
We identify the noise generation techniques to instantiate functions \centrnoise and \partnoise (Alg.~\ref{alg:cmp}).
We present the techniques in order of increasing cryptographic overhead in Sec.~\ref{subsec:centralnoise}--\ref{subsec:distributed-noise-sampling}.
Namely,
centralized sampling for \centrnoise (i.e., sampling in clear, Sec.~\ref{subsec:centralnoise}),
\partialnoise for \partnoise (mainly, secure aggregation, Sec.~\ref{subsec:partialnoise}),
and distributed sampling for \centrnoise (MPC sampling, Sec.~\ref{subsec:distributed-noise-sampling}).
All techniques can realize both \perturbgrad and \perturbout.
}
%
\xx{
Next, we consider semi-honest adversaries, who sample and add noise correctly, but try to reduce the overall privacy guarantees, \eg{removing their noise contributions to the aggregated result}.
Additional considerations for malicious adversaries are given in Sec.~\ref{sec:analysis}.
We overview common noise distributions 
\todo{Overview dist in tab, group them (Perturb Option , continous discrete, PNoise, CNoise, then details later) (Just few senteces)}
\todo{reoder them, group}
for DP in Tab.~\ref{tab:noisedistributions},
and give more details in App.~\ref{app:noisedistributions}.
To illustrate who is involved in the noise sampling,
we overview noise sampling scenarios in Fig.~\ref{fig:dp_noise_generation}.
}

\xx{
\para{Obliviousness}
We denote noise sampling as \emph{oblivious} to highlight that the resulting noise remains hidden from all \pls. Specifically, we differentiate between \emph{oblivious noise selection} and \emph{oblivious noise sampling}. 

In oblivious selection, \eg{a client}, locally samples a list of candidate noises. Then, those noises are encrypted and shuffled. Another client selects one of the noises from the encrypted list. In oblivious sampling, the noise is sampled via cryptographic protocols, \eg{MPC}, and none of the \pls knows the noise value.


}

%
\xx{
For noise distributions in Tab.~\ref{tab:noisedistributions}, we discuss their suitability for noise type (expanded and formalized in Alg.~\ref{alg:mpc_noise}), perturbation options and their DP guarantees. 
}

\xx{

\para{Noise Mechanisms}
\todo{Integrate this sentence}
\tocheck{
The noise mechanisms are reported following systematized works, and the trends is using Gaussian-like perturbations for \perturbgrad, and Laplace-like for \perturbout (Tab.~\ref{tab:summary}).
}
For noise distributions in Tab.~\ref{tab:noisedistributions}, we next discuss their suitability for noise type, perturbation options and their DP guarantees. We formalize sampling algorithms (\sample) in Alg.~\ref{alg:mpc_noise} (Sec.~\ref{subsec:distributed-noise-sampling}) and provide conversion between DP and distributions parameters in {App.~\ref{app:noisedistributions}}.
%
Laplace 
is suitable for \perturbout in \centrnoise, as it is applied only once, and provides pure DP (\enx{\delta=0}).
Distributed Laplace for \partnoise is realized as the difference of \enx{N} i.i.d. \emph{gamma} random variables 
\cite{mugunthan2019smpai,byrd2022collusion}. 

Gaussian is more suitable for \perturbgrad, even though it satisfies ($\eps,\delta$)-DP, 
since the noise is applied over multiple epochs, and Gaussian is closed under summation.
Moreover, its tails decrease faster than Laplace for the same standard deviation, and the standard deviation of Gaussian noise is proportional to the \enx{l_2} sensitivity, which is often smaller than \enx{l_1} for Laplace \cite{mironov2017renyi}.
Gaussian can be sampled for \partnoise and \centrnoise since its distributed version can be obtained by summing \enx{N} i.i.d. Gaussians \enx{\mathcal{N}_i(\mu_i,\sigma^2_i)}, getting \enx{\mathcal{N}(\sum_{i=1}^{N}\mu_i, \sum_{i=1}^{N}\sigma^2_i)} \cite{truex2019hybrid}.
Continuous noise needs quantization as cryptography typically considers integers or fixed-point numbers \cite{agarwal2018cpsgd}.
Alternatively, discrete distributions enable sampling from discrete domains directly.
Here we introduce discrete distributions that mimic the Gaussian guarantees for \perturbgrad,
which systematized works use with
\partnoise \cite{agarwal2021skellam,kairouz2021practical,agarwal2018cpsgd}.
%
%
%
The \emph{binomial} mechanism \cite{agarwal2018cpsgd,pea2023ruan} satisfies \enx{(\eps, \delta)}-DP, and it is closed under summation, i.e., \enx{\mathsf{Bin}(m,p) = \sum_{i}^{n}\mathsf{Bin}(m_i,p)}.
The \emph{Discrete Gaussian} \cite{kairouz2021distributed}, is not closed under summation, but the sum of discrete Gaussian still satisfies \enx{(\alpha, \eps)}-RDP.
The \emph{Skellam} mechanism \cite{agarwal2021skellam,bao2022skellam} satisfies \enx{(\alpha, \eps)}-RDP. It can be computed as the difference of two independent Poisson random variables, and is closed under summation.
All mechanisms in Tab.~\ref{tab:noisedistributions} add noise to the gradients
except for the Poisson-Binomial 
\cite{chen2022poisson} which encodes local gradients into a parameter of the binomial. 
Since this mechanism is optimized for \enx{l_\infty} geometry, each client computes the Kashin representation of \enx{g_i} \cite{lyubarskii2010uncertainty}, \ie{transforms the geometry of the data from the \enx{l_2} to \enx{l_\infty}}. 
The \slong reverts the Kashin representation to \enx{l_2} geometry after the aggregation.
}
\xx{
\para{Basic Sampling Algorithms}
We formalize 
noise sampling algorithms in Alg.~\ref{alg:mpc_noise} (Sec.~\ref{subsec:distributed-noise-sampling}) and discuss basic sampling concepts next.
\emph{Inverse transform sampling} (ITS) allows sampling random variables \enx{\mathcal{Z} \sim f{(x)}} from any probability distribution \enx{f(x)} based on the uniform distribution. ITS computes \enx{\mathcal{Z} = CDF^{-1}(u)}, where \enx{CDF^{-1}(\cdot)} is the inverse \emph{cumulative density function} of \enx{f(x)}, and \enx{u \sim U(0,1)} is a uniformly distributed random variable in \enx{(0,1)} \cite{law2007simulation}.
For example, \xspace{\LaplaceITS} (Alg.~\ref{alg:mpc_noise}) computes \enx{-\frac{1}{\lambda}\; \mathsf{sign}(u-0.5) \log u}
with sign function $\mathsf{sign}(x)$ outputting $+1$ (resp., $-1$) for positive (resp., negative) inputs $x$.
%
However, the Gaussian's \enx{CDF^{-1}} does not have a closed form and needs to be approximated, e.g., via Taylor polynomials \cite{sabater2022private}. An alternative method is the {\BoxMuller} \emph{transform} \cite{box1958note}
(Alg.~\ref{alg:mpc_noise})
which receives two uniformly random variables 
and outputs two samples from the standard Gaussian \enx{\mathcal{N}(0,1)}.\footnote{
    Note that \enx{\mathcal{N}(0,1)} can be transformed to desired $\sigma^2,\mu$, as \enx{\mathcal{N}(\mu,\sigma^2) = \sigma\cdot\mathcal{N}(0,1)+\mu}.
}

}

\xx{
In Tab.~\ref{tab:summary}, we systematize analyzed works in top-down fashion.
First, we group works by learning paradigm. Then, we further distinguish by noise type, sampling technique, perturbation option, and DP mechanism.
{
Noise generation techniques are defined by these noise columns. For example, {\citet{agarwal2018cpsgd} sample \partnoise locally for \perturbgrad with the binomial mechanisms; the same applies for \cite{kairouz2021distributed} but using the discrete Gaussian mechanism}.
}
We also indicate whether gradient and model secrecy is ensured, and if the noise sampling is oblivious. Finally, we list cryptographic techniques and details about malicious adversaries and collusion threshold.
From Tab.~\ref{tab:summary}, we observe that \fl is overwhelmingly used when combining DP and cryptography, as it is very efficient (\ie{only lightweight cryptographic operations}).
\fl with \partnoise, however, offers no model/gradient secrecy (apart from \perturbout) and requires more noise to tolerate \enx{t} colluding \cls, \ie{no noise obliviousness (Sec.~\ref{subsec:partialnoise})}.
\fl with \centrnoise offers noise obliviousness and gradient secrecy.
Collaborative labeling is only used in one work \cite{choquette2021capc} with centralized sampling.
}

\section{Related Works}\label{sec:related}
Only a few works explore the combination of cryptography and DP for \cl. 
Yang et al.~\cite{yang2023privatefl} evaluate optimizations for DP FL including partial noise aggregation but omit, e.g., \ot, multi-server \fl and distributed noise sampling. Chatel et al.~\cite{chatel2021sok} systematize \cl for tree-based models focusing mainly on cryptographic or DP solutions. 
Other works do not focus on \cl. 
Wagh et al.~\cite{wagh2021dp} survey DP and cryptography for distributed analytics, while Fu and Wang~\cite{fu2024benchmarking} benchmark secure sampling protocols 
for continuous and discrete Laplace and Gaussian. 
Meisingseth and Rechberger~\cite{meisingseth2024sok} systematize computational DP definitions 
for multi-party settings.
Recent SoKs focus on encrypted CL without DP. Cabrero-Holgueras and Pastrana~\cite{cabrero2021sok} and Ng and Chow~\cite{sp/NgC23} analyze cryptographic training, while Mansouri et al.~\cite{mansouri2023sok} systematize SecAgg, and Mo et al.~\cite{mo2022machine} systematize training with enclaves. 
Other works focus only on DP ML. Jayaraman and Evans~\cite{jayaraman2019evaluating} evaluates utility and privacy of different DP mechanisms. 
Ponomareva et al.~\cite{ponomareva2023dp} provide guidelines on how to implement DP in centralized and FL settings.

\section{Research Directions}\label{sec:observation}
Building on our systematization, identified gaps ({\doubledash}) and potential enhancements (\enhancemark) from Tab.~\ref{tab:summary}, as well as key observations \obsn{\#}, 
we propose avenues for future research.
Tab.~\ref{tab:research_directions} (App.~\ref{app:obs_mapping}) maps these directions to our observations.

\research{Enhance privacy and performance via pre-processing}
Despite its benefits \obsn{3}, local pre-processing is underexplored in systematized works.
Only PEA \cite{pea2023ruan} \cls perform local computations, i.e., feature extraction and global model initialization via local pre-training, improving efficiency without compromising accuracy.
Similarly, \fl benefits from \cls performing many local iterations, e.g., in DP-FedAvg \cite{mcmahan2017learning}, reducing communication overhead.
%
Incorporating public data in pre-processing can enhance feature learning and improve model accuracy while providing strong privacy~\cite{tramer2020differentially}.

\research{Provide cryptographic building blocks for DP in \ol}
\ot guarantees gradient and model secrecy, preventing GIAs \obsn{1}.
However, few works explore \ol \cite{pea2023ruan,pentyala2022training,das2025communication}, and none provides open-source implementations.
%
%
%
Clipping is the main cryptographic bottleneck of \ol \obsn{2}, yet ML optimizations, e.g. ghost clipping \cite{ding2023dpformer,li2021large} are missing in cryptographic works. Future efforts should focus on efficient, open-source DP building blocks for \ol \obsn{9}, e.g., sampling, clipping, to foster further research and broader adoption.

\research{Develop crypto-friendly discrete \boldmath{\centrnoise} sampling for \epcl}
Several works \cite{bao2022skellam,agarwal2021skellam,agarwal2018cpsgd,kairouz2021distributed,wang2020d2p,stevens2022efficient} implement discrete \partnoise mechanisms \eg{Skellam and discrete Gaussian}, to improve utility and performance \obsn{6}. 
However, distributed sampling of discrete \centrnoise remains unexplored in \epcl, due to challenges like constant runtime \obsn{8}, and high overhead \cite{fu2024benchmarking, das2025communication}.
%
%
Existing MPC sampling for discrete distributions \cite{keller2024secure, fu2024benchmarking, wei2023securely, meisingseth2025practical} do not focus on \cl which needs high variance and large number of samples.
Their high runtimes make them impractical for \epcl \cite{fu2024benchmarking, das2025communication} and calls for optimizations of crypto-friendly discrete sampling for \epcl, potentially integrated with quantization-aware training~\cite{jacob2018quantization}.


%
%

%

\research{Embed DP outside \boldmath{\noise}}
To ensure DP, noise addition is not the only avenue, as DP can also be satisfied during quantization or by noise-based cryptography (e.g., LWE) \obsn{7}. 
Future research should explore how to apply DP in other phases 
For example, randomized rounding satisfies RDP by leveraging DP-focused quantization analysis \cite{youn2023randomized}. 

\research{Propose strong user-level DP algorithms for multi-user datasets}
Privacy laws like GDPR focus on individuals rather than single records \cite{salem2023sok}. In distributed settings, where users may contribute multiple records, it is crucial to protect all user data. 
While for \fl standard user-level DP algorithms exist \cite{mcmahan2017learning}, enforcing strong user-level DP when multiple users' data are pooled in a global dataset is challenging \obsn{4}. Thus, proposing strong user-level DP algorithms for multi-user datasets is an aspect for future research. 
\section{Conclusions}\label{sec:conclusions}
We systematized the landscape of encrypted and differential private collaborative learning (Tab.~\ref{tab:summary}), analyzing how to combine cryptography and DP to guarantee input confidentiality and output privacy.
Our comprehensive framework (Sec.~\ref{sec:framework}) formalized common phases across \epcl paradigms (i.e., federated and outsourced learning), identifying noise sampling as the foundational phase.
We analyzed noise types, secure sampling techniques, and mechanisms, implementing distributed sampling in LAN and WAN. 
We evaluated accuracy and runtime overhead of \epcl paradigm across noise sampling techniques, analyzing the performance, accuracy, and privacy trade-offs. 
Throughout, we identified gaps and possible enhancements in the literature, and highlighted key observations deriving future research directions. 

\section*{Acknowledgments}
The authors would like to thank the anonymous reviewers for their valuable feedback. This work has received funding from the European Union's Horizon Europe research and innovation program under grant agreement No~101070141~(GLACIATION).

\bibliographystyle{IEEEtran}
\bibliography{bibliography}

\appendices

{
\setlength{\abovedisplayskip}{0.7em}
\setlength{\belowdisplayskip}{0.7em}
\section{Systematic Search}\label{app:methodology}
%
Here we detail the methodology for our preliminary and systematic search, and the selection of relevant works. 
{
As a starting point, we considered top-tier conferences and journals in security and ML, based on empirical rankings \cite{securityranking1,securityranking2,mlranking1}. Those rankings measure either the h-5 index, i.e., the largest number \enx{h} such that \enx{h} papers have at least \enx{h} citations in the last 5 years, or the Conference Impact Factor defined in \cite{securityranking2}.
}
For security, we considered the joint set of the top-5 venues from \cite{securityranking1} and \cite{securityranking2}, including cryptography conferences: 
IEEE Symposium on Security and Privacy, 
USENIX Security Symposium, 
ACM Conference on Computer and Communications Security, 
Network and Distributed System Security Symposium, 
IEEE Transactions on Information Forensics and Security (Journal), 
Computer \& Security (Journal), 
International Cryptology Conference, Cryptology-EUROCRYPT.
For ML, we selected top-5 venues from \cite{mlranking1}:
Neural Information Processing Systems, 
International Conference on Learning Representations, International Conference on Machine Learning,
AAAI Conference on Artificial Intelligence, IEEE Transactions on Systems.
The query for Google Scholar and BASE was:
distributed OR collaborative "multi-party" OR multiparty OR homomorphic  OR "privacy-preserving"  OR secure  "differential-privacy" OR "differential privacy"  learning OR training source:"Conference Name". Here "Conference Name" is replaced by each of the 13 venues listed above.
{
Notably, our full selection of works, i.e., including also cited by or citing found works, includes relevant papers beyond these initial venues.

}

\xx{
    From the systematic search, we obtained 650 papers. After checking the abstracts, we narrowed down the selection to 61 papers which contain reference to cryptography and DP for distributed ML training. After carefully analyzing the contributions of those papers, we selected 11 works uniquely combining cryptography and DP in collaborative learning. 
    We expanded our search via snowball sampling, and by applying the same criteria mentioned above to forward and backward references, we added 8 more works.
}


\section{Summary of Notation}\label{app:notation}
\begin{table}[htbp]
    \centering
    \renewcommand{\arraystretch}{1.2}
    \small
    \caption{Summary of Notation and Symbols.}
    \label{tab:notation}
    \begin{tabular}{l p{6cm}}
        \textbf{Symbol} & \textbf{Description} \\
        \midrule
        \multicolumn{2}{l}{\textit{\textbf{Learning Paradigms}}} \\
        CL & Collaborative learning (Sec.~\ref{sec:intro}). \\
        CPCL & Cryptographic and differentially private collaborative learning (Sec.~\ref{sec:intro}).\\
        \fl & Federated learning (Sec.~\ref{subsec:fl}).\\
        \ol & Outsourced learning (Sec.~\ref{subsec:ot}).\\
        \midrule
        \multicolumn{2}{l}{\textit{\textbf{Parties \& Sets (Sec.~\ref{sec:crypto})}}} \\
        $\mathcal{C}$ & Set of clients which are input and output parties (optionally can be also computation parties). \\
        $\mathcal{S}$ & Set of servers which are computation parties. \\
        $\mathcal{O}$ & Set of output parties, which includes clients and optionally servers, i.e., $\clset \subseteq \mathcal{O} \subseteq \clset \cup \slset$. \\
        $C_i, S_j$ & The $i$-th client and $j$-th server. \\
        $S'$ & Semi-trusted server with no access to computation outputs.  \\
        $n$ & Number of clients. \\
        $m$ & Number of servers.\\
        $t$ & Collusion threshold (maximum number of colluding parties). \\
        
        \midrule
        \multicolumn{2}{l}{\textit{\textbf{Data \& Learning (Sec.~\ref{sec:DP})}}} \\
        $D, D_i$ & Global dataset and local dataset of client $C_i$. \\
        $\theta, \theta^{(k)}$ & Model parameters, parameters at step $k$. \\
        $g, \mathbf{G}$ & Gradient vector and Set of gradients. \\
        $\tilde{g}$ & Noisy gradient ($g + \psi$). \\
        $K$ & Gradient clipping threshold. \\
        $\eta$ & Learning rate. \\
        $B$, $b$   & Batch, mini-batch size.\\
        
        \midrule
        \multicolumn{2}{l}{\textit{\textbf{Differential Privacy \& Noise Sampling}}} \\
        $(\epsilon, \delta)$ & DP budget parameters (Sec.~\ref{sec:DP}).\\
        $\psi$ & Noise sample drawn from a distribution in Tab.~\ref{tab:noisedistributions} (Sec.~\ref{sec:DP}). \\
        \perturbgrad & Noise added to gradient updates (Sec.~\ref{sec:DP}). \\
        \perturbout & Noise added to trained parameters  (Sec.~\ref{sec:DP}). \\
        \partnoise & Partial, non-DP noise sampled locally. The aggregation of \partnoise satisfies DP  (Sec.~\ref{sec:secureandprivateML}).\\
        \centrnoise & Centralized noise term sampled globally which satisfies CDP (Sec.~\ref{sec:secureandprivateML}). \\
        $\tau$  & Tuple of noise parameters which defines noise type (\partnoise/\centrnoise), noise mechanism (Tab.~\ref{tab:noisedistributions}, Sec.~\ref{sec:noisegen}) and distribution parameters (e.g., \enx{\sigma^2} for (\enx{\eps, \delta})-DP) (Sec.~\ref{sec:framework}). \\

        \midrule
        \multicolumn{2}{l}{\textit{\textbf{Cryptography \& Formatting (Sec.~\ref{sec:crypto})}}} \\
        $[x]$ or $[x]_s$ & Encrypted or secret-shared value of $x$.  \\
        \glob{Phase} & Computation on global and joint encrypted data (e.g., via MPC). \\
        \underline{\perturb} & Centralized sampling via a semi-trusted server~$S'$.  \\
    \end{tabular}
    
\end{table}

Tab.~\ref{tab:notation} provides a summary of notation and symbols with references to sections which introduce those symbols.

\section{Relaxed DP definitions}\label{app:relaxedDP}
{
Dwork et al.~\cite{dwork2014algorithmic} introduced \emph{advanced composition} to offer tighter privacy bounds, but the composition bound for (\enx{\eps, \delta})-DP. However, the composition bound is still loose, requiring relaxed DP definitions.
}
Next, we formalize Rényi DP, zero-concentrated DP and their conversion lemmas to \enx{(\eps,\delta)}-DP. Table \ref{tab:dpcomparison} reports the conversion formulas and compares the privacy guarantee achieved after \enx{k}-fold composition.

\begin{table*}[tb]
    \centering
    \small
    \caption{Comparison of different relaxed DP definitions from Jayaraman and Evans~\cite[Tab.~1]{jayaraman2019evaluating}}
	\label{tab:dpcomparison}
    \vspace{-0.5em}
    \def\invdelta{\frac{1}{\delta}}
    \begin{tabular}{lccc}
        &
		Advanced composition \enx{(\eps,\delta)}-DP &
		\enx{\rho}-zCDP &
        \enx{(\alpha,\eps)}-RDP
		\\
        \midrule
		Conversion to \enx{(\eps,\delta)}-DP  &
        - &
        \enx{(\rho + 2\sqrt{\rho\log{\invdelta}},\delta)}-DP &
        \enx{(\eps+ \log{\invdelta}/(\alpha-1),\delta)}-DP\\

        Composition of \enx{k} \eps-DP mech. &
        \enx{(\eps \sqrt{2k\log{\invdelta}} + k\eps (e^{\eps} -1), \delta)}-DP &
        \enx{(\eps \sqrt{2k\log{\invdelta}} + k\eps^2/2,\delta)}-DP &
        \enx{(4\eps \sqrt{2k\log{\invdelta}}, \delta)}-DP
        \\
   \end{tabular}%
   \vspace{-1.2em}
\end{table*}

\vspace{0.5em}
\begin{definition}
    (Rényi Differential Privacy)\cite{mironov2017renyi} RDP is a generalization of the notion of differential privacy based on the Rényi divergence:
    {\small
    \begin{displaymath}
        D_\alpha(P||Q) \triangleq \frac{1}{\alpha -1} \ln E_{x \thicksim Q} \Biggl( \frac{P(x)}{Q(x)} \Biggr)^\alpha,
    \end{displaymath}
    }
    for two probability distributions $P$ and $Q$. A randomized mechanism $\mathcal{M}$ satisfies \enx{(\alpha, \eps)}-RDP if for any neighboring dataset $D_1, D_2$ it holds:
    {\small
        \begin{equation}
        D_\alpha(\mathcal{M}(D_1)|| \mathcal{M}(D_2)) \le \eps.
    \end{equation}}
\end{definition}

\begin{lemma}
    (Conversion from \enx{RDP} to \enx{(\eps, \delta)}-DP) \cite{mironov2017renyi}
    The RDP definition can be easily converted to \enx{(\eps,\delta)}-DP when \enx{0 < \delta < 1}. If a randomized mechanism $\mathcal{M}$ holds for \enx{(\alpha, \eps)}-RDP then $\mathcal{M}$ is \enx{(\eps+ \log(1/\delta)/(\alpha-1),\delta)}-DP.
\end{lemma}

\vspace{0.5em}
\begin{definition}
    (zero-Concentrated Differential Privacy) \cite{bun2016concentrated} A randomized mechanism $\mathcal{M}$ satisfies \enx{(\xi, \rho)}-zCDP if for any neighboring dataset $D_1, D_2$ and for all \enx{\alpha \in (1, \infty)}: 
    {\small\begin{equation}
        D_\alpha(\mathcal{M}(D_1)|| \mathcal{M}(D_2)) \le \xi + \rho\alpha,
    \end{equation}}
    where \enx{D_{\alpha}(\cdot)} is the \enx{\alpha}-Rényi divergence. Bun et al.~\cite{bun2016concentrated} define \enx{(0, \rho)}-zCDP as \enx{\rho}-zCDP.
\end{definition}

\begin{lemma}
    (Conversion from \enx{\rho}-zCDP to \enx{(\eps,\delta)}-DP) \cite{bun2016concentrated} If a randomized mechanism \enx{M} provides \enx{\rho}-zCDP, then \enx{M} is \enx{(\rho + 2\sqrt{\rho\log{(1/\delta)}},\delta)}-DP for any \enx{\delta  > 0}
\end{lemma}

Both Rényi DP  and (zero) concentrated DP uses the Rényi divergence to quantify the privacy loss. The Rényi divergence carries a parameter \enx{\alpha}, which define the moments of the divergence. The main difference between RDP and zCDP lies in the fact that RDP applies one moment at a time, whereas zCDP requires a linear bound on all positive moments of a privacy loss variable \cite{mironov2017renyi}. Having the privacy loss value linked to one moment allows for more accurate numerical analysis \cite{mironov2017renyi}.
Notably, while (\enx{\eps, \delta})-DP defines the worst-case privacy loss, which can be infinite with a non-zero probability \enx{\delta} \cite{kairouz2021distributed,mironov2017renyi}, zCDP and RDP do not allow for such privacy breaches as they are defined on the average privacy loss \cite{desfontainesblog20220220}.

Ponomareva et al.~\cite[Tab.~2]{ponomareva2023dp} show the evolution of DP-training bounds with \enx{T} iteration of DP-SGD, with a sampling parameter \enx{q}. Using moments accountant on RDP and converting in \enx{(\eps, \delta)}-DP, the bound on \eps decreases by a factor of \enx{\sqrt{T}}, whereas the bound on \enx{\delta} reduces by \enx{qT}.


\subsection{Local DP} \label{app:localdp}

To avoid relying on a trusted third party as in CDP (Def.~\ref{def:epsdDP}), in \emph{local} DP (LDP) \cite{kasiviswanathan2011can}, each data owner applies the perturbation independently.
\vspace{0.5em}
\begin{definition}
    Formally, $\mathcal{M}$ satisfies (\enx{\eps,\delta})-local DP if:
    {\small
    \begin{equation}
        Pr[\mathcal{M}(x_1)\in \mathcal{S}] \le \exp(\eps) \times Pr[\mathcal{M}(x_2)\in \mathcal{S}]  + \delta
    \end{equation}}
    for any pair of values \enx{x_1, x_2 \in D} and all 
    $\mathcal{S} \subseteq Range\mathcal{(M)}$.
\end{definition}

\subsection{Computational DP} \label{app:compdp}
While CDP and LDP consider an unbounded adversary, cryptography bounds the adversary's capabilities. 

\emph{Computational DP} \cite{dwork2014algorithmic} adapts Def.~\ref{def:epsdDP} to a bounded polynomial-time adversary modelling the probability to break a cryptographic scheme.
\vspace{0.5em}
\begin{definition}
    A randomized algorithm \enx{ \mathcal{C}_{\kappa}: \mathcal{X}_n \to Y} is $\epsilon$\textit{-computationally differentially private} if and only if for all the neighboring datasets \enx{(D_1, D_2)} and for all nonuniform polynomial (in \enx{\kappa}) algorithms T,
    {\small
        \begin{equation}
        Pr[T(\mathcal{C}_{\kappa}(D_1)) = 1] \le \exp(\eps) \times Pr[T(\mathcal{C}_{\kappa}(D_2)) = 1] + \delta(\kappa).
    \end{equation}}
    Here, \enx{\delta(\cdot ) = \delta + \text{negl}(\cdot)}, where \enx{\text{negl}(\cdot)} is any function that grows slower than the inverse of any polynomial. The algorithm \enx{\mathcal{C}_{\kappa}} runs in time polynomial in \enx{n}, \enx{\log|X|}, and \enx{\kappa}. The term \enx{\text{negl}(\kappa)} represents the probability of violating a cryptographic scheme, e.g., guessing the decryption key~\cite{dwork2014algorithmic}.
\end{definition}


\section{Privacy Amplification}\label{app:privacy_amplification}
Randomly sampling a subset of the data for each iteration amplifies the privacy guarantee of DP-SGD. The uncertainty of a sample being selected in a training iteration makes it harder for an adversary to infer the presence of a specific record in the dataset.
Subsampling amplification allows DP-SGD to achieve strong privacy-utility trade-offs, as it can reduce the noise variance for the same privacy guarantee \cite{abadi2016deep}. 
For each iteration \enx{B} samples are selected out of \enx{N}, \ie{each with a probability \enx{q=B/N}}.
Analytically the amplification is {$(O(q(\exp(\eps)-1)), O(q\delta))$-DP} \cite{ponomareva2023dp}.
For small values of \enx{\eps} (\enx{\eps \le 1}), the amplification is of a factor of \enx{q}; this factor decreases as \enx{\eps} increases.
The batch size \enx{B} is crucial, as larger \enx{B} values reduce the noise variance but require more gradient computations, and smaller \enx{B} values increase the noise variance but reduce the computation overhead.
According to \cite{abadi2016deep}, the best batch size \enx{B} is roughly \enx{\sqrt{N}}.

\para{Subsampling Techniques}
The subsampling amplification requires true sampling and not shuffling \cite{ponomareva2023dp}. The two main subsampling techniques are \emph{Poisson} \cite{wang2019subsampled} and \emph{Uniform} \cite{balle2018privacy}. 
The two techniques rely on a different notion of neighboring datasets in Def.~\ref{def:epsdDP}, i.e., {\emph{add/remove} or \emph{replace}}. The add/remove notion is used for Poisson subsampling and consists of adding or removing a sample from a dataset \enx{D_1} to obtain the neighboring one \enx{D_2}. This leads to a variable dataset size that translates into a variable batch size for each training iteration with the Poisson subsampling. The replace notion is used for the uniform subsampling and consists of replacing a sample in \enx{D_1} with another to obtain \enx{D_2}. This leads to a fixed batch size for each training iteration with uniform subsampling~\cite{ponomareva2023dp}.

\para{Subsampling in \fl and \ot}
In \fl, the subsampling amplification happens on \cls, \ie{for each iteration the \sls select a subset of \cls}. This approach is particularly challenging in federated settings where the availability of the \cls can be different at each round. To solve this issue Kairouz et al.~\cite{kairouz2021practical} propose a DP variant of the follow-the-regularized-leader (FTRL) algorithm, \ie{DP-FTRL}, that does not rely on privacy amplification, but on a tree aggregation trick.
%
%
In \ot, the subsampling happens on the global joint dataset and needs to be implemented in MPC, \ie{the \sls need to collaboratively sample Poisson random variable with rate \enx{q} and selected the indexes of the samples to be used in the training iteration.}
From the analyzed works, PEA~\cite{pea2023ruan} implements a resharing-based oblivious shuffling protocol \cite{laur2011round}, but this approach does not satisfy either Poisson or Uniform amplification \cite{ponomareva2023dp}.
To perform subsampling in our evaluation of DP overhead in \ot, reported in Tab.~\ref{tab:combined-dp-overhead} (Sec.~\ref{sec:noisetradeoff}), we implement the \emph{Poisson} subsampling by letting one \slong to perform the subsampling locally and then reshare the indexes to the other \sls. This approach is efficient, \ie{the sampling requires less than \enx{0.1\%} of the \compgrad time}, but leaks the indexes of the selected samples which are secret shared among the \sls.
An alternative solutions could be to let each client subsample its own dataset locally and secret share the subsampled data for each iteration, but this would require additional communication rounds. 

\section{Cryptographic techniques}

\begin{table*}[ht]
    \centering
	\def\colwcomm{0.18\textwidth}
	\def\colwcomp{0.1\textwidth}
    \scriptsize
    \renewcommand{\arraystretch}{1.1}
    \setlength{\tabcolsep}{0.7em}
    \caption{Total comp(utation) and comm(unication) complexity of cryptographic techniques across phases (for non-optimized implementations).} 
	\label{tab:complexity_phases}
    \vspace{-0.5em}
    \begin{tabular}{lcc|cc|cc}
    \multirow{2}{*}{\textbf{Technique}} & \multicolumn{2}{c}{\boldmath{\setup} ({by} {\clset})} & \multicolumn{2}{c}{\boldmath{\protection} ({by} {\clset})} & \multicolumn{2}{c}{\boldmath{\reveal} {(by either {\clset} or {\slset})}} \\ 

	& \textbf{Comp} & \textbf{Comm} & \textbf{Comp} & \textbf{Comm} & \textbf{Comp} & \textbf{Comm} \\ \toprule

	MPC & $O(1)$ & $O(1)$ & $O(n m)$ & $O(nm)$ & $O(m)$ & \specialcell{$O(n m)$ if \Recon by \clset \\ $O(m^2 + n)$ if \Recon by \slset} \\ \midrule
	HE & $O(n)$ & \specialcell{$O(n^2)$ if distributed key gen \cite{bindschaedler17star} \\ $O(n)$ via trusted dealer \cite{sebert2023combining}} & $O(n
    )$ &  $O(n)$  & $O(1)$ & \specialcell{$O(n^2)$ if distributed decryption by \clset \cite{sebert2023combining} \\ $O(n)$ if server act as relay (\Dec by \clset or \slset)\cite{truex2019hybrid}}  \\ \midrule
	
	Masking & $O(n)$  & \specialcell{$O(n^2)$ if pair-wise masking \cite{bonawitz2017practical} \\ $O(n \log n)$ via $k$-regular graphs \cite{bell2020secure} \\ $O(1)$ if LWE-based \cite{stevens2022efficient}} &  $O(n)$ & $O(n)$ & \specialcell{$O(1)$ \cite{agarwal2018cpsgd}} &  \specialcell{$O(1)$ if pair-wise masking \cite{agarwal2018cpsgd} \\ $O(n)$ if dropout resistant \cite{bonawitz2017practical} \\ $O(n)$ if LWE-based \cite{stevens2022efficient}}  \\ 
	
	\end{tabular}
	\vspace{-1.5em}
\end{table*}

Next, we expand on cryptographic techniques briefly introduced in Sec.~\ref{sec:crypto}. In particular, we provide more details for garbled circuits and learning with errors, and we discuss zero-knowledge proofs which can be used as mitigation for privacy attacks discussed in App.~\ref{app:privacyattacks}. 
Additionally, we provide a complexity analysis of the core phases of cryptographic techniques summarized in Tab.~\ref{tab:complexity_phases}.

\subsection{Garbled Circuits}\label{app:gc}

\emph{Garbled circuits} (\gc)
\cite{yao1986generate} is a two-party protocol to securely evaluate a function \enx{f} represented as a Boolean circuit \enx{C}.
One party (garbler), runs \Garble, creating a \gc whose truth table entries consists of random labels.
Another party (evaluator), runs \Eval on the \gc, evaluating the circuit.
To let the evaluator only learn random labels corresponding to its input while hiding the evaluator's input from the garbler,
they run a cryptographic protocol called \emph{oblivious transfer} \cite{rabin2005exchange}.
%
GC requires \emph{correctness}, \ie{outputs of garbled \enx{C} and \enx{f} match}, and \emph{secrecy}, \ie{\Eval reveals only the \gc output}.



\subsection{Learning with Errors}\label{app:lwe}
Learning with Errors (LWE) \cite{regev2009lattices} can be seen as the problem of decoding from a random linear code. The LWE problem is formalized as finding an unknown vector \enx{s \in \mathbb{Z}_q^n} such that \enx{ b = As + e \in \mathbb{Z}_q^m}. Where \enx{A \in \mathbb{Z}_q^{m \times n}} is a matrix built from \enx{m} random vectors \enx{a_i \in \mathbb{Z}_q^n}, and \enx{ e \in \mathbb{Z}_q^m} is the error vector. The challenge of LWE comes from the \enx{e} entries, which are sampled from a suitable probability distribution on \enx{\mathbb{Z}_q} disrupting the linear relation among the equations.
Notably, cryptographic schemes based on LWE are additively homomorphic for both the key and the message, i.e., given two messages (\enx{m_1, m_2}), an encryption algorithm \enx{Enc} and two keys \enx{(s_1,s_2)}, \enx{Enc(s_1, m_1) + Enc(s_2, m_2) = Enc(s_1 + s_2, m_1 + m_2)} \cite{bell2022acorn}.

\para{LWE in \fl} 
In Stevens et al.~\cite{stevens2022efficient}, each client \c{i} samples its secret \enx{s_i} and the error vector \enx{e_i} from a discrete Gaussian distribution. \c{i} masks the local update \enx{v_i} with \enx{b_i}, resulting in \enx{h_i = v_i+b_i} where \enx{b_i = As_i + e_i}, and sends \enx{h_i} to the server \s{}. \s{} computes \enx{h_{sum} = b_{sum} + v_{sum}} where \enx{b_{sum} = As_{sum} + e_{sum}}. To remove the term \enx{As_{sum}} and retrieve the noisy global update \enx{v_{sum} + e_{sum}}, each client secret shares, via \emph{packed Shamir secret sharing} \cite{franklin1992communication}, its \enx{s_i} to all the \cls\footnote{Bell et al.~\cite{bell2022acorn} stated that the reconstruction protocol for the secret \enx{s} could reveal some information about \enx{s}, breaking confidentiality.}. A set of \enx{t} \cls reconstruct \enx{s_{sum}} and send the value to \s{}.

{

\subsection{Zero-Knowledge Proofs}\label{app:zkp}
A zero knowledge proof (ZKP) allows a prover to convince a verifier that it knows a secret \enx{x} satisfying a public predicate \enx{C(x) = 1}, without revealing \enx{x} itself \cite{evans2018pragmatic}. Maliciously secure protocols use ZKPs to prove the validity of computations over private inputs without exposing those inputs. 
For example, Drynx \cite{froelicher2020drynx} leverages ZKPs in FL to verify both local updates and their aggregation. Specifically, \cls provide range proofs \cite{camenisch2008efficient} to ensure their local model updates fall within valid bounds, while \sls generate ZKPs to prove the correctness of the aggregation process.
}

\subsection{Complexity Analysis of Cryptographic Techniques}\label{app:crypto_complexity}

We now detail how specific cryptographic techniques implement the core phases, highlighting the trade-offs in communication and computation for non-optimized version of protocols summarized in Tab.~\ref{tab:complexity_phases}.

\para{\boldmath{\setup}}
Cryptographic techniques have different \setup requirements. MPC requires multiple non-colluding servers but no key generation for \cls, since \Share uses local randomness. However, MPC typically requires servers to exchange cryptographic material in the offline phase.
In contrast, HE and masking operate with a single server and require a secret key known by (or distributed to) \cls. 
In HE, \cls agree on a set of keys \enx{(pk, sk)}. Key generation can be done via a trusted dealer in $O(n)$ messages sent to clients \cite{sebert2023combining}, or via distributed key generation requiring  clients interaction in $O(n^2)$ messages \cite{truex2019hybrid,bindschaedler17star}.  
Masking requires \cls to exchange pairwise seeds for mask generation, in \enx{O(n^2)} messages. 
Communication can be reduced to \enx{O(n\log n)} via \enx{k}-regular graphs, i.e., exchanging seeds with \enx{k=\log n} neighbors \cite{bell2020secure}, or via LWE-based masking \cite{regev2009lattices} where matrix $A$ is publicly shared ($O(1)$ messages).

\para{\boldmath{\protection}}
In MPC, each client secret shares (\Share) data among \enx{m} servers, in $O(m)$ messages. Instead, in single-server settings, \cls leverage HE or masking to encrypt (\Enc) or mask the data sending $O(1)$ messages to the server.

\para{\boldmath{\reveal}}
MPC typically incurs $O(mn)$ total messages if \cls reconstruct ({\Recon}) the result by receiving $O(m)$ messages from \sls (e.g., summing shares in additive SS), or $O(m^2 + n)$ if \sls reconstruct and send results to \cls. 
%
%
With HE, clients can use a common secret key for decryption (\Dec), risking that a client and the server can collude to decrypt other \cls' data \cite{sebert2023combining}. A safer option is distributed decryption via threshold HE, where \cls communicate directly in $O(n^2)$ messages \cite{sebert2023combining} or the \sls act as a relay ($O(n)$ messages) \cite{bindschaedler17star, truex2019hybrid}. 
For pair-wise masking \reveal is implicit as masks cancel out in {\aggr} (\enx{O(1)} messages). In LWE-based masking, however, clients reconstruct the aggregated \enx{\sum_{i}^{n}s_i} in \enx{O(n)} messages, with a new key per epoch.



\section{Framework Generality}\label{app:framework_generality}
Figure~\ref{fig:flows_per_paradigm} illustrates the generalization of our \epcl framework across different collaborative learning paradigms as introduced in Sec.~\ref{sec:framework}.  
We detail the execution flows for \fl (including both client- and server-side noise sampling) and \ot, as formalized in Alg.~\ref{alg:cmp}. Furthermore, we demonstrate the framework's modular extensibility to peer-to-peer (P2P), split learning (SL) \cite{pereteanu2022split, khan2023love, kanpak2024cure}, and vertically partitioned data in \fl (VFL) \cite{xu2021achieving,zhu2021pivodl} and \ot (VOL) \cite{gascon2016secure}. 


Specifically, in \emph{peer-to-peer} learning \cls act as \sls \enx{(\mathcal{C = S})}, throughout the training.
In \emph{split learning}~\cite{wang2023privacy,thapa2022splitfed} the model is partitioned across \cls and \sls, which perform \compgrad on their respective partitions. Here, \cls apply \protection on intermediate activations and send them to \sls. \aggr operates at layer level rather than on full-model gradients, and gradients are not combined across clients. Hence, each client can execute \perturb locally adding \centrnoise to its gradients. Meanwhile, for model partitions trained by servers, \perturb can either sample \partnoise locally or \centrnoise via MPC or a semi-trusted party.
%
%
In \emph{vertical federated learning} \cite{liu2024vertical} clients hold different features for the same samples. Thus, \setup also includes private entity alignment to link records across datasets \cite{liu2024vertical}. Clients train their local model and execute \protection on intermediate activations. Servers perform \aggr on local activations, and then execute \compgrad on the global loss. The gradients are then sent back to clients. As in split learning, \aggr is executed on intermediate results and the same considerations on \perturb apply.
In contrast, for \emph{vertical outsourced learning}, clients perform \protection locally, while servers  merge the vertically partitioned datasets in \aggr before proceeding with horizontal training as in \ol (Alg.~\ref{alg:cmp}).

\input{figures/flow_diagrams_paradigms.tex}

\section{Collaborative Labeling}
\label{app:collab-labeling}
CAPC \cite{choquette2021capc} augments the DP approach Private Aggregation of Teacher Ensembles (PATE) by Papernot et al.~\cite{papernot2018scalable} with cryptography. In PATE, a \emph{student} (\enx{St}) queries a set of \emph{teachers} (\enx{T_i}) to collaboratively label its dataset using their local models. The goal of collaborative labeling is not to train a DP model but to label a dataset for local training. The protocol output is the predicted label set via majority voting.
The labeling needs to be executed securely, \ie{with encrypted data}, since the \enx{St} data need to remain private, and the teachers' models need to be protected from inference attacks (App.~\ref{app:privacyattacks}), \ie{applying \perturblabel}.
Specifically,
in CAPC, \enx{St} encrypts its data and performs secure inference
with each \enx{T_i}
via a 2-party mixed-protocol \cite{boemer19}, \ie{combining FHE \cite{cheon2017homomorphic} and MPC \cite{demmler2015aby}}.
From the inference, each \enx{T_i} receives a homomorphically encrypted label,
masks it (i.e., adds random value), and sends to \enx{St}.
Thus, the label is secret shared with \enx{St},
\ie{\enx{St} learns the masked value after decryption and \enx{T_i} knows the mask}.
Now, \enx{St} and \enx{T_i} compute the \emph{one-hot encoding} of the label via GC \cite{emp-toolkit}, secret sharing the result.
Afterwards, each \enx{T_i} sends its share of the (one-hot encoded) label to a third party, called \emph{privacy guardian} (PG),
which applies \perturblabel,
and runs GC \cite{emp-toolkit} with \enx{St} to compute the label with the most votes.

\section{Quantization techniques}\label{sec:quantization}
Here, we describe the advantages and disadvantages of employing quantization and working with discrete values. Quantization approximates \enx{d}-dimensional vectors from \enx{\mathbb{R}^d} to \enx{\mathbb{Z}^d}. This can reduce the size of messages exchanged among \pls, decreasing the communication costs.
On the other side, quantization introduces an approximation error that leads to a trade-off between communication and accuracy, since larger number of bits mean more accuracy but also more communication.
Next, we provide a brief description of the quantization methods applied to secure aggregation, by two of the analyzed works. Agarwal et al.~\cite{agarwal2018cpsgd} apply stochastic \enx{k}-bit quantization composed with binomial mechanism. The data are preprocessed with random rotation to reduce the leading quantization error term according to Suresh et al.~\cite{suresh2017distributed}. They show that this technique not only reduces the approximation error, but also improves the privacy guarantee.
In \cite[Corollorary 3]{agarwal2018cpsgd} they state that their protocol achieves the same error and privacy of the full precision Gaussian mechanism with a total communication costs of \enx{O(nd\log{\log{(nd/\delta\eps)}})} when \enx{d=O(n\eps^2)} instead of \enx{O(nd\log{nd})} for \enx{d} variable and \enx{n} \cls.
%
%
Kairouz et al.~\cite{kairouz2021distributed} observe that the communication cost depends on the  dimensionality \enx{d} of inputs and \enx{\log{m}}, which is the number of bits per coordinate. Since they can not control \enx{d}, the only factor that can be reduced is \enx{\log{m}}.
Kairouz et al.~\cite{kairouz2021distributed} scale and clip the input vectors, through rotations/reflections, flatten the vectors to reduce the distortions due to modular wrap around, i.e. this is caused by modulo operations in secure aggregation \cite{bonawitz2017practical}.
Each element of the input vector is randomly and independently rounded to one of the two the nearest integers.
The scaling and flattening are undone on the server side during the \aggr.
They combine quantization with discrete Gaussian noise aggregation, and show that 16 bits per coordinate are sufficient to nearly match the utility of the Gaussian baseline with fixed precision of 32 bits.

\section{Handling Dropouts}\label{sec:dropouts}
Dropouts are \pls that prematurely leave the protocol during its execution.
In single-server architectures, handling dropouts is crucial, as \cls are actively involved in the protocol.
With multiple servers, \cls are either \cpls or they only outsource their data to \cpls (e.g., secret sharing).
Dropouts cause the protocol to stop, and in distributed settings this can frequently happen, e.g., mobile devices in secure aggregation can easily drop out \cite{bonawitz2017practical}. We can distinguish dropout resistant protocols with or without additional \cls interaction.
Stevens et al.~\cite{stevens2022efficient} and Bindschaedler et al.~\cite{bindschaedler17star} guarantee dropout resistance without additional \cls interaction since both protocols have a distinct phase for secret key reconstruction based on threshold secret sharing which can terminate successfully without the local updates of the dropped \cls, \ie{if less than threshold \cls drop out}.

Secure aggregation protocols based on Bonawitz et al.~\cite{bonawitz2017practical}, \eg{Kairouz et al.~\cite{kairouz2021distributed}}, require additional communication rounds since the \cls can dropout after the \setup phase, i.e., sharing of the random masks. To ensure confidentiality, each client adds the masks of all the \cls including the dropouts, which do not cancel out in the \aggr phase. To ensure dropouts resistance, each client secret shares its mask with the other \cls, so the \slong can ask the \cls to reconstruct the masks of the dropouts which is removed after \aggr. This causes a privacy issue for the dropouts, since the \slong can maliciously mark a client as dropped out and later retrieve the local update of that client removing the reconstructed mask. Bonawitz et al.~\cite{bonawitz2017practical} implement a double masking to protect the privacy of dropouts. Each client generates two masks which are both secret shared among the \cls, \ie{one for dropout resistance and one to protect privacy}. The \slong can asks for the reconstruction of only one mask per client.
For a complete description of the protocol we refer to \cite{bonawitz2017practical}.

{
Recent advancements \cite{taiello2024let, karakocc2024fault} leverage a threshold version of the Joye-Libert (TJL) AHE scheme \cite{mansouri2022learning} to avoid the double masking.
In Karako\c{c} et al.~\cite{karakocc2024fault} a key dealer distributes a secret key for each client and the decryption key to the server. The decryption key works only if all the clients send their encrypted update. To handle dropout, during setup, the clients secret share the encryption of zero via a \tnss scheme with other clients. If a client drops out, the other clients can collaboratively encrypt a zero on behalf of the dropped client and the server can decrypt the sum of the updates.
Taiello et al.~\cite{taiello2024let} extend Karako{\c{c}} et al.~\cite{karakocc2024fault} with a secure aggregation protocol which relies only on online clients. 
Each client generates a per-round JL key and protects it with a TJL key which each client secret shares among selected clients. The server reconstructs the per-round aggregation key with the collaboration of at least \enx{t} online clients. The model parameters are computed using the aggregation key which is the sum of the per-round keys of online clients. 


}

Dropouts are also an issue for \partialnoise, since the remaining \cls have to ensure that the aggregated noise achieves CDP.
As discussed in Section \ref{subsec:partialnoise}, to also tolerate up to $s$ dropouts, the denominator of Eq. \eqref{eq:partgaussvariance} can be set to \enx{n - (t + s)}. For examples, Dwork et al.~\cite{dwork2006our} guarantees Byzantine robustness, with \enx{2/3} of remaining \cls.

%
\section{Noise distributions}\label{app:noisedistributions}
Next, we provide details on how to convert privacy parameters to distribution parameters for different noise mechanisms.

\para{Distributed Laplace}
The distributed Laplace mechanism is realized combining gamma-distributed random variables in the following way:
\enx{\mathsf{Lap}(0, \lambda) =  \sum_{k=1}^{N}  \mathsf{Gamma}_k - \mathsf{Gamma}_{k}'}, where the scale parameter \enx{\lambda = \Delta_1/\eps}, and each gamma-distributed value \enx{\mathsf{Gamma}_k, \mathsf{Gamma}_{k}'} is sampled from:
{\small\begin{displaymath}
    {\mathsf{Gamma}(x;1/N,\lambda) = \frac{1}{\Gamma(1/N)\lambda^{1/N}}x^{1/N-1}\exp\biggl(-\frac{x}{\lambda}\biggr)},
\end{displaymath}}
where \enx{\Gamma(1/N) = \int_{0}^{\infty}x^{1/N-1}\exp(-x)\, dx}. 

\para{Gaussian}
For the Gaussian distribution, the noise variance \enx{\sigma_{DP}}, for \enx{\eps \le 1}, is computes as:
\enx{\sigma_{DP} \leftarrow \sqrt{2\ln(1.25/\delta)} \Delta_2/\eps}
For \partialnoise, \enx{\sigma_{i}} is computed according to Eq.~\eqref{eq:partgaussvariance} considering \enx{\sigma_{DP}} as target variance.

\para{Binomial} For the multidimensional binomial distribution, \ie{data with \enx{d} dimensions}, the number of coin flip (\enx{p=0.5}) to achieve DP guarantees close to the Gaussian, can be reduced to \enx{m \ge 8 \log(2/\delta)/\eps^2}.
The equivalence with the discrete Gaussian is for \enx{ \sigma = s \sqrt{Np(1-p)}} Here, \enx{s = 1/j, j\in \mathbb{N}} is a quantization scale, set to \enx{s \le \sigma/(c\sqrt{d})} \cite{agarwal2018cpsgd}.

\para{Discrete Gaussian}
The value of \enx{\sigma} for the discrete Gaussian has to optimize the following \cite{kairouz2021distributed}:
{\small
\begin{displaymath}
    \eps = \min \biggl( \sqrt{\frac{\Delta_2^2}{n \sigma^2} + \frac{1}{2}\tau d},  \sqrt{\frac{\Delta_2^2}{n \sigma^2} + 2\frac{\Delta_1}{\sqrt{n} \sigma}\tau + \tau^2 d}, \frac{\Delta_2}{\sqrt{n} \sigma} + \tau \sqrt{d} \biggr),
\end{displaymath}
}
where \enx{d} is the data dimensionality, and $ \tau = 10\cdot \sum_{k=1}^{n-1} \exp(-2\pi^2\sigma^2 \frac{k}{k+1})$\;
The \enx{\sigma} has to be scaled according to the quantization scale \cite{kairouz2021distributed}, as discussed in Sec.~\ref{subsec:distributed-noise-sampling}.

\para{Skellam}
A Skellam random variable can be sampled as the difference of two Poisson distributed random variable sampled from the Poisson distribution:
{\small\begin{displaymath}
    {\mathsf{Poisson}\Bigl(x;\frac{\mu}{2}\Bigr) = \Bigl( \frac{\mu}{2}\Bigr)^{x} \frac{\exp(-x)} {x!}}.
\end{displaymath}}

The variance \enx{\mu} for the Skellam distribution, to achieve \enx{(\alpha, \eps)}-RDP, optimizes the following \cite{agarwal2021skellam}:
{\small
\begin{displaymath}
    \eps(\alpha) \le \frac{\alpha\Delta_2^2}{2\mu} + \min\Bigl( \frac{(2\alpha - 1) \Delta_2^2}{4s^2\mu^2} + \frac{3\Delta_1}{2s^3\mu^2}, \frac{3\Delta_1}{2 s\mu} \Bigr).
\end{displaymath}}

Here, \enx{s} is the scaling factor, and \enx{\Delta_1} and \enx{\Delta_2} are the \enx{l_1} and \enx{l_2} sensitivities. The variance \enx{\mu} needs to be rescaled according to the quantization scale \cite{agarwal2021skellam}, as discussed in Sec.~\ref{subsec:distributed-noise-sampling}.

\para{Poisson-Binomial}
All mechanisms in Tab.~\ref{tab:noisedistributions} add noise to the gradients except for the Poisson-Binomial mechanism for \partnoise \cite{chen2022poisson} which encodes local gradients into a parameter of the binomial distribution.
{\small
\begin{displaymath}
    \mathsf{PoiBin}(x;\mathbf{g},b,p) =  \mathsf{Bin}(x; b, \mathbf{g} \theta/K + 1/2).
\end{displaymath}}

Here, \enx{\theta \in [0,1/4]}, and \enx{b} is the output bit-width. The mechanism first applies a rescaling of the input gradient \enx{\mathbf{g}}, and then uses the result as the prob of binomial.
Since this mechanism is optimized for \enx{l_\infty} geometry, each client computes the Kashin representation of \enx{g_i} \cite{lyubarskii2010uncertainty}, \ie{transforms the geometry of the data from the \enx{l_2} to \enx{l_\infty}}.
The \slong reverts the Kashin representation to \enx{l_2} geometry after the aggregation.

\subsection{Noise Sampling Algorithms}\label{app:sampling_alg}
We formalize sampling algorithms in Alg.~\ref{alg:mpc_noise}. Specifically, we provide the pseudocode to sample in cleartext \BoxMuller, \LaplaceITS, \Skellam, {\DiscLaplace} and {\DiscGaussian} distributions with related subroutines, i.e., \Poisson, \textsf{Geom} and \textsf{Bern}. Before proceeding, we detail basic sampling concepts.

\para{Basic Sampling Algorithms}
%
\emph{Inverse transform sampling} (ITS) allows sampling random variables \enx{\Psi \sim f{(x)}} from a probability distributions \enx{f(x)} using  a uniformly distributed random variable \enx{u \sim U(0,1)} in \enx{(0,1)} \cite{law2007simulation}. ITS sample \enx{\Psi} via the inverse \emph{cumulative density function} of \enx{f(x)} \enx{\Psi = CDF^{-1}(u)}. 
For example, to sample from a Laplace distribution \xspace{\LaplaceITS} (Alg.~\ref{alg:mpc_noise}) computes \enx{-\frac{1}{\lambda}\; \mathsf{sign}(u-0.5) \log u}
with sign function $\mathsf{sign}(x)$. 
%
However, the Gaussian's \enx{CDF^{-1}} has no closed form and needs to be approximated, e.g., via Taylor polynomials \cite{sabater2022private}. Alternatively, the {\BoxMuller} \emph{transform} \cite{box1958note}
(Alg.~\ref{alg:mpc_noise}) receives two \enx{u_1,u_2 \sim U(0,1)} and outputs two samples from the standard Gaussian \enx{\mathcal{N}(0,1)}.

{
\setlength{\textfloatsep}{0.5em}
\setlength{\floatsep}{0.5em}
\setlength{\intextsep}{0.5em}
\renewcommand{\baselinestretch}{0.95}

\begin{algorithm*}[t!]
    \SetAlgoNlRelativeSize{-2}
    
    \SetAlgoNlRelativeSize{-2}
    
    \SetAlgoNlRelativeSize{-2}
    \SetInd{0.7em}{0.7em} 
    \def\dg{\ensuremath{s}}
    \def\mpsep{0.22\textwidth}
    \scriptsize
    \SetInd{0.5em}{0.5em} 
    \caption{
        Noise sampling algorithms (\sample, Alg.~\ref{alg:cmp}). DP to distribution parameters conversion in Tab.~\ref{tab:noisedistributions} and App.~\ref{app:noisedistributions}. 
    }\label{alg:mpc_noise}
    \DontPrintSemicolon
    \KwIn{
        Tuple of noise parameters $\tau$. For each mechanisms, the parameters are:
        $u,u_1,u_2\thicksim U(0,1)$,
        scale $\lambda$ 
        (Tab.~\ref{tab:noisedistributions}), 
        mean $\mu$ and scaling factor $\dg$. 
    }
    \SetKwFunction{BoxMullerFn}{\BoxMuller}
    \SetKwFunction{LapITS}{\LaplaceITS}
    \SetKwFunction{Poisson}{\textsf{Poisson}}
    \SetKwFunction{SkellamFn}{\Skellam}
    \SetKwFunction{Bernoulli}{\textsf{Bern}}
    \SetKwFunction{DiscreteLaplace}{\DiscLaplace}
    \SetKwFunction{Geometric}{\textsf{Geom}}
    \SetKwFunction{DiscreteGaussian}{\DiscGaussian}
    \SetKwFunction{DistributedNoiseSampling}{\mpcnoise}

    \SetKwFunction{NoiseType}{\textsf{NoiseType}}
    \SetKwFunction{Mechanism}{\textsf{Mechanism}}
    \SetKwFunction{params}{\textsf{params}}
    \SetKwFunction{parse}{\textsf{Parse}}

    \SetKwProg{Fn}{Mech}{:}{}
    \SetKwProg{Fna}{Func}{:}{}
    \SetKwProg{Helper}{Func}{:}{}
    \SetKwProg{Pr}{Protocol}{:}{}
    \SetKwRepeat{Do}{do}{while}

    \SetKwFunction{PartialNoise}{\textcolor{blue!60!black}{PartNoise}}
    \SetKwFunction{MPCNoise}{\textcolor{blue!60!black}{CentrNoise}}

    \SetKwFunction{Sample}{{\sample}}

    \Fna{\Sample{$\tau$}}{
        \scriptsize
        $\NoiseType, \Mechanism, \params \leftarrow \parse{$\tau$}$ \tcp{From $\tau$ parse the tuple of noise parameters, e.g., $\tau = (\partnoise, \DiscGaussian, (s, \sigma))$}
        \vspace{-0.05em}
        \Return{\inlineIfElse{\NoiseType $==$ \partnoise}{{\Mechanism}({\params})}{\glob{{\Mechanism}(\params)}}} \tcp{Call the sampling function, e.g., \DiscGaussian{($ s^2 \sigma^2$)}}
    }

        \begin{minipage}[t]{0.215\textwidth}
            \Fn{\BoxMullerFn{$u_1,u_2, \sigma$}}{
                $ {\mathcal{N}_{1}} \leftarrow \sqrt{-2 \log(u_1)} \cos(2 \pi u_2)$ 
                $ {\mathcal{N}_2} \leftarrow \sqrt{-2 \log(u_1)} \sin(2 \pi u_2)$ 
                \Return{${\sigma\mathcal{N}_1}, {\sigma\mathcal{N}_2}$}
            }
            \vspace{-0.4em}

            \vspace*{1ex}
            \Fn{\LapITS{$\lambda, u$}}{
                \Return{$- \frac{1}{\lambda}\,\mathsf{sign}(u-0.5)\log u$}
            }
        \end{minipage}
\hfill
        \begin{minipage}[t]{0.17\textwidth}
            \Fn{\DiscreteGaussian{$\dg^2 \sigma^2$}}{
                $b \leftarrow 2{\dg^2\sigma^2}$\;
                \Do{$\Bernoulli{$a, b$} == 1$}{
                    $l \leftarrow \DiscreteLaplace{$\dg \sigma$}$\;
                    $a \leftarrow (|l| - \dg\sigma)^2 $\;
                }
                \Return{$l$}
            }
        \end{minipage}
\hfill
        \begin{minipage}[t]{\mpsep}
            \Fn{\DiscreteLaplace{$\lambda$}}{
                $g_1, g_2 \thicksim \Geometric{$1-\exp(-\frac{1}{\lambda})$}$\;
                \Return{$g_1-g_2$}
            }

            \vspace*{1ex}
            \Fn{\SkellamFn{$s^2\mu$}}{
                $p_1, p_2 \thicksim \Poisson{$s^2 \mu/2$}$\;
                \Return{$p_1 - p_2$}
            }
        \end{minipage}
\hfill
        \begin{minipage}[t]{0.175\textwidth}
            \Helper{\Poisson{$\mu$}}{
                $prod \leftarrow 1, p \leftarrow -1$\;
                \While{$ prod > \exp(-1/\mu)$}{
                    $p \leftarrow p + 1$\;
                    $u \thicksim U(0,1)$ \;
                    $prod \leftarrow prod  \cdot u$\;
                    }
                \Return{p}
            }
        \end{minipage}
\hfill
    \begin{minipage}[t]{0.15\textwidth}
    \Helper{\Bernoulli{$a,b$}}{
        $u \thicksim U(0,1)$ \;
        \Return{$u < \exp(-\frac{a}{b})$}
    }
    \Helper{\Geometric{$p$}}{
        $u \thicksim U(0,1)$ \;
        \Return{$\lfloor \frac{\log{(1-u)}}{\log{(1-p)}} \rfloor$}
    }
    \end{minipage}
\end{algorithm*}
\renewcommand{\baselinestretch}{1.0}
\begin{figure*}
    \vspace{-2.5em}
\end{figure*}
}

\section{Noise Sampling Scenarios}
In Fig.~\ref{fig:dp_noise_generation}, we depict the noise sampling scenarios for \centrnoise and \partnoise from Alg.~\ref{alg:cmp}. The four noise scenarios are instantiations  of the three noise sampling techniques analyzed in Sec.~\ref{subsec:centralnoise}, \ref{subsec:partialnoise}, \ref{subsec:distributed-noise-sampling}.

\begin{figure}[t]
    \footnotesize

    \def\clientSize{1.05em}
	\def\tikzScale{0.8}
	\def\serverheight{2.7em}
    \def\serverwidth{2.7em}
	\def\ydistance{5.7em}
	\def\xdistance{5.8em}
	\def\minfontsize{1.5em}

    \tikzset{
        client/.style={
			fill=\clientfillcolor,thick,
			draw=\clientbordercolor,
			minimum width=\clientSize,
			minimum height=\clientSize,
			circle,
			font = \small
		},
		server/.style={
			draw,thick,rounded corners=0.5em,
			inner sep=0.5em,
			minimum width=\serverwidth,
			minimum height=\serverheight,
			font = \small
		},
		single/.style={
			fill=\mylightergreen,
			draw=\mydarkergreen
		},
		agg/.style={
			fill=\mylighterred,
			draw=\mydarkerred
		},
		mpc/.style={
			fill=\mylighterblue,
			draw=\mydarkerblue
		},
		arrow/.style={->,
			>=latex,
			shorten >=1.5pt,
			shorten <=3pt,
			\arrowColor,
			text=black
		},
		doublearrow/.style={<->,
			>=latex,
			shorten >=1.5pt,
			shorten <=3pt,
			\arrowColor,
			text=black
		},
		optarrow/.style={->,
			dashed,
			>=latex,
			shorten >=1.5pt,
			shorten <=3pt,
			\arrowColor,
			text=black
		},
		doubleoptarrow/.style={<->,
			dashed,
			>=latex,
			shorten >=1.5pt,
			shorten <=3pt,
			\arrowColor,
			text=black
		},
		selfarrow/.style={->,
		out=-60,
		in=-120,
		looseness=4,
		dotted,
		>=latex,very thick,
		orange,
		},
		selfsidearrow/.style={->,
		out=20,
		in=-10,
		looseness=4,
		dotted,
		>=latex,very thick,
		orange,
		},
		selfsidearrowleft/.style={->,
		out=160, 
    	in=200, 
		looseness=4,
		dotted,
		>=latex,very thick,
		orange,
		},
	}

	\begin{minipage}[b]{0.47\columnwidth}
		\centering
		\subfloat[Client-side \partnoise in \fl]{
			\begin{tikzpicture}[scale=\tikzScale, every node/.style={transform shape}]

				\node[client] (ci) [] {\c{i}};
				\node[client] (c1) [left of=ci, node distance=\xdistance] {\c{1}};
				\node[client] (cn) [right of=ci, node distance=\xdistance] {\c{n}};


				\draw[selfsidearrow,looseness=4]
					(cn.45) to  node[midway, left] {\textcolor{black}{$\npartial{n}$}} (cn.315);
				\draw[selfsidearrow,looseness=4]
					(c1.45) to  node[midway, left] {\textcolor{black}{$\npartial{1}$}}(c1.315);
				\draw[selfsidearrow,looseness=4]
					(ci.45) to node[midway, left] {\textcolor{black}{$\npartial{i}$}} (ci.315);

			\label{fig:pnoisefl}
			\end{tikzpicture}
		}
	\end{minipage}
	\hspace*{\fill}
	\begin{minipage}[b]{0.47\columnwidth}
		\centering
		\subfloat[\underline{Centralized} \centrnoise] 
		{
			\begin{tikzpicture}[scale=\tikzScale, every node/.style={transform shape}]


				\node[server,agg] (s2) [] {\s{}'};
				\node[server, mpc, white] (s1) [right of=s2, node distance=\xdistance] {\s{1}};
				\node[server, mpc, white] (s3) [left of=s2, node distance=\xdistance] {\s{2}};

				\draw[selfsidearrow,looseness=3]
					(s2.35) to node[midway, left] {\textcolor{black}{$\ncentral$}} (s2.325);
				\label{fig:cnoisecentr}
			\end{tikzpicture}
		}
	\end{minipage}
	\begin{minipage}[b]{0.47\columnwidth}
		\centering
		\subfloat[Server-side \partnoise]{

			\begin{tikzpicture}[scale=\tikzScale, every node/.style={transform shape}]

				\node[server,mpc] (s1) at (0,0) {\s{1}};
				\node[server,mpc] (s2) [below right of=s1, node distance=\xdistance] {\s{m}};
				\node[server,mpc] (s3) [below left of=s1, node distance=\xdistance] {\s{i}};


                \draw[selfsidearrow,looseness=3]
					(s2.35) to node[midway, left, xshift=0.24em] {\textcolor{black}{$\npartial{m}$}} (s2.325);
				\draw[selfsidearrow,looseness=3]
					(s1.35) to node[midway, left] {\textcolor{black}{$\npartial{1}$}} (s1.325);
				\draw[selfsidearrowleft, looseness=3]
					(s3.145) to node[midway, right] {\textcolor{black}{$\npartial{i}$}} (s3.215);


				\label{fig:pnoiseot}
			\end{tikzpicture}
		}
	\end{minipage}
	\hspace*{\fill}
    \begin{minipage}[b]{0.47\columnwidth}
		\centering
		\subfloat[\glob{Distributed} \centrnoise]{
			\begin{tikzpicture}[scale=\tikzScale, every node/.style={transform shape}]
				\node[server,mpc] (s1) at (0,0) {\s{1}};
				\node[server,mpc,white] (sx) [below right of=s1, node distance=\xdistance,xshift=3.5em] {\s{m}};
				\node[server,mpc,white] (sy) [below left of=s1, node distance=\xdistance,xshift=-3.5em] {\s{m}};
				\node[server,mpc] (sm) [below right of=s1, node distance=\xdistance] {\s{m}};
				\node[server,mpc] (si) [below left of=s1, node distance=\xdistance] {\s{i}};
				\draw[doublearrow, dotted, orange,
					very thick]
					(s1.275)
					to [bend left=30]
					(sm.165);

				\draw[	doublearrow, dotted, orange,
						very thick ]
					(s1.275)
					to [bend right=30]
					(si.15);
				\draw[	doublearrow, dotted, orange,
						very thick]
					(sm.195)
					to [bend left=30]
					(si.345);

				\node at (0,-\xdistance/1.7) {$MPC(\ncentral)$};

				\label{fig:cnoisempc}
			\end{tikzpicture}
		}

	\end{minipage}

	\caption{
		Noise sampling scenarios from Alg.~\ref{alg:cmp}.
		Dotted arrows (\protect\tikz[baseline=-0.25em]{\protect\draw[optarrow,dotted,orange,very thick](0,0)--(2.0em,0);}) indicate noise sampling. 
	}
	\label{fig:dp_noise_generation}
	\vspace{-2.0em}
\end{figure}

\section{Details on Evaluation Setup}\label{app:evalsetup}
For all the evaluations (Sec.~\ref{subsec:distributed-noise-sampling}) we used AWS t2.xlarge instances with 4 vCPUs (based on Intel Xeon) and 16GB RAM. In the LAN setting, with the two machines in Frankfurt (Germany), we have about $1$ Gbps bandwidth and $27.6$ ms latency. In the WAN setting, with one machine in Frankfurt (Germany) and the other in California (USA), we have about $74$ Mbps bandwidth and $287.7$ ms latency.

\para{MPC overhead (Sec.~\ref{subsec:distributed-noise-sampling})}
For the evaluation of distributed \centrnoise sampling with MP-SPDZ \cite{spdz} in Sec.~\ref{subsec:distributed-noise-sampling}, we used $32$-bit fixed-point protocols for continuous distributions, \ie{{\LaplaceITS} and Gaussian via {\BoxMuller}}, and $16$ bit integer protocols for discrete ones, \ie{{\DiscGaussian} and {\Skellam}}, since according to the evaluation of Agarwal et al.~\cite{agarwal2021skellam} 16 bits are enough to match the accuracy of continuous Gaussian.
We used standard MP-SPDZ parameters, \ie{$40$ bit statistical security and $128$ bit security parameter.}

\para{Accuracy Evaluation (Tab.~\ref{tab:accuracy}, Sec.~\ref{sec:noisetradeoff})}
We used the 3-layers neural network from \cite{das2025communication,abadi2016deep} with \enx{768-100-10} neurons for a total of 79510 parameters. We used ReLU activation and cross-entropy loss with DP-SGD optimizer with no momentum.
We evaluate the accuracy on two common benchmarking datasets for image classification: MNIST \cite{lecun1998gradient} and Fashion MNIST \cite{xiao2017fashion}, which contain grayscale images of handwritten digits and clothing, respectively. Both datasets  have $60$k training data images of size \enx{28\times28} with 10 classes.
We rely on Poisson subsampling and the privacy loss accounting tool from the PFL library \cite{granqvist2024pfl}.
For \ot, we select \enx{B=500}, \enx{K=4.0} and \enx{\eta=0.1} according to  Abadi et al.~\cite{abadi2016deep}. For \fl, we subsample \enx{n=100} clients from \enx{1000} clients for each round. From our hyperparameters search in App.~\ref{app:hyperparameter_fl}, we found that the best parameters for \fl are \enx{K \in \{0.3,0.5\}}, \enx{\eta=0.1} and \enx{B=60}.


%

{
    }
%
%
%
%
%
%

%

\section{Evaluation of Basic MPC Protocols}\label{app:basic_eval}
{Table~\ref{tab:eval_basics_wan} reports our evaluation for basic MPC protocols with the MP-SPDZ framework \cite{spdz}
used in Alg.~\ref{alg:mpc_noise} (App.~\ref{app:sampling_alg}) with 32-bit fixed-point protocols.
We evaluated in the same setup of Sec.~\ref{subsec:distributed-noise-sampling} (detailed in App.~\ref{app:evalsetup}), in both LAN and WAN settings, with 2-party maliciously secure Mascot\cite{keller2016mascot}.
The offline phase requires \enx{10^3\times} more communication than the online one, as well as \enx{10^3\times} the running time for LAN. The square root is the most expensive function to compute since it requires the highest amount of communication for online and offline phases and has the longest runtime. The trigonometric functions, \ie{\enx{\sin} and \enx{\cos}}, have the lowest runtime in LAN, which is half of the square root one. The exponentiation base two, \ie{\enx{2^{x}}}, requires less communication than the other functions for the online phase, which motivates why for LAN it has the same runtime of \enx{\log_2}, whereas for the WAN setting it has the lowest online runtime, \ie{\enx{1/3} of \enx{\log_2} one}.
We evaluate \enx{\log_2} and \enx{2^{(\cdot)}}
as MP-SPDZ
uses them as building blocks to
compute 
\enx{\log_b(x) = \log_b(2) \cdot \log_2(x)} for a base $b$,
and \enx{x^y = 2^{y \log_2(x)}}.

\begin{table}[tb]
    \scriptsize
    \def\seps{/}
    \def\colw{0.16\columnwidth}
    \def\colwfirst{0.12\columnwidth}
    \setlength{\tabcolsep}{0.2em}
	\renewcommand{\arraystretch}{1}
	\centering
    \caption{
        Cost of online and offline phase for basic MPC protocols
        as running time (in ms/s for LAN/WAN, resp.)
        and communication per party (same unit for LAN/WAN).
	}
   \label{tab:eval_basics_wan}
   \vspace{-0.5em}

  	\begin{tabular}{p{\colwfirst}C{\colw}C{\colw}C{\colw}C{\colw}C{\colw}}
		\textbf{Online} &
        \enx{\log_2(\cdot)} &
        \enx{2^{(\cdot)}} &
        \enx{\sin(\cdot)} &
        \enx{\cos(\cdot)} &
        \enx{\sqrt{\cdot}}
        \\
        \midrule
        \textbf{LAN/ms}&
        $278.6 \pm 5.7$ &
		$278.3 \pm 5.6$&
        $154.6 \pm 6.5$ &
		$154.5 \pm 5.4$ &
        $351.7 \pm 11.3$\\
        \textbf{WAN/s} &
        $14.14 \pm 0.03 $ &
		 $5.89 \pm 0.01 $ &
         $8.13 \pm 0.01 $ &
		 $8.11 \pm 0.02 $ &
        $19.56 \pm 0.04 $ \\
        \textbf{KB} &
        $20.9$  &
		$5.6 $ &
        $6.3 $ &
		$6.3 $ &
        $21.2 $ \\
   	\end{tabular}%

    \vspace*{0.7em}
  	\begin{tabular}{p{\colwfirst}C{\colw}C{\colw}C{\colw}C{\colw}C{\colw}}
		\textbf{Offline} &
        \enx{\log_2(\cdot)} &
        \enx{2^{(\cdot)}} &
        \enx{\sin(\cdot)} &
        \enx{\cos(\cdot)} &
        \enx{\sqrt{\cdot}}
        \\
        \midrule
        \textbf{LAN/s}&
        $ 2.80 \pm 0.01 $ &
        $ 3.00 \pm 0.03 $ &
        $ 1.49 \pm 0.01 $ &
		$ 1.49 \pm 0.01 $ &
        $ 3.33 \pm 0.02$ \\
        \textbf{WAN/s} &
        $  16.50 \pm 0.76 $ &
		$  17.43 \pm 0.65 $ &
        $  10.90 \pm 0.48$ &
		$  10.59 \pm 0.26$ &
        $  18.93 \pm 0.04 $ \\
        \textbf{MB} &
        $ 54.54 $ &
		$ 55.86 $ &
        $ 27.43 $ &
		$ 27.43 $ &
        $ 64.50$ \\
   	\end{tabular}%

   \vspace{-2.0em}
\end{table}

%


\section{Evaluation Details for MPC Noise Sampling}\label{app:mpc_sampling}
Next, we provide more details on the evaluation of MPC noise sampling algorithms (Sec.~\ref{subsec:distributed-noise-sampling}) including how we set the loop count for sampling {\DiscGaussian} and \Poisson with minimal failure probability, the online and offline phases in LAN and WAN settings, and the communication overhead.
\subsection{Sampling Threshold}\label{app:samplingthreshold}
To ensure MPC constant runtime, we replace all \textbf{while} loops of Alg.~\ref{alg:mpc_noise} (App.~\ref{app:sampling_alg}) with \textbf{for} loops with a fixed number of iterations. To find a loop count satisfy an empirical failure probability of at most \enx{10^{-6}},
we iteratively increased the counts,
until sampling succeeded without any failure for $10^6$ runs.
Given such a count candidate,
we repeated the sampling (with $10^6$ runs) 100 times and computed the average number of failures with 95\% confidence interval.
For {\DiscGaussian} we fixed the number of iterations to \enx{10}, obtaining on average \enx{0.78 \pm 0.21} failures, whereas {\Poisson} with \enx{\mu=10}  requires at least 29 iterations with \enx{0.74 \pm 0.20} failures on average, and we set it to \enx{30}.




\subsection{Online \& Offline Phases in LAN/WAN}\label{app:online}

Recall, MPC has a data-independent offline phase, to pre-compute material for the online phase, and a data-dependent online phase. Next, we report the online and offline runtimes for the noise sampling techniques implemented in MP-SPDZ from Alg.~\ref{alg:mpc_noise} (App.~\ref{app:sampling_alg}).
\begin{table}[t]
    \scriptsize
    \def\seps{/}
    \def\colwfirst{0.24\columnwidth}
    \def\colw{0.17\columnwidth}
    \def\colwalt{0.19\columnwidth}
	\renewcommand{\arraystretch}{1}
	\centering
    \caption{MPC online runtime for LAN/WAN in ms/seconds.}
    \label{tab:eval_online}
    \vspace{-0.5em}
    {
  	\begin{tabular}{p{\colwfirst}C{\colw}C{\colw}C{\colwalt}}
        \textbf{LAN, ms} &
        {\LaplaceITS} &
        {\BoxMuller} &
        {\DiscGaussian} 
        \\ 
        \midrule
    Shamir
    & $ 68.9 \pm 9.7 $
    & $ 186.8 \pm 20.3 $
    & $ 2,866 \pm 128 $ 
        \\
    Malicious Shamir
	& $ 73.4 \pm 14.5 $
    & $ 167.3 \pm 24.4 $
	& $ 2,422 \pm 132 $
        \\
    Mascot
	& $ 143.0 \pm 7.29 $
    & $ 425.4 \pm 13.4 $
	& $ 5,882 \pm 55$ 
        \\ 
   	\end{tabular}%

   \vspace*{.7em}
  	\begin{tabular}{p{\colwfirst}C{\colw}C{\colw}C{\colwalt}}
        \textbf{WAN, s}&
        {\LaplaceITS} &
        {\BoxMuller} &
        {\DiscGaussian} 
        \\ 
        \midrule
    Shamir
        & $ 5.76 \pm 0.02$ 
        & $ 17.90 \pm 0.05$ 
        & $ 253.76 \pm 0.32$ 
        \\
    Malicious Shamir
		& $ 5.39 \pm 0.04$ 
        & $ 16.04 \pm 0.07 $ 
		& $ 219.59 \pm 0.21 $ 
        \\
    Mascot
		& $ 17.19 \pm 0.02$ 
        & $53.23 \pm 0.09$ 
		& $ 703.58 \pm 1.92$ 
        \\ 
   	\end{tabular}%
    }
   	
\end{table}

\para{Online Phase}
Tab.~\ref{tab:eval_online} reports the online runtimes in LAN and WAN settings for the noise sampling techniques implemented in MP-SPDZ from Alg.~\ref{alg:mpc_noise} (App.~\ref{app:sampling_alg}). Semi-honest and malicious Shamir have comparable runtimes, while Mascot is at least \enx{2\times} slower (for \DiscGaussian~in LAN). The continuous sampling techniques are at least \enx{10\times} faster than the discrete ones. The LAN runtime for all protocols is \enx{10^2\times} smaller than the WAN due to lower communication latency in a LAN setting.
We omit {\Skellam} in the table, as its online runtime is disproportionate, \ie{about \enx{20} minutes in a LAN}.

\para{Offline Phase}
Tab.~\ref{tab:eval_offline} reports the offline runtimes per party in LAN and WAN settings for the noise sampling technique implemented in MP-SPDZ from Alg.~\ref{alg:mpc_noise} (App.~\ref{app:sampling_alg}).
The offline runtime of the continuous sampling techniques is at least \enx{10\times} faster than the discrete ones, as they are not based on iterative methods. 
The LAN runtime for all protocols is \enx{10^2\times} smaller than the WAN due to lower communication latency in a LAN. 

We omit {\Skellam} in the table, due to its comparatively much larger overhead.
In the LAN setting, {\Skellam} has the slowest offline phase, taking on average, \enx{43.5} seconds for Shamir, \enx{3.5} minutes for Malicious Shamir (Mal.\,Sh.) and about \enx{9} hours for Mascot, due to the high number of iterations required in our implementation.
\begin{table}[t]
    \scriptsize
    \def\colwfirst{0.24\columnwidth}
    \def\colw{0.17\columnwidth}
    \def\colwalt{0.19\columnwidth}
	\renewcommand{\arraystretch}{1}
	\centering
    \caption{MPC offline runtimes for LAN/WAN in ms/seconds per party}
   \label{tab:eval_offline}
   \vspace{-0.5em}
    {
  	\begin{tabular}{p{\colwfirst}C{\colw}C{\colw}C{\colwalt}}
        \textbf{LAN, ms} &
        {\LaplaceITS} &
        {\BoxMuller} &
        {\DiscGaussian } 
        \\
        \midrule
    Shamir
    & $ 4.02 \pm 1.60 $
    & $ 9.94 \pm 2.52 $
    & $ 147.07\pm 8.68$ 
        \\
    Mal. Shamir
	& $ 42.02 \pm 7.43 $
    & $ 52.80 \pm 4.79 $
	& $ 627.1 \pm 21.88 $ 
        \\
    Mascot
	& $ 3335  \pm 26  $
    & $ 10142 \pm 35  $
	& $ 176,427 \pm 210  $ 
        \\
   	\end{tabular}%

   \vspace*{.7em}
  	\begin{tabular}{p{\colwfirst}C{\colw}C{\colw}C{\colwalt}}
        \textbf{WAN, s}&
        {\LaplaceITS} &
        {\BoxMuller} &
        {\DiscGaussian} 
        \\
        \midrule
    Shamir
    & $ 0.45 \pm 0.001$
    & $ 0.85 \pm 0.05$
    & $ 7.88 \pm 0.18 $
        \\
    Mal. Shamir
	& $ 1.97 \pm 0.03$
    & $ 2.53 \pm 0.07 $
	& $ 12.99 \pm 0.14$
        \\
    Mascot
	& $ 17.72 \pm 0.40$
    & $53.97 \pm 1.56$
	& $ 934.53 \pm 7.55$
        \\
   	\end{tabular}%
    }
   \vspace{-2.0em}
\end{table}

\subsection{Communication}\label{app:communication}
    Tab.~\ref{tab:comm} shows the per-party communication for the online and offline phase for the noise sampling techniques implemented in MP-SPDZ from Alg.~\ref{alg:mpc_noise} (App.~\ref{app:sampling_alg}).
    The discrete sampling techniques require at least \enx{10\times} more communication than continuous ones.
    The technique that requires more communication for both online and offline phases is {\Skellam}, i.e., at least \enx{10\times} more than{ \DiscGaussian} and \enx{10^3\times} more than the continuous techniques. {\LaplaceITS} is the most communication efficient technique.

\begin{table}[t]
    \scriptsize
    \def\seps{/}
    \def\colwfirst{0.19\columnwidth}
    \def\colw{0.145\columnwidth}
	\renewcommand{\arraystretch}{1}
	\centering
    \caption{MPC communication per party for online/offline phase in KB/MB, resp.}
    \label{tab:comm}
    \vspace{-0.5em}
    {
  	\begin{tabular}{p{\colwfirst}C{\colw}C{\colw}C{\colw}C{\colw}}
        \textbf{Online (KB)} &
        {\LaplaceITS} &
        {\BoxMuller} &
        {\DiscGaussian } &
        {\Skellam }
        \\
        \midrule
    Shamir
    & $ 11.01$ 
    & $ 25.74$ 
    & $ 411.4 $ 
	& $ 93,112 $ 
        \\
    Mal.\,Shamir
	& $ 41.31$
    & $ 93.54 $
	& $ 1,494 $
	& $ 346,099$ 
        \\
    Mascot
	& $ 23.20 $
    & $ 55.44 $
	& $  862.72 $
	& $  183,515 $
        \\
   	\end{tabular}%

   \vspace*{.7em}
  	\begin{tabular}{p{\colwfirst}C{\colw}C{\colw}C{\colw}C{\colw}}
        \textbf{Offline (MB)}&
        {\LaplaceITS} &
        {\BoxMuller} &
        {\DiscGaussian} &
        {\Skellam}
        \\
        \midrule
    Shamir
    & $ 0.08$ 
    & $ 0.26$ 
    & $ 1.49$ 
	& $ 642.08$ 
        \\
    Mal.\,Shamir
	& $ 0.50$ 
    & $ 1.20 $ 
	& $ 19.92$ 
	& $ 3,395 $ 
        \\
    Mascot
	& $ 60.74$
    & $ 184.75$
	& $ 3,219 $
	& $ 32,616 $ 
        \\
   	\end{tabular}%
    }
   	
   \vspace{-2.0em}
\end{table}

\section{Accuracy Evaluation}
In this section, we report the results of the hyperparameter search for \fl experiments on the MNIST dataset and the impact of different hyperparameters for \ol with a CNN model on the EMNIST dataset. Finally, we evaluate how the different collusion thresholds affect the accuracy of \partnoise on the EMNIST dataset.

\subsection{FL Hyperparameter Search}\label{app:hyperparameter_fl}
Next, we report the results from our hyperparameter search for \fl training on the MNIST dataset. Compared to plain training (i.e., no DP or cryptography), \fl introduces three additional hyperparameters: the number of local iterations per client, the local clipping threshold, and the number of clients per iteration.
First, we describe the evaluation setup and then analyze the relationship between learning rate and clipping threshold. Finally, we discuss how accuracy varies with the number of local iterations.

\para{Evaluation Setup}
We trained the 3-layers neural network from \cite{das2025communication,abadi2016deep} and used the DP-SGD optimizer described in App.~\ref{app:evalsetup} on the MNIST dataset. Following systematized works \cite{agarwal2021skellam,bao2022skellam,kairouz2021distributed}, we leverage the federated averaging algorithm (FedAvg) \cite{mcmahan2017communication} which allows clients to perform multiple local iterations and send only the delta over their local model parameters to the server. We split the MNIST dataset among 1000 clients, resulting in \enx{60} samples for each client. We subsample \enx{n=100} clients per \fl iteration and vary the number of local iterations in \enx{\{1,2,5\}}. We run the training for \enx{100} epochs, with clipping threshold \enx{K \in \{4.0, 1.0, 0.5, 0.3, 0.1, 0.03, 0.01\}} and learning rate \enx{\eta \in \{0.1, 0.05, 0.01, 0.005, 0.001\}}. In this analysis, we focus on \enx{\eps=1.0} with \centrnoise, i.e., no additional noise variance introduced to mitigate collusion as with \partnoise. 

\para{Impact of Clipping Threshold and Learning Rate}
Fig.~\ref{fig:confusion_matrix_k_lr} shows how accuracy varies across different combinations of \enx{\eta} and \enx{K} while fixing the number of local iterations to 5.
The results highlight that the learning rate significantly impacts the model accuracy, with a \enx{\eta= 0.1} achieving the best accuracy over all clipping thresholds. A too small \enx{\eta=0.001} leads to accuracy drops of over $70$\pp compared to \enx{\eta=0.1}, with accuracy close to random guessing across all clipping thresholds.
The clipping threshold reduces the impact of noise on the gradient updates since the DP noise variance scales with \clipparam, i.e., \enx{\sigma_{{\text{DP}}} = \sigma\clipparam}. Too large {\enx{K=4.0}} or too small \enx{K=0.01} lead to poor performance since either the noise variance is too high or the gradient update is too small to ensure convergence. Both show $65$\pp lower accuracy than the best clipping threshold for both \enx{K=4.0} and \enx{K=0.01}.
\enx{K \in \{0.5, 0.3\}} and \enx{\eta=0.1} achieve the best result, with less than $2$\pp accuracy difference between the two \clipparam.

\para{Impact of Local Iterations and Clipping Threshold}
Fig.~\ref{fig:confusion_matrix_k_local_iterations} reports the accuracy for different numbers of local iterations and clipping thresholds while fixing \enx{\eta=0.1}, which is the best-performing learning rate according to Fig.~\ref{fig:confusion_matrix_k_lr}.
By varying the number of local iterations, the best performing clipping thresholds remain \enx{\clipparam \in \{0.5,0.3\}}, since they balance noise variance and gradient updates to ensure convergence.
Increasing the number of local iterations improves accuracy and reduces \fl communication bottleneck due to aggregation, as analyzed in Sec.~\ref{sec:noisetradeoff}. More local iterations, as in FedAvg, allow for more refined local updates and reduce the impact of noise since noise is added over multiple iterations.
We achieve the best accuracy with 5 local iterations, i.e., \enx{87.09\%} accuracy, which is $10$\pp higher than 2 local iterations and $27$\pp higher for 1 local iteration.
Higher accuracy with fewer local iterations requires more \fl iterations, increasing communication overhead. 
For example, with 1 local iteration, accuracy improves to \enx{80.82\%} after 1000 \fl iterations, still $6$\pp lower than using 5 local iterations with only 100 \fl iterations. Similarly, with 2 local iterations, accuracy reaches \enx{83.99\%} after 500 \fl iterations, $3$\pp below the best setting.



\begin{figure}[t]
    \centering
    \def\tikzScale{0.7}
    \begin{tikzpicture}[scale=\tikzScale, every node/.style={transform shape}]
        \begin{axis}[
            colorbar,
            colormap/cool,
            point meta min=0,    
            point meta max=100,
            xlabel={Clipping threshold $K$},
            ylabel={Learning rate $\eta$},
            xtick={0,1,2,3,4,5,6},  
            ytick={0,1,2,3,4},      
            xticklabels={4.0, 1.0, 0.5, 0.3, 0.1, 0.03, 0.01}, 
            yticklabels={0.1, 0.05, 0.01, 0.005, 0.001},       
            enlargelimits=false,
            axis on top,
            nodes near coords,
            point meta=explicit,
            legend style={draw=none}
        ]
        \addplot[
            matrix plot*,
            mesh/cols=7,  
            mesh/rows=5
        ] table [meta=TestScore] {
            x y TestScore
            0 0 20.42
            1 0 73.55
            2 0 85.50
            3 0 87.09
            4 0 78.11
            5 0 36.86
            6 0 20.79
            0 1 14.55
            1 1 65.81
            2 1 80.55
            3 1 79.63
            4 1 68.34
            5 1 33.07
            6 1 20.46
            0 2 14.16
            1 2 32.73
            2 2 50.30
            3 2 46.25
            4 2 20.34
            5 2 18.73
            6 2 9.89
            0 3 14.12
            1 3 21.23
            2 3 39.50
            3 3 31.68
            4 3 17.42
            5 3 11.53
            6 3 9.80
            0 4 14.16
            1 4 16.49
            2 4 15.95
            3 4 13.77
            4 4 13.25
            5 4 9.82
            6 4 9.80
        };
        \end{axis}
    \end{tikzpicture}
    \caption{\fl accuracy on MNIST for different $K$ and $\eta$ fixing the number of local iterations to 5, epochs to 100, and \enx{\eps=1.0}.}
    \label{fig:confusion_matrix_k_lr}
    \vspace{-1.6em}
\end{figure}

\begin{figure}
    \centering
    \def\tikzScale{0.7}
    \begin{tikzpicture}[scale=\tikzScale, every node/.style={transform shape}]
        \begin{axis}[
            colorbar,
            colormap/cool,
            point meta min=0,    
            point meta max=100,
            xlabel={Clipping threshold $K$},
            ylabel={Local iterations},
            xtick={0,1,2,3,4,5,6},  
            ytick={0,1,2},      
            xticklabels={4.0, 1.0, 0.5, 0.3, 0.1, 0.03, 0.01}, 
            yticklabels={1, 2, 5},       
            enlargelimits=false,
            axis on top,
            nodes near coords,
            point meta=explicit,
            legend style={draw=none}
        ]
        \addplot[
            matrix plot*,
            mesh/cols=7,  
            mesh/rows=3
        ] table [meta=TestScore] {
            x y TestScore
            0 0 14.53
            1 0 47.39
            2 0 58.76
            3 0 60.64
            4 0 41.17
            5 0 21.17
            6 0 13.98
            0 1 14.69
            1 1 60.74
            2 1 77.42
            3 1 78.37
            4 1 61.93
            5 1 25.19
            6 1 20.16
            0 2 19.65
            1 2 20.42
            2 2 85.50
            3 2 87.09
            4 2 78.11
            5 2 36.86
            6 2 20.79
        };
        \end{axis}
    \end{tikzpicture}
    \caption{\fl accuracy on MNIST for different $K$ and local iterations fixing \enx{\eta=0.1}, epochs to 100, and \enx{\eps=1.0}.}
    \label{fig:confusion_matrix_k_local_iterations}
    \vspace{-2.0em}
\end{figure}


\subsection{\ol Evaluations with CNN}\label{app:hyperparameter}
Next, we evaluate how the clipping threshold and the learning rate affect the accuracy of simulated \ol training. First, we introduce the evaluation setup and then discuss~results.

\para{Evaluation Setup}
We conduct the hyperparameter search in this section by simulating \ol training with DP-SGD via PyTorch.
We use the CNN model for image classification from LEAF with 6603710 parameters, as it is a standard benchmark for distributed settings \cite{caldas2018leaf}.
We conduct our evaluation on the EMNIST dataset \cite{cohen2017emnist}, as \cite{agarwal2021skellam,kairouz2021distributed}, to classify handwritten characters, \ie{62 unbalanced classes}.
The EMNIST dataset contains $N=814255$ samples with different handwriting styles, which can simulate collaborative settings, \ie{about 3500 users 
\cite{caldas2018leaf}}.
We applied the same transformation as LEAF \cite{caldas2018leaf} and Agarwal et al.~\cite{agarwal2021skellam} to the EMNIST dataset, \ie{rescaling the pixel values from (0,255) to (0,1) and setting 1 for the background and 0 for black pixels}. 
We run the training over 300 epochs with a batch size of \enx{B=836} according to \cite{abadi2016deep} {(B=\enx{\sqrt{N}})}. We used Gaussian noise with Poisson subsampling. 
We vary the learning rate \enx{\eta \in \{0.15, 0.1, 0.05, 0.01, 0.001\}} and the clipping threshold \enx{K \in \{10, 5, 3, 1\}}.

\para{Impact of Clipping Threshold and Learning Rate} 
Fig.~\ref{fig:confusion_matrix_ol} reports the accuracy on the EMNIST dataset for different clipping thresholds and learning rates. 
We achieved the best accuracy with a learning rate of \enx{0.15} and a clipping threshold of \enx{K=3}.
The results show that by fixing $\eta$, the \enx{K} has a significant impact on the model accuracy, since a too large clipping threshold, (\enx{K=10}), leads to a model with poor performance, \ie{accuracy of \enx{5\%}} since the noise standard deviation is too high, covering the gradient signal. On the other hand, when the clipping threshold is smaller (\enx{K=1}), the model performance starts to decrease, \ie{\enx{2\%} lower compared to \enx{K=3}}, since we clip too much of the gradient signal.

By fixing the number of training epochs, the learning rate has a significant impact since a learning rate that is too small leads to poor performance. For example, with a learning rate of \enx{0.001}, the model accuracy is \enx{5.81\%}, which is \enx{71.34\%} lower compared to the best learning rate of \enx{0.15}, with \enx{K=3}.
Notably, decreasing the learning rate for \enx{K=10} does not achieve the expected results. Since it achieves the best accuracy (\enx{76.39 \%}) with \enx{\eta=0.01}, which is more than \enx{70 \%} higher than the accuracy with \enx{\eta=0.15}. Then for \enx{\eta=0.001} the accuracy drops to \enx{46.59 \%}.
This is probably because a smaller learning rate compensates for the large noise variance.

\begin{figure}
    \centering
    \def\tikzScale{0.7}
    \begin{tikzpicture}[scale=\tikzScale, every node/.style={transform shape}]
        \begin{axis}[
            colorbar,
            colormap/cool,
            point meta min=0,    
            point meta max=100,
            xlabel={Clipping threshold $K$},
            ylabel={Learning rate $\eta$},
            xtick={0,1,2},  
            ytick={0,1,2,3},      
            yticklabels={0.15, 0.1, 0.01, 0.001}, 
            xticklabels={1.0, 3.0, 10.0},       
            enlargelimits=false,
            axis on top,
            nodes near coords,
            point meta=explicit,
            legend style={draw=none}
        ]
        \addplot[
            matrix plot*,
            mesh/cols=3,  
            mesh/rows=4
        ] table [meta=TestScore] {
            x y TestScore
            0 0 79.15
            1 0 81.56
            2 0 5.37
            0 1 75.88
            1 1 81.21
            2 1 5.40
            0 2 47.33
            1 2 60.39
            2 2 76.39
            0 3 5.81
            1 3 25.65
            2 3 46.59
        };
        \end{axis}
    \end{tikzpicture}
    \caption{Accuracy on EMNIST for different learning rates and clip thresholds for $\eps=1.0$ over 300 epochs}\label{fig:confusion_matrix_ol}
    \vspace{-1.0em}
\end{figure}
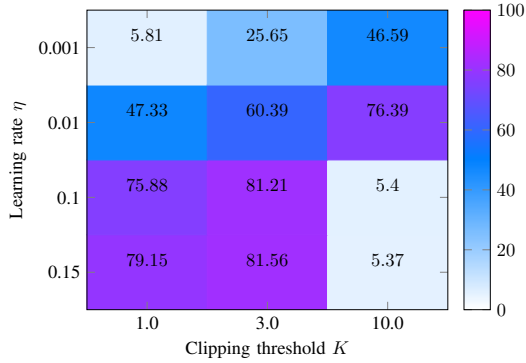
\subsection{Collusion Effect on \boldmath{\partnoise}}\label{app:collusion}
Next, we evaluate the impact of collusion on \partnoise training by varying the collusion threshold \enx{t} and analyzing its effect on accuracy and noise variance. First, we describe our evaluation setup and then present our results.

\para{Evaluation Setup}
We trained the same CNN as in App.~\ref{app:hyperparameter} on the EMNIST dataset using DP-SGD. We vary the collusion threshold $t \in \{0,m/4,m/3,m/2,2m/3, 3m/4, m-1\}$. We set \enx{\eps=1.0} and use the best hyperparameters from App.~\ref{app:hyperparameter}. We simulate a \partnoise setting with \enx{m=100} using PyTorch \cite{paszke2019pytorch}. Fig.~\ref{fig:tradeoff-collusion} reports the average accuracy over 10 runs by taking the maximum for each run over the last five epochs to account for randomness in noise sampling. 

\para{Accuracy Impact}
Fig.~\ref{fig:tradeoff-collusion} shows the impact of varying \enx{t} on accuracy and noise standard deviation \enx{\sigma} according to Eq.~\eqref{eq:partgaussvariance}. 
Interestingly, the accuracy remains stable until \enx{t = 9m/10}, despite a \enx{3 \times} increase in \enx{\sigma}. Although \enx{\sigma} is greater than \enx{K}, meaning the noise has a higher probability of overshadowing gradient updates, the results suggest that gradient directions remain largely unaffected.
This may explain why none of the systematized works implement oblivious \partnoise protocols with \perturbgrad. 
However, when \enx{t=m-1}, accuracy drops by approximately \enx{10\%} compared to \enx{t=0}.
%
%

\begin{figure}[t]
    \centering
    \normalsize
    \def\tikzScale{0.7}
    \begin{tikzpicture}[scale=\tikzScale, every node/.style={transform shape}]
    \begin{axis}[
        xlabel={$t$},
        ylabel={Accuracy},
        xmin=0, xmax=1,
        ymin=70, ymax=85,
        xtick={0,1/4,1/3,1/2,2/3,3/4,9/10,1},
        xticklabels={$0$,$\frac{m}{4}$,$\frac{m}{3}$,$\frac{m}{2}$,$\frac{2m}{3}$,$\frac{3m}{4}$,$\frac{9m}{10}$,$m-1$},
        ytick={70,75,80,85},
        ymajorgrids=true,
        grid style=dashed,
        no markers,
        width=1.1*\linewidth, 
        height=0.46*\linewidth, 
    ]
    \addplot+[only marks, fill=blue, mark=*, mark size=0.12em, thick, draw=none] 
        plot[error bars/.cd, y dir=both, y explicit, error bar style={}] 
        coordinates {
        (0,81.35) +-(0,0.3)
        (1/4,81.09) +-(0,0.4)
        (1/3,81.14) +-(0,0.3)
        (1/2,81.0) +-(0,0.2)
        (2/3,81.11) +-(0,0.5)
        (3/4,80.81) +-(0,0.4)
        (9/10,80.56) +- (0,0.3)
        (1,75.48) +-(0,0.6)
        }; \label{plot_one}
    \end{axis}

    \begin{axis}[
        ylabel={Noise \enx{\sigma}},
        xmin=0, xmax=1,
        ymin=0, ymax=30,
        axis y line*=right,
        axis x line=none,
        ytick={0,3,10,20,30},
        legend pos=outer north east,
        legend style={at={(1.15,0.75)},anchor=north west},
        ymajorgrids=true,
        grid style=dashed,
        width=1.1*\linewidth, 
        height=0.46*\linewidth, 
    ]
    \addplot+[color=red, mark=triangle*, mark options={fill=red},thick] 
        coordinates {
        (0,2.83) 
        (1/4,3.26) 
        (1/3,3.46) 
        (1/2,4.00) 
        (2/3,4.90) 
        (3/4,5.48) 
        (9/10, 8.93) 
        (1,28.3) 
        };
        \addlegendentry{$\sigma$}
        \addlegendimage{/pgfplots/refstyle=plot_one}\addlegendentry{Accuracy}
    \addplot+[color=green,thick, dotted, no markers] 
        coordinates {
        (0,3.0)
        (1/4,3.0)
        (1/3,3.0)
        (1/2,3.0)
        (2/3,3.0)
        (3/4,3.0)
        (1,3.0)
        };
        \addlegendentry{$K$}
    \end{axis}
    \end{tikzpicture}
    \vspace{-2.5em}
    \caption{Accuracy of \ol with \partnoise under varying collusion thresholds (\enx{t}). Simulation over 300 training epochs with $\eps=1$, $\eta=0.15$, $K=3.00$, $B=836$, $m=100$, with Gaussian noise (starting from $\sigma=2.83$).}
    \label{fig:tradeoff-collusion}
    \vspace{-2.0em}
\end{figure}
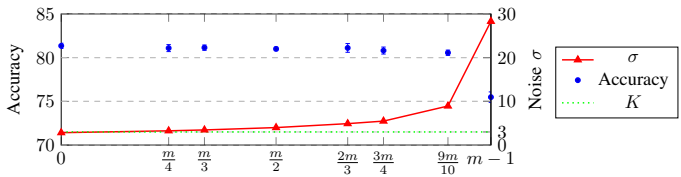

\section{Privacy Attacks in \cl}\label{app:privacyattacks}
Next, we discuss privacy risks of 
cryptography-only CL in various threat models, focusing on membership inference (MIA), gradient inversion (GIA), and DP noise tampering attacks.
We overview DP and cryptographic mitigations, while addressing DP side-channels.
We focus on attribute inference attacks in App.~\ref{app:attributeinference}.

\para{Threat Models} 
\ol operates in black-box settings, where adversaries access only model outputs and datasets of corrupted \clset, since \slset train on \encrypted data. 
Here, active attackers can tamper with local datasets and manipulate noise protocols. 
\fl operates in white-box settings, where \clset and \slset access model parameters and gradients. Here, an active attacker can also alter gradients.
Next, we discuss black-box attacks and then risks of revealing gradients.

\para{MIA}
Membership inference attacks (MIA) aim to infer whether a specific record (or user) was part of the training data by exploiting differences in model behavior, e.g., confidence scores \cite{shokri2017membership}. 
%
Black-box attacks typically use a shadow dataset (i.e., similar in distribution to the target dataset) to train an attack model to distinguish members \cite{nasr2019comprehensive}.
White-box access to gradient patterns can improve attack accuracy by up to \enx{7\%} \cite{nasr2019comprehensive}.
%
%
Active adversaries can perform \emph{poisoning MIA} (PMIA) to amplify MIA's effects by corrupting target models \cite{zhang2023agrevader}. 
In \ot, \clset can inject mislabelled target samples, causing the model to treat correctly labeled samples as outliers. 
Reportedly, \enx{8} poisoned samples in CIFAR-10 can improve MIA's true positive rate by $52\pp$ \cite{tramer2022truth}.
In \fl, corrupted \clset can alter local parameters to increase the loss on a target record \enx{x}. If another client has \enx{x}, the tampered local SGD sharply reduces its gradient norm \cite{nasr2019comprehensive, zhang2023agrevader}, boosting attack accuracy by $10\pp$ over black-box \cite{zhang2023agrevader}. 
Poisoning can also embed privacy-backdoors in pretrained models to facilitate MIA. During pretraining, models memorize target samples, and reinforce or forget them in fine-tuning, i.e., depending on their presence in the victim's dataset \cite{wen2024privacy}.

\para{Mitigating MIA/PMIA via DP and Cryptography}
DP-SGD effectively mitigates MIA and PMIA by bounding the influence of any data point, even a poisoned one, on the model \cite{tramer2022truth, aerni2024evaluations}.
For example, a small amount of noise (e.g., \enx{\eps > 5000}, \enx{K = 10}) reduces MIA effectiveness to random guessing \cite{carlini2022membership}. 
However, mitigating PMIA requires stricter privacy guarantees leading to accuracy drop, \eg{by \enx{15\pp} with 1000 poisoned samples (\enx{\eps \approx 3}) \cite{chen2022amplifying}}.
DP-SGD is also effective against privacy-backdoors, but requires strong privacy guarantees (\enx{\eps \le 8}) \cite{feng2024privacy}. 
%
While relying solely on DP introduces a privacy-accuracy trade-off, {augmenting DP with cryptography can enhance model utility and robustness}. 
In \fl, zero-knowledge proofs (ZKP, App.~\ref{app:zkp}) can ensure gradients are within valid ranges without revealing the values \cite{evans2018pragmatic}. 
Alternatively, robust secure aggregation identifies malicious updates, down-weighting them during aggregation; e.g., CAFCOR \cite{allouah2025towards} identifies updates that disproportionately increase the collective variance. 
In \ot, \sls can implement MPC variants of outlier detection, regularization techniques \cite{chen2022amplifying}, or poisoning-resistant training algorithms \cite{chaudhari2022safenet}.

\para{Noise Tampering and Cryptographic Mitigations}
Active adversaries can manipulate noise protocols, weakening DP guarantees.  
Non-oblivious \partnoise approaches pre-adjust variance to account for up to \enx{t} zero-noise additions (Sec.~\ref{subsec:partialnoise}), at utility cost (Tab.~\ref{tab:accuracy}).
To preserve model utility and mitigate noise tampering {oblivious protocols need to be combined with verifiable noise sampling}.
Verifiable sampling 
requires \cls (or \sls) to provide proofs of correct sampling of \partnoise (or \centrnoise), \eg{via ZKPs \cite{sabater2022private}}.
However, cryptographic proofs introduce significant overhead, e.g. up to hours to prove one sample via {\BoxMuller} \cite{sabater2022private}. 
Discrete sampling methods incur even higher costs due to their iterative nature.
For distributed \centrnoise sampling, maliciously secure protocols ensure correct computation but incur additional overhead, e.g., up to \enx{7\times} slower than semi-honest in a WAN (Tab.~\ref{tab:eval}). 
Verifiable and oblivious noise selection can be achieved via verifiable shuffling \cite{froelicher2020drynx}, or ZKPs. In client-side protocols, ZKPs ensure selection of one sample from a proposed list, while in server-aided protocols, ZKPs verify that exactly one sample per client has been selected \cite{bindschaedler17star}.

\para{DP Side-Channels and Cryptographic Mitigations}
DP implementations can introduce side-channels undermining DP guarantees. 
Sampling continuous noise on finite-precision machines can lead to \emph{floating-point attacks} \cite{jin2022we, mironov2012significance}. 
The intuition is that finite-precision machines can not represent all possible values, revealing holes in distributions \cite{tumult22}. As a result, some DP outputs occur only for certain inputs. 
Similarly, \emph{timing attacks} infer the noise magnitude from the sampling runtime \cite{jin2022we}. 
Fixed-point and quantized integer representations typically used in {cryptographic protocols for DP inherently mitigate DP side-channels} \cite{mironov2012significance}.
Thus, \epcl works are secure against those attacks. 
Furthermore, constant-time MPC protocols for \centrnoise \cite{iwahana2022spgc,chase2017,jayaraman2019evaluating} mitigates timing attacks, by avoiding data-dependent leaks (Sec.~\ref{subsec:distributed-noise-sampling}). 
Caching offline-sampled noise is a possible mitigation for \partnoise and centralized \centrnoise. 
Moreover, \epcl samples multiple noise values per epoch, attackers only observe cumulative times, reducing attacks' effectiveness \cite{jin2022we}.

\para{GIA in \fl and Cryptographic Mitigations}
White-box access enables gradient inversion attacks, where adversaries reconstruct training samples by optimizing a loss function to match observed gradients \cite{wu2023learning}.
Furthermore, an active attacker who corrupts the \slong can recover training samples by injecting convolutional layers that enable separation of client gradients after aggregation \cite{zhao2023loki}.
While DP mitigates MIA, GIAs can still be feasible even with LDP \cite{wu2023learning}: 
a corrupted \slong can reconstruct training samples with a model trained on auxiliary data (\enx{\eps \approx 10} and \enx{n \le 4}). 
Notably, large \enx{\eps \approx 10^3} is ineffective against a malicious \slong, while small \enx{\eps \approx 0.1} reduce the success to near-random guess \cite{zhao2023loki}. 
To mitigate GIA, \fl can leverage cryptographic techniques to guarantee gradient and model secrecy against \slset. HE with a single server \cite{sebert2023combining} or SS with multiple servers ensure that \slset operate on encrypted data, and only \clset learn gradient updates. 
This guarantee can be extended to active adversaries by using ZKPs to prove correct aggregation \cite{froelicher2020drynx}. 

\subsection{Attribute Inference Attacks}\label{app:attributeinference}


In attribute inference attacks (AIA), an adversary with partial knowledge of a record attempts to infer unknown attributes using a model trained on that record \cite{jayaraman2022attributeinferenceattacksjust}. Unlike MIA, which determine whether a record was in the training set, AIAs leverage statistical correlations in the data to infer sensitive attributes \cite{jayaraman2022attributeinferenceattacksjust}. 
AIAs pose an additional risk as ML models can learn unintended information beyond their original task. For example, a model trained to predict age from profile photos may inadvertently learn to infer race \cite{song2019overlearning, liu2022ml}. These attacks are particularly effective in white-box settings, where the adversary has not only access to model parameters, but also to intermediate representations of a target sample, e.g., embeddings \cite{liu2022ml}. For example, in scenarios where the model is split into embeddings and a classifier, the clients can locally compute embeddings and send the result to a server for classification. However, the adversary can exploit the embeddings to launch an AIA, as the embeddings may over-learn sensitive features during training~\cite{song2019overlearning}.

\para{Discussion on DP mitigations for AIA}
Theoretically, DP protects against AIA, as the DP adversary is stronger and assumes access to more information than typical AIA adversaries, who only have partial knowledge of a record \cite{salem2023sok}. Specifically, a DP adversary seeks to distinguish between models trained on neighboring datasets (i.e., differing in at most a record/user), while an AIA adversary uses partial knowledge of a record to infer unknown attributes.

Empirical evaluations show that black-box AIAs rarely learn more than an adversary could infer from prior knowledge alone. However, white-box AIAs, where an attacker has access to model parameters, can identify records with a sensitive attribute more effectively \cite{jayaraman2022attributeinferenceattacksjust}. DP does not directly mitigate this risk because its guarantees focus on record- or user-level indistinguishability rather than protecting against statistical leakage from the distribution itself. In fact, DP aims to maintain population-level information while protecting individual records or users.
Furthermore, while DP noise is injected into gradients to reduce the impact of a single record, the noise can also amplify correlations between neurons and sensitive attributes in some cases \cite{jayaraman2022attributeinferenceattacksjust}. Thus, while strong privacy guarantees (\enx{\eps < 1}) may mitigate AIAs to some extent \cite{fredrikson2014privacy}, 
DP does not inherently prevent attackers from leveraging the statistical properties of the training distribution, since learning those properties is the goal of ML model training.


\para{Cryptographic mitigations for white-box AIA}
The use of cryptography in \epcl inherently mitigates white-box AIA based on intermediate model representations. 
The model trained in \ol can be kept encrypted during inference, preventing the adversary from accessing intermediate representations. In \fl, clients have access to local models but do not know the embeddings of other clients, preventing white-box AIA attacks \cite{liu2022ml}.
In settings where the model is split into embeddings and a classifier, the clients can encrypt the embeddings before sending them to the server for classification. This prevents the adversary from exploiting the embeddings to launch an AIA, as the embeddings are encrypted and the server works over encrypted data and model. 

\xx{
In an attribute inference attack, an adversary has partial knowledge
of some training records and access to a model trained on those
records, and infers the unknown values of a sensitive feature of
those records \cite{jayaraman2022attributeinferenceattacksjust}.

Previous attribute inference methods do not reveal
more about the training data from the model than can be inferred by
an adversary without access to the trained model, but with the same
knowledge of the underlying distribution as needed to train the
attribute inference attack; (2) black-box attribute inference attacks
rarely learn anything that cannot be learned without the model; but
(3) white-box attacks, which we introduce and evaluate in the paper,
can reliably identify some records with the sensitive value attribute
that would not be predicted without having access to the model.
Furthermore, we show that proposed defenses such as differentially
private training and removing vulnerable records from training
do not mitigate this privacy risk \cite{jayaraman2022attributeinferenceattacksjust}.

Differential privacy mechanisms typically limit the exposure of individual records by introducing noise in the training process. Differential privacy has been shown to both theoretically bound and experimentally mitigate membership inference risks. However, the theoretical guarantees provided by differential privacy re at the level of individual records and not at the level of statistical properties in a distribution.
Differential privacy done at the level of individual records can provide a bound on the difference in inference
accuracy between training and non-training records, but provides
no assurances about inferences regarding the ability of an adversary
to infer an attribute from distributional information leaked by the
model. While the noise distorts the model parameters to reduce the
impact of any single training record, it may inadvertently increase
the correlation of a subset of neurons to the sensitive value.
Differential privacy ensures that an individual training record cannot be
distinguished from a non-training record, but provides no bound
on inferences made based on statistical information revealed about
the training distribution \cite{jayaraman2022attributeinferenceattacksjust}.

}

\section{Summary of Research Directions and Observations}
\label{app:obs_mapping}

Tab.~\ref{tab:research_directions} summarizes the mapping between the key observations identified in our systematization and the corresponding research directions proposed in Sec.~\ref{sec:observation}.
\begin{table*}[tb]
    \centering
    \def\colwres{0.4\textwidth}
    \def\colwobs{0.5\textwidth}
    
    \scriptsize
    \renewcommand{\arraystretch}{1.4} 
    \setlength{\tabcolsep}{1em}
    \caption{Mapping of Key Observations to Future Research Directions.} 
    \label{tab:research_directions}

    \begin{tabular}{p{\colwres} p{\colwobs}}
    \textbf{Research Direction} & \textbf{Driving Observation(s)} \\ \midrule
    
    \researchmanual{1}{}Enhance privacy and performance via pre-processing.
    & 
    \obsmanual{3}{}Client local pre-processing in OL reduces cryptographic overhead while maintaining high accuracy. \\ \midrule
    
    \researchmanual{2}{}Provide cryptographic building blocks for DP in OL.
    & 
    \obsmanual{1}{}OL inherently guarantees gradient and model secrecy against both clients and servers, preventing GIA. \par\medskip
    \obsmanual{2}{}There are no secure per-example clipping optimizations via cryptographic protocols. \par\medskip
    \obsmanual{9}{}Efficient and accurate CPCL solutions require careful design considerations for learning paradigms, cryptographic protocols, and noise sampling techniques \\ \midrule
    
    \researchmanual{3}{}Develop crypto-friendly discrete CNoise sampling.
    & 
    \obsmanual{6}{}The choice of noise distribution significantly affects utility and performance \par\medskip
    \obsmanual{8}{}Sampling algorithms for discrete distributions have non-constant runtimes. \\ \midrule
    
    \researchmanual{4}{}Embed DP outside Perturb. 
    & 
    \obsmanual{7}{}DP can be embedded in phases besides \perturb. \\ \midrule
    
    \researchmanual{5}{}Propose strong user-level DP algorithms for multi-user datasets. 
    & 
    \obsmanual{4}{}No established best practice exists for user-level DP in multi-user datasets. \\ 
    \end{tabular}
\end{table*}

}

\end{document}